\documentclass[aps,prd,reprint,superscriptaddress,preprintnumbers,nofootinbib,floatfix,longbibliography]{revtex4-2}

\usepackage[T1]{fontenc}
\usepackage[utf8]{inputenc}
\usepackage[english]{babel}
\usepackage{amsmath,amsthm,amssymb,amsfonts,mathrsfs,amsbsy,bm}
\usepackage{tensor}
\usepackage{slashed}
\usepackage{esint}
\usepackage{capt-of}
\usepackage[a4paper, margin=1.2cm]{geometry}
\usepackage{cancel}
\usepackage{graphicx}
\usepackage{multirow}
\usepackage{array}
\usepackage{booktabs}
\usepackage{makecell}
\usepackage{xcolor}
\colorlet{BLUE}{blue}
\usepackage{tikz}
\usetikzlibrary{quotes,angles,arrows,decorations.markings,decorations.pathmorphing}
\usepackage{hyperref}
\hypersetup{colorlinks=true,breaklinks=true,citecolor=blue,linkcolor=[rgb]{0,0.5,0.9},urlcolor=blue}
\graphicspath{{figures/}}

\newcommand{\be}{\begin{equation}}
\newcommand{\ee}{\end{equation}}
\newcommand{\Be}{\begin{eqnarray}}
\newcommand{\Ee}{\end{eqnarray}}

\newcommand{\mincir}{\raise-3.truept\hbox{\rlap{\hbox{$\sim$}}\raise4.truept\hbox{$<$}\ }}
\newcommand{\magcir}{\raise-3.truept\hbox{\rlap{\hbox{$\sim$}}\raise4.truept\hbox{$>$}\ }}

\providecommand{\U}[1]{}
\newcommand{\ie}{\begin{equation}}
\newcommand{\fe}{\end{equation}}
\newcommand{\se}{\begin{eqnarray}}
\newcommand{\ff}{\end{eqnarray}}

\begin{document}
\emergencystretch=4em

%%%%%%%%%%%%%%%%%%%%%%%%%%%%%%%%%%%%%%%%%%%%%%%%%%%%%%%%%%%%%%%%%%%%%%%%%%%%%%%%%%%%%%%%%%%%%%%%%%%%%%%%%%%%%%%%%%%%%%%%%%%%%%%%%%%%%%%%%%%%%%%%%%%%%%%%%%%%%%%%%%%%%%%%%%%%%%%%%%%%%%%%%%%%%%%%%%%%%%%%%%%%%%%%%%%%%%%%%%%%%%%%%%%%%%%%%%%%%%%%%%%%%%%%%%%%%%%%%%%%%%%%%%%%%%%%%%%%%%%%%%%%%%%%%%%%%%%%%%%%%%%%%%%%%%%%%%%%%%%%%%%%%%%%%%%%%%%%%%%%%%%%%%%%%%%%%%%%%%%%%%%%%%%%%%%%%%%%%%%%%%%%%%%%%%%%%%%%%%%%%%%%%%%%%%%%%%%%%%%%%%%%%%%%%%%%%%%%%%%%%%%%%%%%%%%%%%%%%%%%

\title{Quantum geometric signatures in neutrino dynamics around a holonomy black hole}

\author{A. A. Ara\'{u}jo Filho}
\email{dilto@fisica.ufc.br}
\affiliation{Departamento de F\'isica, Universidade Federal da Para\'iba, Caixa Postal 5008, 58051--970, Jo\~ao Pessoa, Para\'iba, Brazil.}
\affiliation{Departamento de F\'isica, Universidade Federal de Campina Grande, Caixa Postal 10071, 58429--900 Campina Grande, Para\'iba, Brazil.}
\affiliation{Center for Theoretical Physics, Khazar University, 41 Mehseti Street, Baku, AZ-1096, Azerbaijan.}

%%%%%%%%%%%%%%%%%%%%%%%%%%%%%%%%%%%%%%%%%%%%%%%%%%%%%%%%%%%%%%%%%%%%%%%%%%%%%%%%%%%%%%%%%%%%%%%%%%%%%%%%%%%%%%%%%%%%%%%%%%%%%%%%%%%%%%%%%%%%%%%%%%%%%%%%%%%%%%%%%%%%%%%%%%%%%%%%%%%%%%%%%%%%%%%%%%%%%%%%%%%%%%%%%%%%%%%%%%%%%%%%%%%%%%%%%%%%%%%%%%%%%%%%%%%%%%%%%%%%%%%%%%%%%%%%%%%%%%%%%%%%%%%%%%%%%%%%%%%%%%%%%%%%%%%%%%%%%%%%%%%%%%%%%%%%%%%%%%%%%%%%%%%%%%%%%%%%%%%%

\date{\today}

%%%%%%%%%%%%%%%%%%%%%%%%%%%%%%%%%%%%%%%%%%%%%%%%%%%%%%%%%%%%%%%%%%%%%%%%%%%%%%%%%%%%%%%%%%%%%%%%%%%%%%%%%%%%%%%%%%%%%%%%%%%%%%%%%%%%%%%%%%%%%%%%%%%%%%%%%%%%%%%%%%%%%%%%%%%%%%%%%%%%%%%%%%%%%%%%%%%%%%%%%%%%%%%%%%%%%%%%%%%%%%%%%%%%%%%%%%%%%%%%%%%%%%%%%%%%%%%%%%%%%%%%%%%%%%%%%%%%%%%%%%%%%%%%%%%%%%%%%%%%%%%%%%%%%%%%%%%%%%%%%%%%%%%%%%%%%%%%%%%%%%%%%%%%%%%%%%%%%%%%%%%%%%%%%%%%%%%%%%%%%%%%%%%%%%%%%%%%%%%%%%%%%%%%%%%%%%%%%%%%%%%%%%%%%%%%%%%%%%%%%%%%%%%%%%%%%%%%%%%%%%%%%%%%%%%%%%%%%%%%%%%%%%%%%%%%%%%%%%%%%%%%%%

\begin{abstract}

We investigate three flavor neutrino phenomena in an effective holonomy corrected Schwarzschild geometry. We derive the weak--deflection angle through second post--Minkowskian order, obtain the semiclassical phases along radial and nonradial trajectories, and formulate the two-image flavor probability by including magnifications, Fermat phases, and wave packet overlap. The cross path phase acquires a logarithmic holonomy contribution and can retain information about the absolute neutrino mass scale. We also characterize flavor mode correlations and trajectory--flavor entanglement through the balance among path predictability, interference visibility, and I-concurrence. Neutrino--antineutrino annihilation outside an effective neutrinosphere is examined, with holonomy entering the integrated power through the radial proper volume measure. For a solar mass lens and $10\,\mathrm{MeV}$ neutrinos, the numerical results show displaced oscillation fringes and a geometry dependent redistribution among the electron, muon, and tau channels. The electron--neutrino trajectory--flavor concurrence reaches approximately $0.23$, with an entropy near $0.10$ bit, while remaining weakly sensitive to the mass ordering in the selected configuration. By contrast, the annihilation power increases monotonically, reaching an increase of approximately $39.3\%$ relative to Schwarzschild spacetime (for a particular configuration, i.e., $\lambda=2$ and $M/R_{\nu}=1/3$). Furthermore, at fixed source parameters and $M/R_\nu=1/3$, an assumed maximum excess of $10\%$ in the integrated annihilation power relative to Schwarzschild yields the conditional bounds $a/(2M)\leq0.2879$ and $|\lambda|\leq0.6358$.

\end{abstract}

\maketitle

\tableofcontents

%%%%%%%%%%%%%%%%%%%%%%%%%%%%%%%%%%%%%%%%%%%%%%%%%%%%%%%%%%%%%%%%%%%%%%%%%%%%%%%%%%%%%%%%%%%%%%%%%%%%%%%%%%%%%%%%%%%%%%%%%%%%%%%%%%%%%%%%%%%%%%%%%%%%%%%%%%%%%%%%%%%%%%%%%%%%%%%%%%%%%%%%%%%%%%%%%%%%%%%%%%%%%%%%%%%%%%%%%%%%%%%%%%%%%%%%%%%%%%%%%%%%%%%%%%%%%%%%%%%%%%%%%%%%%%%%%%%%%%%%%%%%%%%%%%%%%%%%%%%%%%%%%%%%%%%%%%%%%%%%%%%%%%%%%%%%%%%%%%%%%%%%%%%%%%%%%%%%%%%%%%%%%%%%%%%%%%%%%%%%%%%%%%%%%%%%%%%%%%%%%%%%%%%%%%%%%%%%%%%%%%%%%%%%%%%%%%%%%%%%%%%%%%%%%%%%%%%%%%%%%%%%%%%%%%%%%%%%%%%%%%%%%%

\section{Introduction }

The holonomy corrected Schwarzschild black hole offers a particularly instructive setting for examining how quantum geometry can influence gravitational phenomena beyond the region where the classical singularity is resolved. Its construction is motivated by loop quantum gravity, in which holonomies of the connection and fluxes of the densitized triad provide the elementary variables of the quantum description~\cite{AshtekarBianchi2021,Perez2017}. Earlier investigations of spherical black holes established different realizations of singularity resolution within this framework~\cite{GambiniPullin2013,AshtekarOlmedoSingh2018}. Among the effective geometries subsequently proposed, the model developed by Alonso-Bardaj\'i, Brizuela, and Vera is distinguished by a closed constraint algebra and a covariant interpretation of the resulting spacetime~\cite{AlonsoBardaji2022PLB,AlonsoBardaji2022PRD}. The Schwarzschild singularity is replaced by a regular spacelike surface of minimum area, through which the trapped interior continues into an antitrapped region. The same construction has also been extended to charged configurations in the presence of a cosmological constant~\cite{AlonsoBardaji2023}.

An appealing feature of the uncharged solution is the selective manner in which the correction enters its exterior geometry. The temporal metric component retains its Schwarzschild form, whereas the radial component acquires an additional factor controlled by the holonomy parameter. As a consequence, the horizon radius, the photon sphere radius, and the critical impact parameter preserve their classical expressions at fixed Schwarzschild parameter, even though the radial evolution of a ray is modified~\cite{AlonsoBardaji2022PRD,Soares2023}. This separation gives physical significance to observables accumulated along a trajectory.

The gravitational signatures of this solution have already been investigated from several complementary perspectives. The scalar quasinormal spectrum exhibits changes in the oscillation frequencies and damping rates, together with a nontrivial dependence of the overtones on the holonomy correction~\cite{Moreira2023}. Subsequent studies have considered scalar, electromagnetic, and Dirac perturbations, including massive fields with arbitrarily long lived modes~\cite{Bolokhov2024}. Analyses of axial gravitational perturbations have further addressed the corresponding quasinormal frequencies, greybody factors, and Hawking emission~\cite{YangEtAl2024}. More recently, calculations of transmission and absorption for several field spins have shown how the deformation affects the radiative response, with the reduced Hawking temperature suppressing the total emission in the configurations examined~\cite{LutfuogluEtAl2026}.

Furthermore, the gravitational lensing supplies a direct connection between these geometric modifications and the apparent properties of a distant source \cite{Filho:2024tgy,Filho:2024zxx,Filho:2024isd,Ahmed:2025cuz,Al-Badawi:2025ipr,Ahmed:2025did,Ahmed:2025vww}. For the holonomy corrected Schwarzschild solution, both weak and strong deflection analyses have established changes in image positions, magnifications, angular separations, and relative arrival times~\cite{Soares2023}. Such calculations draw on the general theory of gravitational lenses and on systematic treatments of the strong deflection limit~\cite{SchneiderEhlersFalco1992,VirbhadraEllis2000,Bozza2002,Tsukamoto2017}, while the optical geometry provides an alternative description through its curvature~\cite{GibbonsWerner2008,Qiao:2022hfv,Qiao:2022jlu}. Investigations of related geometries have also included topological charge and a surrounding cloud of strings, with the latter extending the discussion to accretion disk emission~\cite{SoaresEtAl2024Topology,AhmedKala2025}.

On the other hand, neutrinos introduce another physical ingredient into this discussion: the signal transported along a gravitationally deflected trajectory possesses an internal quantum structure. The connection between flavor and mass, anticipated in the early development of neutrino mixing~\cite{Pontecorvo1968,MakiNakagawaSakata1962}, is supported by atmospheric, solar, and reactor observations~\cite{FukudaEtAl1998,AhmadEtAl2002,AnEtAl2012}. Within the three flavor description, the interference among mass eigenstates determines the probability of detecting an electron, muon, or tau neutrino, with the mixing parameters constrained through global analyses~\cite{EstebanEtAl2024}. Ordinary vacuum oscillations depend on differences of squared masses, while their coherent description requires that production and detection do not distinguish the participating mass components~\cite{GrossmanLipkin1997,AkhmedovSmirnov2009}. Their weak interactions and sensitivity to accumulated phases make neutrinos a natural subject for studying propagation in curved spacetime.

The relevance of neutrinos near compact objects also extends to energy transfer. Hot accretion flows can radiate substantial neutrino luminosities, and the reaction $\nu+\bar\nu\rightarrow e^-+e^+$ can deposit part of this energy into a pair plasma outside the emitting region~\cite{PophamEtAl1999}. Relativistic calculations have established the roles of trajectory bending, gravitational redshift, and source geometry in determining this deposition~\cite{SalmonsonWilson1999,AsanoFukuyama2000,BirklEtAl2007}. The process has consequently been examined in modified gravitational backgrounds~\cite{LambiaseMastrototaro2020}, including studies that relate neutrino phenomena to other properties of noncommutative \cite{AraujoFilho:2024mvz,AraujoFilhoEtAl2026Neutrinos} and other Lorentz--violating black holes \cite{Shi:2025rfq,Shi:2025plr,Shi:2025ywa}. In this manner, for a geometry whose temporal and radial components respond differently to the correction, it is particularly useful to distinguish the local reaction rate from the power obtained after integration over proper volume.

The gravitational influence on flavor evolution is described through the phase accumulated by each mass eigenstate between emission and detection. The covariant treatment of matter wave phases~\cite{Stodolsky1979} motivated early studies of neutrino oscillations in gravitational fields~\cite{AhluwaliaBurgard1996,CardallFuller1997,FornengoEtAl1997}. These investigations also brought attention to the comparison of phases at common spacetime events and to the distinction between coordinate distances and locally measured quantities~\cite{BhattacharyaHabibMottola1999}. Here, we consider minimal coupling to a static effective metric and propagation through a vacuum exterior. Coherent forward scattering in matter, responsible for an additional modification of flavor evolution~\cite{Wolfenstein1978,MikheyevSmirnov1985}, is excluded from the propagation model. The holonomy dependence enters through the spacetime geometry traversed by the neutrino.

When a gravitational lens connects the source and detector through more than one trajectory, flavor oscillations become part of a richer interference problem. The amplitudes associated with different mass eigenstates must then be combined with those associated with the gravitational images~\cite{CrockerGiuntiMortlock2004,AlexandreClough2018}. Magnifications determine their relative weights, while travel times and image parity contribute to the phases familiar from wave optics~\cite{NakamuraDeguchi1999}. A notable consequence is that interference between unequal paths can retain dependence on the individual neutrino masses, even when ordinary vacuum oscillations along a single path are sensitive only to squared mass differences~\cite{SwamiEtAl2020}. Related investigations in deformed geometries have shown how this dependence is accompanied by a response to the spacetime parameters~\cite{ChakrabartyEtAl2022,AraujoFilhoEtAl2026Neutrinos}.

The survival of this interference nevertheless depends on the physical preparation and measurement of the signal. Wave packet separation, unequal image arrival times, and averaging over the source or the detected energy can suppress the terms coupling distinct propagation alternatives~\cite{GiuntiKimLee1998,AkhmedovSmirnov2009,SwamiEtAl2021}. Studies of neutrino wave packets in curved spacetime have made explicit the influence of gravity on the coherence conditions~\cite{ChatelainVolpe2020,SwamiEtAl2021}.

In addition, quantum information gives a further description of the same propagating state. When the Hilbert space is partitioned according to flavor occupation, a single neutrino in a coherent flavor superposition can exhibit entanglement among its occupation modes~\cite{BlasoneEtAl2008,BlasoneEtAl2009}. This interpretation has been developed within both quantum mechanical and field theoretical treatments of mixing~\cite{BlasoneEtAl2014}, and extended to the analysis of three flavor correlations~\cite{BanerjeeEtAl2015,BittencourtEtAl2024}. Coherence and entanglement capture different properties of the density operator, with their meaning fixed by the chosen basis and partition~\cite{HorodeckiEtAl2009,BaumgratzCramerPlenio2014,StreltsovEtAl2017}. Investigations in Schwarzschild spacetime have already explored gravitational modifications of neutrino coherence, entanglement, and nonclassical correlations~\cite{EttefaghiEtAlGravity,WangEtAl2024Gravity}.

Lensing also allows a distinct partition in which the trajectory and flavor are retained as separate degrees of freedom. If the two images carry different internal flavor states, their coherent superposition can exhibit entanglement between the path and flavor sectors. The associated reduction of interference visibility is naturally related to path predictability and to the distinguishability of the internal states~\cite{Englert1996,RungtaEtAl2001,JakobBergou2007}. This concerns the degrees of freedom of one particle, as in other forms of single particle interferometry\footnote{It must be distinguished from flavor occupation entanglement and from correlations between separate neutrinos. In the coherent limit considered here, this distinction also has a useful consequence: the lensed flavor probability can depend on the absolute mass scale, whereas pure state entanglement between trajectory and flavor is invariant under a common shift of all squared masses.}~\cite{HasegawaEtAl2003}.

Motivated by these developments, we investigate neutrino dynamics in the holonomy corrected Schwarzschild exterior within a common geometric description. We obtain the weak deflection angle through second order in the gravitational strength, evaluate the semiclassical phases along radial and nonradial trajectories, and construct the lensed three flavor probability with image magnifications, Fermat phases, and wave packet overlap. The analysis follows the resulting flavor mode correlations and the entanglement between trajectory and flavor, together with neutrino pair energy deposition outside an effective neutrinosphere. Our purpose is to establish which features arise from the modified radial geometry, how they enter the different neutrino observables, and under which coherence assumptions the associated interference can survive. The comparison with Schwarzschild spacetime also allows us to examine the degeneracies with source, lens, and neutrino parameters that must be controlled before these effects can be used to constrain the holonomy correction.

%%%%%%%%%%%%%%%%%%%%%%%%%%%%%%%%%%%%%%%%%%%%%%%%%%%%%%%%%%%%%%%%%%%%%%%%%%%%%%%%%%%%%%%%%%%%%%%%%%%%%%%%%%%%%%%%%%%%%%%%%%%%%%%%%%%%%%%%%%%%%%%%%%%%%%%%%%%%%%%%%%%%%%%%%%%%%%%%%%%%%%%%%%%%%%%%%%%%%%%%%%%%%%%%%%%%%%%%%%%%%%%%%%%%%%%%%%%%%%%%%%%%%%%%%%%%%%%%%%%%%%%%%%%%%%%%%%%%%%%%%%%%%%%%%%%%%%%%%%%%%%%%%%%%%%%%%%%%%%%%%%%%%%%%%%%%%%%%%%%%%%%%%%%%%%%%%%%%%%%%%%%%%%%%%%%%%%%%%%%%%%%%%%%%%%%%%%%%%%%%%%%%%%%%%%%%%%%%%%%%

\section{The holonomy corrected Schwarzschild geometry and weak gravitational lensing}
\label{sec:holonomy_geometry}

Before addressing neutrino oscillations, we establish the geometric quantities that determine both the propagation phase and the weak gravitational deflection. In loop quantum gravity, holonomies replace the connection as elementary variables, and effective constructions based on them provide a controlled way of investigating quantum geometric corrections to classical black hole spacetimes \cite{AshtekarBianchi2021,Perez2017}. The background considered here follows from a holonomy modification of the spherically symmetric Hamiltonian of general relativity. The constraints are combined so that their algebra remains anomaly free, which permits the resulting phase space solution to be interpreted as a covariant spacetime geometry \cite{AlonsoBardaji2022PLB,AlonsoBardaji2022PRD}.

In Schwarzschild--like coordinates, the exterior line element is written as \cite{AlonsoBardaji2022PLB,AlonsoBardaji2022PRD}
\begin{equation} 
\mathrm{d}s^{2} = -F(r)\,\mathrm{d}t^{2}+\frac{\mathrm{d}r^{2}}{F(r)H(r)} + r^{2}\left(\mathrm{d}\vartheta^{2}+\sin^{2}\!\vartheta\,\mathrm{d}\varphi^{2}\right), 
\label{eq:holonomy_metric} 
\end{equation}
where the two metric functions and the holonomy length are defined  by
\begin{equation} 
F(r)=1-\frac{2M}{r}, \quad H(r)=1-\frac{a}{r}, \quad a\equiv r_{0}=\frac{2M\lambda^{2}}{1+\lambda^{2}}. 
\label{eq:holonomy_functions} 
\end{equation}
Here, $M>0$ denotes the ADM mass and $\lambda$ is the dimensionless polymerization parameter. It is useful to introduce the bounded quantity
\begin{equation} 
\delta\equiv\frac{a}{2M}=\frac{\lambda^{2}}{1+\lambda^{2}}, \qquad 0\leq\delta<1,
\label{eq:delta_definition} 
\end{equation}
in terms of which the Schwarzschild geometry is recovered for $\delta\rightarrow0$. Although the temporal component remains identical to its Schwarzschild counterpart, the radial component contains the additional factor $H(r)$. In other words, as it is straightforward to see, quantities determined exclusively by $g_{tt}$ preserve their classical form, whereas radial distances, propagation phases, and deflection angles retain an explicit dependence on the holonomy parameter.

On the other hand, the event horizon remains at
\begin{equation} 
r_{h} = 2M, 
\label{eq:event_horizon} 
\end{equation}
while the hypersurface $r=a$ always lies in the black hole interior because $a<2M$. Rather than representing a curvature singularity, this surface is a regular spacelike transition surface of minimum area connecting trapped and antitrapped regions. The maximal extension contains a black hole/white hole interior bounded by two asymptotically flat exterior regions, with the curvature invariants remaining finite throughout the geometry \cite{AlonsoBardaji2022PLB,AlonsoBardaji2022PRD}. The exterior causal boundary is, on the other hand, unchanged, but the radial sector already carries the information associated with the quantum correction, which can influence both field propagation and optical observables \cite{Moreira2023,Soares2023}.

The neutrinos studied below are assumed to be ultrarelativistic. Their nonzero masses determine the relative oscillation phases, while the mass dependent deviations of their trajectories are suppressed by $m_i^{2}/E^{2}$. In this manner, to the leading order relevant here, the different mass eigenstates propagate along a common null trajectory of the geometry in Eq.~\eqref{eq:holonomy_metric}. Restricting the motion to the equatorial plane, $\vartheta=\pi/2$, the null condition follows from
\begin{equation} 
2\mathcal{L} = -F(r)\dot{t}^{\,2}+\frac{\dot{r}^{\,2}}{F(r)H(r)}+r^{2}\dot{\varphi}^{\,2}=0, 
\label{eq:null_lagrangian} 
\end{equation}
where the dot denotes differentiation with respect to an affine parameter. The temporal and axial Killing symmetries give the constants of motion
\begin{equation} 
E = F(r)\dot{t}, \qquad L=r^{2}\dot{\varphi}, \qquad b\equiv\frac{L}{E}, 
\label{eq:null_constants} 
\end{equation}
where $b$ is the impact parameter. Substituting Eq.~\eqref{eq:null_constants} into Eq.~\eqref{eq:null_lagrangian} gives, without introducing an independent effective potential,
\begin{equation} 
\begin{split}
\dot{r}^{\,2} & = E^{2}H(r)\left[1-\frac{b^{2}F(r)}{r^{2}}\right], \,\, \text{and}\\
& \left(\frac{\mathrm{d}r}{\mathrm{d}\varphi}\right)^{2}=\frac{r^{4}H(r)}{b^{2}}\left[1-\frac{b^{2}F(r)}{r^{2}}\right]. \label{eq:null_radial_equation} 
\end{split}
\end{equation}
If $r_{c}$ denotes the radial turning point, the condition $\dot r(r_c)=0$ yields
\begin{equation} 
b^{2}=\frac{r_{c}^{2}}{F(r_{c})}=\frac{r_{c}^{3}}{r_{c}-2M}. 
\label{eq:turning_point} 
\end{equation}
Here, notice that $H(r)$ changes the radial evolution along the trajectory but does not modify the algebraic relation between $b$ and $r_c$. This distinction is particularly useful in the weak--field regime because the quantum correction enters the accumulated angle along the path even though the turning point relation remains Schwarzschild--like.

The same result can be organized geometrically by introducing the optical metric \cite{Qiao:2022hfv,Qiao:2022jlu,Qiao:2024qsf,Pereira:2026lyt,AraujoFilho:2024mvz,Heidari:2024bkm}. Setting $\mathrm{d}s^{2} = 0$ in Eq.~\eqref{eq:holonomy_metric} and remaining on the equatorial plane gives
\begin{equation} 
\mathrm{d}t^{2} = \frac{\mathrm{d}r^{2}}{F(r)^{2}H(r)}+\frac{r^{2}}{F(r)}\,\mathrm{d}\varphi^{2}\equiv\gamma_{ab} \,\mathrm{d}x^{a}\mathrm{d}x^{b}, \label{eq:optical_metric} 
\end{equation}
whose area element is
\begin{equation} \mathrm{d}S = \sqrt{\det\gamma}\,\mathrm{d}r\,\mathrm{d}\varphi=\frac{r}{F(r)^{3/2}\sqrt{H(r)}}\,\mathrm{d}r\,\mathrm{d}\varphi. \label{eq:optical_surface} \end{equation}
The Gaussian curvature associated with this two-dimensional optical geometry is
\begin{equation} 
K(r)=-\frac{4M+a}{2r^{3}}+\frac{3M^{2}+\tfrac{9}{2}aM}{r^{4}}-\frac{6aM^{2}}{r^{5}}.
\label{eq:gaussian_curvature} 
\end{equation}
Eq.~\eqref{eq:gaussian_curvature} reduces  to the Schwarzschild optical curvature when $a=0$, while its leading asymptotic contribution shows that a positive value of $a$ increases the focusing of distant trajectories. In this representation, the weak bending angle may be obtained by integrating $K$ over the exterior optical domain through the Gauss--Bonnet construction \cite{GibbonsWerner2008,Ovgun:2026vcl,Pantig:2026xjj,Ovgun:2025ctx}.

For a ray emitted in the asymptotic region, reaching $r_c$, and arriving at a distant observer, the bending angle is determined by
\begin{equation} 
\widehat{\alpha}(b) = 2\int_{r_{c}}^{\infty}\frac{b\, \mathrm{d}r}{r^{2}\sqrt{H(r)}\sqrt{1-b^{2}F(r)/r^{2}}}-\pi, 
\label{eq:exact_deflection} 
\end{equation}
with $r_c$ related to $b$ by Eq.~\eqref{eq:turning_point} \cite{Soares2023}. Since the neutrino lensing configuration considered in this work belongs to the weak--field sector, we take $b\gg M$ and expand Eq.~\eqref{eq:exact_deflection} at fixed $\delta=a/(2M)$. The resulting post--Minkowskian expression is
\begin{equation} 
\widehat{\alpha}(b)=2(2+\delta)\frac{M}{b}+\frac{3\pi}{4}\left(5+2\delta+\delta^{2}\right)\frac{M^{2}}{b^{2}}+\mathcal{O}\!\left(\frac{M^{3}}{b^{3}}\right), 
\label{eq:weak_deflection} 
\end{equation}
whose leading contribution can equivalently be expressed as \cite{Soares2023}
\begin{equation} 
\widehat{\alpha}(b) = \frac{4M+a}{b}+\mathcal{O}\!\left(\frac{M^{2}}{b^{2}},\frac{Ma}{b^{2}},\frac{a^{2}}{b^{2}}\right). 
\label{eq:weak_deflection_leading} 
\end{equation}
As it is straightforward to see, the Schwarzschild result $\widehat{\alpha}=4M/b$ is naturally recovered for $a=0$, whereas a positive holonomy scale strengthens the weak deflection. Moreover, Eq.~\eqref{eq:weak_deflection} retains all terms belonging to the same order in $M/b$, since $a=2\delta M$ and must not be treated as an unrelated large distance scale.

To connect the bending angle with the trajectories contributing to the neutrino signal, let $D_L$, $D_{LS}$, and $D_S$ denote the observer--lens, lens--source, and observer--source angular diameter distances, respectively. If $\beta_s$ represents the unlensed angular position of the source and $\theta$ the angular position of an image, then the small angle relation $b\simeq D_L|\theta|$ transforms the deflection law into the thin lens equation as follows
\begin{equation} 
\beta_s = \theta - \frac{D_{LS}}{D_S}\operatorname{sgn}(\theta)\widehat{\alpha}\!\left(D_L|\theta|\right). 
\label{eq:thin_lens_equation} 
\end{equation}
Using the leading term of Eq.~\eqref{eq:weak_deflection_leading}, the Einstein angular radius follows as
\begin{equation} 
\theta_E^{2} = \frac{(4M+a)D_{LS}}{D_LD_S}, 
\label{eq:einstein_angle} 
\end{equation}
and the corresponding weak--field image positions are given  by
\begin{equation} 
\theta_{\pm} = \frac{1}{2}\left(\beta_s\pm\sqrt{\beta_s^{2}+4\theta_E^{2}}\right). 
\label{eq:weak_images} 
\end{equation}
Here, some comments are worthy to be mentioned: the holonomy correction increases the Einstein scale and shifts the impact parameters of the two weakly deflected trajectories. These are precisely the optical quantities required in the subsequent neutrino analysis: the lens equation selects the admissible paths, while the metric functions determine the phase accumulated by each mass eigenstate along them. In this way, the parameter $a$ enters the flavor-transition probability both through the radial phase integral and through the weak--field geometry of the lensed trajectories. All these aspects (and more) will be shown in the next sections.

%%%%%%%%%%%%%%%%%%%%%%%%%%%%%%%%%%%%%%%%%%%%%%%%%%%%%%%%%%%%%%%%%%%%%%%%%%%%%%%%%%%%%%%%%%%%%%%%%%%%%%%%%%%%%%%%%%%%%%%%%%%%%%%%%%%%%%%%%%%%%%%%%%%%%%%%%%%%%%%%%%%%%%%%%%%%%%%%%%%%%%%%%%%%%%%%%%%%%%%%%%%%%%%%%%%%%%%%%%%%%%%%%%%%%%%%%%%%%%%%%%%%%%%%%%%%%%%%%%%%%%%%%%%%%%%%%%%%%%%%%%%%%%%%%%%%%%%%%%%%%%%%%%%%%%%%%%%%%%%%%%%%%%%%%%%%%%%%%%%%%%%%%%%%%%%%%%%%%%%%%%%%%%%%%%%%%%%%%%%%%%%%%%%%%%%%%%%%%%%%%%%%%%%%%%%%%%%%%%%%%%%%%%%%%%%%%%%%%%%%%%%%%%%%%%%%%%%%%%%%%%%%%%%%%%%%%%%%%%%%%%%%%%%%%%%%%%%%%%%%%%%%%%%%%%

\section{Holonomy signatures on neutrino pair energy transfer}
\label{sec:holonomy_pair_transfer}

The reaction $\nu_{\ell}+\bar{\nu}_{\ell}\rightarrow e^{-}+e^{+}$ converts part of the neutrino luminosity surrounding a compact source into an electron--positron plasma. Such a channel is relevant to the production of energetic outflows because the created pairs can deposit energy in a baryon region above the emitting matter \cite{SalmonsonWilson1999,LambiaseMastrototaro2020}. In the present geometry, the black hole horizon is not regarded as the emitting surface. We instead introduce an effective neutrinosphere of areal radius $R_{\nu}>3M$, which may represent the inner boundary of a hot accretion flow or another optically thick source of thermal neutrinos. Neutrinos and antineutrinos are taken to be emitted isotropically in the local static frame, and their motion outside $R_{\nu}$ is governed by the holonomy corrected metric of Eq.~\eqref{eq:holonomy_metric}, as we could naturally expect.

For each flavor $\ell=e,\mu,\tau$, the local energy transferred to the pair plasma per unit proper time and proper volume is written as \cite{SalmonsonWilson1999}
\begin{equation} 
\begin{split}
&\dot{q}_{\ell}(r)\equiv\frac{\mathrm{d}E_{\mathrm{loc}}}{\mathrm{d}\tau\,\mathrm{d}V_{\mathrm{prop}}} \\
& = 2K_{\ell}G_{\mathrm{F}}^{2}\mathcal{F}(r)\int_{0}^{\infty}\int_{0}^{\infty}n(\epsilon_{\nu})n(\epsilon_{\bar{\nu}})(\epsilon_{\nu}+\epsilon_{\bar{\nu}})\epsilon_{\nu}^{3}\epsilon_{\bar{\nu}}^{3}\,\mathrm{d}\epsilon_{\nu}\,\mathrm{d}\epsilon_{\bar{\nu}}, 
\label{eq:local_pair_deposition} 
\end{split}
\end{equation}
where $G_{\mathrm{F}}$ is the Fermi constant, $n(\epsilon)$ denotes the thermal occupation number, and $\mathcal{F}(r)$ contains the angular overlap of the incident momenta. The weak--interaction coefficients distinguish the electron channel, which receives charged- and neutral current contributions, from the muon and tau channels, for which only the neutral current contributes. They are
\begin{equation} 
\begin{split}
& K_{e} = \frac{1}{6\pi}\left(1+4\sin^{2}\theta_{\mathrm{W}}+8\sin^{4}\theta_{\mathrm{W}}\right), \\
& K_{\mu} = K_{\tau}=\frac{1}{6\pi}\left(1-4\sin^{2}\theta_{\mathrm{W}}+8\sin^{4}\theta_{\mathrm{W}}\right), 
\label{eq:weak_channel_coefficients} 
\end{split}
\end{equation}
with $\theta_{\mathrm{W}}$ denoting the Weinberg angle. For vanishing chemical potential, the local neutrino and antineutrino populations follow the Fermi--Dirac distribution
\begin{equation} 
n(\epsilon) = \frac{2}{h^{3}}\left[\exp\!\left(\frac{\epsilon}{k_{\mathrm{B}}T(r)}\right)+1\right]^{-1}, 
\label{eq:fermi_dirac_distribution} 
\end{equation}
where $h$ and $k_{\mathrm{B}}$ are the Planck and Boltzmann constants. Evaluation of the two energy integrals in Eq.~\eqref{eq:local_pair_deposition}, we obtain
\begin{equation} 
\dot{q}_{\ell}(r)=\frac{21\zeta(5)\pi^{4}}{h^{6}}K_{\ell}G_{\mathrm{F}}^{2}\mathcal{F}(r)\left[k_{\mathrm{B}}T(r)\right]^{9}, 
\label{eq:thermal_pair_deposition} 
\end{equation}
where $\zeta(s)$ is the Riemann zeta function. Notice that the ninth power of the local temperature makes this process highly sensitive to gravitational redshift near the emission region.

The angular factor follows from the relative orientation of the annihilating particles and takes the form
\begin{equation} 
\begin{split}
& \mathcal{F}(r)\equiv\int\int\left(1-\boldsymbol{\Omega}_{\nu}\!\cdot\!\boldsymbol{\Omega}_{\bar{\nu}}\right)^{2}\,\mathrm{d}\Omega_{\nu}\,\mathrm{d}\Omega_{\bar{\nu}}\\
& = \frac{2\pi^{2}}{3}\left[1-x(r)\right]^{4}\left[x(r)^{2} + 4x(r) + 5\right],
\label{eq:annihilation_angular_factor} 
\end{split}
\end{equation}
where $\boldsymbol{\Omega}_{\nu}$ and $\boldsymbol{\Omega}_{\bar{\nu}}$ are unit propagation vectors. To determine $x(r)$, let $\psi(r)$ be the angle between the limiting ray and the outward radial direction as measured by a static orthonormal observer. The local angle and the conserved impact parameter are related by
\begin{equation} 
\sin\psi(r) = \frac{b_{\nu}\sqrt{F(r)}}{r}, \qquad b_{\nu} =  \frac{R_{\nu}}{\sqrt{F(R_{\nu})}}, 
\label{eq:limiting_neutrino_ray} 
\end{equation}
where $b_{\nu}$ corresponds to tangential emission at the neutrinosphere. Defining $x(r)\equiv\cos\psi(r)$ gives
\begin{equation} 
x(r)^{2} = 1 - \frac{R_{\nu}^{2}}{r^{2}}\frac{F(r)}{F(R_{\nu})}. 
\label{eq:holonomy_emission_cone} 
\end{equation}
The factor $H(r)$ does not occur in Eq.~\eqref{eq:holonomy_emission_cone}. Indeed, the locally measured angular aperture depends on the lapse $F(r)$ and on the areal radius, both of which preserve their Schwarzschild forms. By contrast, $H(r)$ controls the radial proper length and will enter when the local deposition density is integrated over space.

Thermal equilibrium in a static gravitational field imposes the Tolman relation
\begin{equation} 
T(r)\sqrt{F(r)} = T_{\nu}\sqrt{F(R_{\nu})}, \qquad T_{\nu}\equiv T(R_{\nu}), 
\label{eq:tolman_neutrino_temperature} 
\end{equation}
whereas the luminosity measured at infinity is related to the local luminosity of the neutrinosphere through
\begin{equation} 
L_{\infty} = F(R_{\nu})L(R_{\nu}). 
\label{eq:redshifted_neutrino_luminosity} 
\end{equation}
For one flavor, including its antiparticle, the thermal luminosity emitted by the spherical surface is
\begin{equation} 
L(R_{\nu}) = \frac{7\pi}{4}a_{\mathrm{r}}cR_{\nu}^{2}T_{\nu}^{4}, 
\label{eq:thermal_neutrino_luminosity} 
\end{equation}
where $a_{\mathrm{r}}$ is the radiation constant and $c$ is the speed of light. Combining Eqs.~\eqref{eq:tolman_neutrino_temperature}--\eqref{eq:thermal_neutrino_luminosity} eliminates $T_{\nu}$ in favor of the luminosity observed at infinity and gives
\begin{equation} 
\left[k_{\mathrm{B}}T(r)\right]^{9} = k_{\mathrm{B}}^{9}\left(\frac{7\pi a_{\mathrm{r}}c}{4}\right)^{-9/4}L_{\infty}^{9/4}R_{\nu}^{-9/2}\frac{F(R_{\nu})^{9/4}}{F(r)^{9/2}}. 
\label{eq:temperature_luminosity_reduction} 
\end{equation}
It is convenient to gather the microscopic constants into
\begin{equation} \mathcal{C}_{\ell}\equiv\frac{21\zeta(5)\pi^{4}}{h^{6}}K_{\ell}G_{\mathrm{F}}^{2}k_{\mathrm{B}}^{9}\left(\frac{7\pi a_{\mathrm{r}}c}{4}\right)^{-9/4}, \label{eq:annihilation_constant} \end{equation}
so that the local deposition density assumes the compact form
\begin{equation} 
\dot{q}_{\ell}(r)=\mathcal{C}_{\ell}L_{\infty}^{9/4}R_{\nu}^{-9/2}F(R_{\nu})^{9/4}\frac{\mathcal{F}(r)}{F(r)^{9/2}}. 
\label{eq:local_deposition_luminosity_form} 
\end{equation}
No explicit $a$ dependence appears in Eq.~\eqref{eq:local_deposition_luminosity_form}. This is a distinctive property of the holonomy corrected solution: at fixed $M$, $R_{\nu}$, and $L_{\infty}$, the lapse, gravitational temperature shift, luminosity redshift, and angular emission cone are identical to those of Schwarzschild spacetime. The quantum geometric contribution emerges solely through the radial measure.

On a hypersurface of constant Killing time, the proper volume of a spherical shell is
\begin{equation} 
\mathrm{d}V_{\mathrm{prop}} = \frac{4\pi r^{2}}{\sqrt{F(r)H(r)}}\,\mathrm{d}r. 
\label{eq:holonomy_proper_volume} 
\end{equation}
The radial distribution of deposited power becomes
\begin{equation} 
\frac{\mathrm{d}\dot{Q}_{\ell}}{\mathrm{d}r}=4\pi\mathcal{C}_{\ell}L_{\infty}^{9/4}R_{\nu}^{-9/2}F(R_{\nu})^{9/4}\frac{r^{2}\mathcal{F}(r)}{F(r)^{5}\sqrt{H(r)}}. 
\label{eq:radial_deposited_power} 
\end{equation}
After integration from the neutrinosphere to spatial infinity, the total rate is
\begin{equation} 
\dot{Q}_{\ell}=4\pi\mathcal{C}_{\ell}L_{\infty}^{9/4}R_{\nu}^{-9/2}F(R_{\nu})^{9/4}\int_{R_{\nu}}^{\infty}\frac{r^{2}\mathcal{F}(r)}{F(r)^{5}\sqrt{H(r)}}\,\mathrm{d}r. 
\label{eq:total_holonomy_deposition} 
\end{equation}
Introducing the dimensionless radius
$y\equiv r/R_{\nu}$, allows Eq.~\eqref{eq:total_holonomy_deposition} to be expressed as
\begin{equation} 
\dot{Q}_{\ell}=\frac{8\pi^{3}}{3}\mathcal{C}_{\ell}L_{\infty}^{9/4}R_{\nu}^{-3/2}F(R_{\nu})^{9/4}\mathcal{I}_{\mathrm{hol}}\!\left(\frac{M}{R_{\nu}},\frac{a}{R_{\nu}}\right), 
\label{eq:dimensionless_total_deposition} 
\end{equation}
where the entire geometric response is encoded in
\begin{equation} 
\mathcal{I}_{\mathrm{hol}}\!\left(\frac{M}{R_{\nu}},\frac{a}{R_{\nu}}\right)\equiv\int_{1}^{\infty}\frac{y^{2}\left[1 - x(y)\right]^{4}\left[x(y)^{2}+4x(y)+5\right]}{F(yR_{\nu})^{5}\sqrt{H(yR_{\nu})}}\,\mathrm{d}y,
\label{eq:holonomy_deposition_integral} 
\end{equation}
with
\begin{equation} 
\begin{split}
& F(yR_{\nu}) = 1 - \frac{2M}{yR_{\nu}}, \,\,\,\, H(yR_{\nu}) = 1-\frac{a}{yR_{\nu}}, \\
& x(y)^{2} = 1 - \frac{F(yR_{\nu})}{y^{2}F(R_{\nu})}. \label{eq:dimensionless_holonomy_functions} 
\end{split}
\end{equation}
If all three flavors carry the same luminosity and temperature, their contributions add according to
\begin{equation} 
\dot{Q}=\sum_{\ell=e,\mu,\tau}\dot{Q}_{\ell}, \qquad K_{\mathrm{tot}}=K_{e}+K_{\mu}+K_{\tau}.
\label{eq:flavor_summed_deposition} 
\end{equation}

The Newtonian limit shows a transparent normalization of the gravitational effect. For $F=H=1$, the limiting cone becomes $x_{\mathrm{N}}(y)^{2}=1-y^{-2}$ and its angular integral obeys
\begin{equation} 
\int_{1}^{\infty}y^{2}\left[1-x_{\mathrm{N}}(y)\right]^{4}\left[x_{\mathrm{N}}(y)^{2}+4x_{\mathrm{N}}(y)+5\right]\,\mathrm{d}y=\frac{1}{3}.
\label{eq:newtonian_angular_integral} 
\end{equation}
The relativistic enhancement relative to this reference rate is
\begin{equation} 
\frac{\dot{Q}}{\dot{Q}_{\mathrm{Newt}}} = 3 F(R_{\nu})^{9/4}\mathcal{I}_{\mathrm{hol}}\!\left(\frac{M}{R_{\nu}},\frac{a}{R_{\nu}}\right). 
\label{eq:holonomy_newtonian_ratio} 
\end{equation}
Because $F(r)$ and $x(r)$ do not depend on $a$, the comparison with the Schwarzschild geometry can be made without altering the thermal or angular factors. Defining the positive weight
\begin{equation} 
\mathcal{W}(y) \equiv \frac{y^{2}\left[1-x(y)\right]^{4}\left[x(y)^{2}+4x(y)+5\right]}{F(yR_{\nu})^{5}},
\label{eq:deposition_weight_function} 
\end{equation}
we get
\begin{equation} 
\frac{\dot{Q}(a)}{\dot{Q}(0)}=\frac{\displaystyle\int_{1}^{\infty}\mathcal{W}(y)\left(1-\frac{a}{yR_{\nu}}\right)^{-1/2}\mathrm{d}y}{\displaystyle\int_{1}^{\infty}\mathcal{W}(y)\,\mathrm{d}y}>1 \,,  
\label{eq:holonomy_schwarzschild_ratio} 
\end{equation}
with $0<a<2M$; in this manner, within the stated emission model, the holonomy parameter produces an increase instead of a suppression of the volume integrated pair power. Notice that the origin of this behavior is genuinely geometric: the factor $H(r)^{-1/2}$ enlarges the proper volume assigned to each radial shell, while leaving its local thermal and angular deposition density unchanged.

In addition, for $a/R_{\nu}\ll1$, the leading correction is exposed by
\begin{equation} 
\frac{\dot{Q}(a)}{\dot{Q}(0)} = 1+\frac{a}{2R_{\nu}}\frac{\displaystyle\int_{1}^{\infty}\mathcal{W}(y)y^{-1}\,\mathrm{d}y}{\displaystyle\int_{1}^{\infty}\mathcal{W}(y)\,\mathrm{d}y} + \mathcal{O}\!\left(\frac{a^{2}}{R_{\nu}^{2}}\right). 
\label{eq:small_holonomy_deposition} 
\end{equation}
The correction fades in the dilute field limit because both $M/R_{\nu}$ and $a/R_{\nu}$ vanish, recovering the Newtonian rate. Near a compact neutrinosphere, however, the proper volume deformation leaves a direct signature on the energy available to the electron--positron plasma and supplies an observable channel complementary to neutrino phases and weak gravitational lensing.

%%%%%%%%%%%%%%%%%%%%%%%%%%%%%%%%%%%%%%%%%%%%%%%%%%%%%%%%%%%%%%%%%%%%%%%%%%%%%%%%%%%%%%%%%%%%%%%%%%%%%%%%%%%%%%%%%%%%%%%%%%%%%%%%%%%%%%%%%%%%%%%%%%%%%%%%%%%%%%%%%%%%%%%%%%%%%%%%%%%%%%%%%%%%%%%%%%%%%%%%%%%%%%%%%%%%%%%%%%%%%%%%%%%%%%%%%%%%%%%%%%%%%%%%%%%%%%%%%%%%%%%%%%%%%%%%%%%%%%%%%%%%%%%%%%%%%%%%%%%%%%%%%%%%%%%%%%%%%%%%%%%%%%%%%%%%%%%%%%%%%%%%%%%%%%%%%%%%%%%%%%%%%%%%%%%%%%%%%%%%%%%%%%%%%%%%%%%%%%%%%%%%%%%%%%%%%%%%%%%%%%%%%%%%%%%%%%%%%%%%%%%%%%%%%%%%%%%%%%%

\section{Quantum geometric phase transport and flavor conversion}
\label{sec:holonomy_neutrino_phase}

Neutrinos are created and detected through weak interactions in flavor states, whereas free propagation is governed by states of definite mass. In vacuum, a flavor eigenstate $|\nu_{\alpha}\rangle$, with $\alpha=e,\mu,\tau$, is  expressed as a coherent superposition of the mass eigenstates $|\nu_{k}\rangle$, with $k=1,2,3$, according to \cite{MakiNakagawaSakata1962,GrossmanLipkin1997}
\begin{equation} 
|\nu_{\alpha}\rangle=\sum_{k=1}^{3}U_{\alpha k}^{*}|\nu_{k}\rangle, 
\label{eq:flavor_mass_relation} 
\end{equation}
where $U$ denotes the Pontecorvo--Maki--Nakagawa--Sakata matrix. Its three mixing angles determine the flavor content of the propagating modes, while the Dirac phase $\delta_{\mathrm{CP}}$ controls CP violation. As argued in the previous sections, we consider the vacuum exterior of the holonomy corrected black hole and assume minimal coupling between the neutrino fields and the physical metric. Accordingly, the parameter $a$ influences flavor evolution through the geometry. It is important to mention that natural units, $\hbar=c=1$, are used throughout this section for the sake of simplicity.

Let $S=(t_{S},r_{S},\varphi_{S})$ and $D=(t_{D},r_{D},\varphi_{D})$ denote the emission and detection events on the equatorial plane. At leading semiclassical order, the $k$--th mass eigenstate evolves as
\begin{equation} 
\begin{split}
&|\nu_{k}(D)\rangle = \exp\!\left[-i\Phi_{k}(S,D)\right]|\nu_{k}(S)\rangle, \\
& \text{with} \, \, \, \, \Phi_{k}(S,D) = -\int_{S}^{D}p_{\mu}^{(k)}\,\mathrm{d}x^{\mu}, \label{eq:covariant_neutrino_phase} 
\end{split}
\end{equation}
where the sign follows from the metric convention adopted in Eq.~\eqref{eq:holonomy_metric} \cite{Stodolsky1979,BhattacharyaHabibMottola1999}. The contraction $p_{\mu}^{(k)}\mathrm{d}x^{\mu}$ is a scalar; in other words, the observable phase cannot depend on the coordinates chosen to describe the trajectory.

The Hamilton--Jacobi function may be separated as
\begin{equation} 
\begin{split}
& S_{k} = -E_{k}t + L_{k}\varphi + S_{r,k}(r), \qquad p_{\mu}^{(k)} = \partial_{\mu}S_{k}, \\
&\Phi_{k} = -\left[S_{k}(D)-S_{k}(S)\right], \label{eq:neutrino_hamilton_jacobi_action} 
\end{split}
\end{equation}
where $E_{k} = -p_{t}^{(k)}$ and $L_{k}=p_{\varphi}^{(k)}$ are the conserved Killing energy and angular momentum. The mass shell condition in the geometry of Eq.~\eqref{eq:holonomy_metric} reads
\begin{equation} 
- \frac{E_{k}^{2}}{F(r)} + F(r)H(r)\left[p_{r}^{(k)}\right]^{2} + \frac{L_{k}^{2}}{r^{2}} = - m_{k}^{2}, 
\label{eq:holonomy_neutrino_mass_shell} 
\end{equation}
from which the radial momentum follows as
\begin{equation} 
p_{r}^{(k)} = \pm\frac{1}{F(r)\sqrt{H(r)}}\sqrt{E_{k}^{2} -F(r)\left(m_{k}^{2} + \frac{L_{k}^{2}}{r^{2}}\right)}. 
\label{eq:holonomy_neutrino_radial_momentum} 
\end{equation}
The two signs distinguish outward and inward radial motion. At leading WKB order, the spin connection transports the spinor amplitude but does not generate a separate mass dependent contribution to the relative eikonal phase \cite{BhattacharyaHabibMottola1999}.

For ultrarelativistic propagation, $m_{k}^{2}/E_{k}^{2}\ll1$, the trajectory may be evaluated along the corresponding null geodesic while the mass dependence is retained in the phase. We adopt the equal energy prescription
\begin{equation} 
E_{k} = E_{0} + \mathcal{O}\!\left(\frac{m_{k}^{2}}{E_{0}}\right), \qquad \frac{L_{k}}{E_{k}} = b_{p} + \mathcal{O}\!\left(\frac{m_{k}^{2}}{E_{0}^{2}}\right), 
\label{eq:equal_energy_neutrino_prescription} 
\end{equation}
where $E_{0}$ is the common Killing energy and $b_{p}$ labels the reference null path $p$. Evaluating all mass eigenstates between the same spacetime events removes the mass independent eikonal contribution and yields \cite{GrossmanLipkin1997,CardallFuller1997}
\begin{equation} 
\Phi_{k}^{(p)} = \frac{m_{k}^{2}}{2E_{0}}\int_{p}\frac{\mathrm{d}r}{\sqrt{H(r)}\sqrt{1-b_{p}^{2}F(r)/r^{2}}}+\mathcal{O}\!\left(\frac{m_{k}^{4}}{E_{0}^{3}}\right). 
\label{eq:master_holonomy_neutrino_phase} 
\end{equation}
Eq.~\eqref{eq:master_holonomy_neutrino_phase} displays the geometric roles of the two metric functions without ambiguity. The Schwarzschild potential $F(r)$ governs the nonradial part of the orbit, whereas the holonomy function $H(r)$ weights every radial interval traversed by the neutrino.

For a radial trajectory, $b_{p}=0$, the phase becomes
\begin{equation} 
\Phi_{k}^{\mathrm{rad}}(r_{S},r_{D}) = \frac{m_{k}^{2}}{2E_{0}}\left|\mathcal{R}_{a}(r_{D}) - \mathcal{R}_{a}(r_{S})\right|, 
\label{eq:exact_radial_holonomy_phase} 
\end{equation}
where the primitive of the holonomy kernel is
\begin{equation} 
\begin{split}
& \mathcal{R}_{a}(r)\equiv\sqrt{r(r-a)}+a\ln\!\left(\sqrt{r}+\sqrt{r-a}\right), \\
&\frac{\mathrm{d}\mathcal{R}_{a}}{\mathrm{d}r}=\frac{1}{\sqrt{1-a/r}}. \label{eq:holonomy_radial_primitive} 
\end{split}
\end{equation}
The logarithm is harmless because only the difference of $\mathcal{R}_{a}$ at the two endpoints enters the phase. In the weak--field region, where $a/r\ll1$, the exact interval reduces to
\begin{equation} 
\mathcal{R}_{a}(r_{D}) - \mathcal{R}_{a}(r_{S})=(r_{D}-r_{S})+\frac{a}{2}\ln\!\left(\frac{r_{D}}{r_{S}}\right)+\mathcal{O}\!\left(\frac{a^{2}}{r_{<}}\right), 
\label{eq:weak_radial_holonomy_interval} 
\end{equation}
where $r_{<}\equiv\min(r_{S},r_{D})$. The relative phase of two mass eigenstates is
\begin{equation} 
\Delta\Phi_{ij}^{\mathrm{rad}}\equiv\Phi_{i}^{\mathrm{rad}} -\Phi_{j}^{\mathrm{rad}}=\frac{\Delta m_{ij}^{2}}{2E_{0}}\left|\mathcal{R}_{a}(r_{D})-\mathcal{R}_{a}(r_{S})\right|, 
\label{eq:radial_holonomy_phase_difference} 
\end{equation}
with $\Delta m_{ij}^{2}\equiv m_{i}^{2}-m_{j}^{2}$. At fixed asymptotic energy and fixed areal endpoints, a positive $a$ enlarges the accumulated phase, as it is straightforward to see. A remark is worthy to be highlithed: this modification is not a constant rescaling: the logarithm retains information about both the source and detector positions.

The distinction between an areal coordinate baseline and a locally measured distance is essential. Along a radial path, the infinitesimal phase and proper length satisfy
\begin{equation} 
\frac{\mathrm{d}\Delta\Phi_{ij}^{\mathrm{rad}}}{\mathrm{d}r}=\frac{\Delta m_{ij}^{2}}{2E_{0}\sqrt{H(r)}}, \qquad \mathrm{d}\ell_{\mathrm{prop}}=\frac{\mathrm{d}r}{\sqrt{F(r)H(r)}}. 
\label{eq:radial_phase_and_proper_length} 
\end{equation}
The corresponding oscillation intervals measured in the areal coordinate and in local proper distance are
\begin{equation} 
\begin{split}
& L_{\mathrm{osc},ij}^{(r)}(r) = \frac{4\pi E_{0}\sqrt{H(r)}}{|\Delta m_{ij}^{2}|}, \\
& L_{\mathrm{osc},ij}^{\mathrm{prop}}(r)=\frac{4\pi E_{\mathrm{loc}}(r)}{|\Delta m_{ij}^{2}|} = \frac{4\pi E_{0}}{|\Delta m_{ij}^{2}|\sqrt{F(r)}}, \label{eq:holonomy_oscillation_lengths} 
\end{split}
\end{equation}
where
$ E_{\mathrm{loc}}(r) = E_{0}/\sqrt{F(r)}$. Here, $H(r)$ changes the phase accumulated between fixed areal radii by altering the proper radial geometry, but it cancels from the local oscillation length expressed in terms of proper distance and locally measured energy.

For a nonradial trajectory that approaches the black hole, reaches a turning point $r_{0,p}$, and then escapes toward the detector, the phase must be evaluated over the two monotonic cases,
\begin{equation} 
\begin{split}
\Phi_{k}^{(p)}  = & \frac{m_{k}^{2}}{2E_{0}}\left[\int_{r_{0,p}}^{r_{S}}\frac{\mathrm{d}r}{\sqrt{H(r)}\sqrt{1-b_{p}^{2}F(r)/r^{2}}} \right. \\
& \left. +\int_{r_{0,p}}^{r_{D}}\frac{\mathrm{d}r}{\sqrt{H(r)}\sqrt{1-b_{p}^{2}F(r)/r^{2}}}\right]. \label{eq:two_branch_holonomy_phase} 
\end{split}
\end{equation}
The turning point is fixed by
\begin{equation} 
b_{p}^{2} = \frac{r_{0,p}^{2}}{F(r_{0,p})}, \qquad r_{0,p} = b_{p}-M + \mathcal{O}\!\left(\frac{M^{2}}{b_{p}}\right). 
\label{eq:neutrino_phase_turning_point} 
\end{equation}
Retaining $r_{0,p}$ during the expansion avoids the spurious divergence that would arise from expanding directly around the flat space turning point. To first order in $M/r_{0,p}$ and $a/r_{0,p}$, one monotonic segment is described by
\begin{equation} 
\begin{split}
& \mathcal{J}(r,r_{0,p}) = \sqrt{r^{2}-r_{0,p}^{2}} + M\sqrt{\frac{r-r_{0,p}}{r+r_{0,p}}} \\
& +\frac{a}{2}\ln\!\left(\frac{r+\sqrt{r^{2}-r_{0,p}^{2}}}{r_{0,p}}\right) + \mathcal{O}\!\left(\frac{M^{2}}{r_{0,p}},\frac{Ma}{r_{0,p}},\frac{a^{2}}{r_{0,p}}\right), 
\label{eq:weak_holonomy_phase_branch} 
\end{split}
\end{equation}
and the complete phase becomes
\begin{equation} 
\Phi_{k}^{(p)}=\frac{m_{k}^{2}}{2E_{0}}\left[\mathcal{J}(r_{S},r_{0,p}) + \mathcal{J}(r_{D},r_{0,p})\right]. 
\label{eq:complete_weak_holonomy_phase} 
\end{equation}
When the source and detector are both far from the lens, $b_{p}\ll r_{S},r_{D}$, Eq.~\eqref{eq:complete_weak_holonomy_phase} assumes the form
\begin{equation} 
\begin{split}
\Phi_{k}^{(p)} \simeq  \frac{m_{k}^{2}}{2E_{0}}&\left[(r_{S}+r_{D})\left(1-\frac{b_{p}^{2}}{2r_{S}r_{D}}\right) \right.\\
& \left. + 2M + \frac{a}{2}\ln\!\left(\frac{4r_{S}r_{D}}{b_{p}^{2}}\right)\right]. 
\label{eq:far_source_holonomy_phase} 
\end{split}
\end{equation}
The first term is the geometric path contribution, the second is the Schwarzschild correction, and the logarithm isolates the leading holonomy correction. For two eigenstates traveling along the same path, the common geometric factor ensures that
\begin{equation} 
\begin{split}
\Delta\Phi_{ij}^{(p)} \simeq \frac{\Delta m_{ij}^{2}}{2E_{0}} & \left[(r_{S}+r_{D})\left(1-\frac{b_{p}^{2}}{2r_{S}r_{D}}\right) \right.\\
& \left. + 2M + \frac{a}{2}\ln\!\left(\frac{4r_{S}r_{D}}{b_{p}^{2}}\right)\right]. 
\label{eq:single_path_holonomy_phase_difference}
\end{split}
\end{equation}
An ordinary single path oscillation remains sensitive only to the squared mass differences. Dependence on the absolute mass scale requires interference between distinct trajectories and will enter only when the weak--lensing amplitudes are combined.

The flavor amplitude at the detector is obtained by projecting the propagated state onto $|\nu_{\beta}\rangle$,
\begin{equation} 
\mathcal{A}_{\alpha\rightarrow\beta}^{(p)}=\sum_{k=1}^{3}U_{\beta k}U_{\alpha k}^{*}\exp\!\left[-i\Phi_{k}^{(p)}\right], \,\,\,\, P_{\alpha\rightarrow\beta}^{(p)}=\left|\mathcal{A}_{\alpha\rightarrow\beta}^{(p)}\right|^{2}. 
\label{eq:single_path_flavor_amplitude} 
\end{equation}
Introducing the rephasing invariant combination
\begin{equation} 
\mathcal{J}_{\alpha\beta}^{ij}\equiv U_{\alpha i}^{*}U_{\beta i}U_{\alpha j}U_{\beta j}^{*}, 
\label{eq:pmns_rephasing_invariant} 
\end{equation}
the three flavor probability may be written as
\begin{equation} 
\begin{split}
P_{\alpha\rightarrow\beta}^{(p)} = & \,\, \delta_{\alpha\beta}-4\sum_{i>j}\operatorname{Re}\!\left(\mathcal{J}_{\alpha\beta}^{ij}\right)\sin^{2}\!\left(\frac{\Delta\Phi_{ij}^{(p)}}{2}\right)\\
&+2\sum_{i>j}\operatorname{Im}\!\left(\mathcal{J}_{\alpha\beta}^{ij}\right)\sin\!\left(\Delta\Phi_{ij}^{(p)}\right). \label{eq:three_flavor_holonomy_probability} 
\end{split}
\end{equation}
For antineutrinos, $U$ is replaced by $U^{*}$, reversing the CP--odd term proportional to $\operatorname{Im}(\mathcal{J}_{\alpha\beta}^{ij})$. In a survival channel, that contribution vanishes and one finds
\begin{equation} 
P_{\alpha\rightarrow\alpha}^{(p)}=1-4\sum_{i>j}|U_{\alpha i}|^{2}|U_{\alpha j}|^{2}\sin^{2}\!\left(\frac{\Delta\Phi_{ij}^{(p)}}{2}\right). 
\label{eq:three_flavor_survival_probability} 
\end{equation}
Within a two flavor reduction characterized by a mixing angle $\vartheta$, the conversion and survival probabilities reduce to
\begin{equation} 
P_{\alpha\rightarrow\beta}^{(p)} = \sin^{2}(2\vartheta)\sin^{2}\!\left(\frac{\Delta\Phi_{21}^{(p)}}{2}\right), \,\,\, P_{\alpha\rightarrow\alpha}^{(p)}=1-P_{\alpha\rightarrow\beta}^{(p)}, 
\label{eq:two_flavor_holonomy_probabilities} 
\end{equation}
where $\alpha\neq\beta$. For radial propagation, the conversion probability is obtained directly from the exact holonomy interval,
\begin{equation} P_{\alpha\rightarrow\beta}^{\mathrm{rad}}=\sin^{2}(2\vartheta)\sin^{2}\!\left[\frac{\Delta m_{21}^{2}}{4E_{0}}\left|\mathcal{R}_{a}(r_{D})-\mathcal{R}_{a}(r_{S})\right|\right]. \label{eq:exact_radial_holonomy_probability} \end{equation}
As we can see, the holonomy parameter shifts the positions of the oscillation maxima and minima without changing the mixing amplitude $\sin^{2}(2\vartheta)$. This separation between geometric phase displacement and intrinsic flavor mixing will play a fundamental role when the weakly lensed paths are superposed in the following section.

%%%%%%%%%%%%%%%%%%%%%%%%%%%%%%%%%%%%%%%%%%%%%%%%%%%%%%%%%%%%%%%%%%%%%%%%%%%%%%%%%%%%%%%%%%%%%%%%%%%%%%%%%%%%%%%%%%%%%%%%%%%%%%%%%%%%%%%%%%%%%%%%%%%%%%%%%%%%%%%%%%%%%%%%%%%%%%%%%%%%%%%%%%%%%%%%%%%%%%%%%%%%%%%%%%%%%%%%%%%%%%%%%%%%%%%%%%%%%%%%%%%%%%%%%%%%%%%%%%%%%%%%%%%%%%%%%%%%%%%%%%%%%%%%%%%%%%%%%%%%%%%%%%%%%%%%%%%%%%%%%%%%%%%%%%%%%%%%%%%%%%%%%%%%%%%%%%%%%%%%%%%%%%%%%%%%%%%%%%%%%%%%%%%%%%%%%%%%%%%%%%%%%%%%%%%%%%%%%%%%%%%%%%%%%%%%%%%%%%%%%%%%%%%%%%%%%%%%%%%

\section{Holonomy flavor interference along lensed neutrino paths}
\label{sec:holonomy_neutrino_interference}

An ultrarelativistic neutrino is deflected by the holonomy corrected black hole along the same null trajectories that govern massless propagation, up to corrections of order $m_{i}^{2}/E_{0}^{2}$. The finite masses may be neglected in the lens equation while remaining indispensable in the quantum phases. When the source, lens, and detector are nearly aligned, two weak--field images can connect the same emission and detection events. If the detector does not distinguish these trajectories and their wave packets still overlap, the observed flavor signal becomes an interference pattern assembled from both the mass and path degrees of freedom \cite{CrockerGiuntiMortlock2004,AlexandreClough2018}.

The weak deflection obtained in Eq.~\eqref{eq:weak_deflection} can be reorganized as
\begin{equation}
\begin{split}
& \widehat{\alpha}(b) = \frac{\mathcal{A}_{\mathrm{h}}}{b}+\frac{\mathcal{B}_{\mathrm{h}}}{b^{2}}+\mathcal{O}\!\left(\frac{M^{3}}{b^{3}}\right), \,\, \mathcal{A}_{\mathrm{h}}\equiv4M+a=2(2+\delta)M, \\
& \mathcal{B}_{\mathrm{h}}\equiv\frac{3\pi}{4}\left(5+2\delta+\delta^{2}\right)M^{2}, \label{eq:holonomy_deflection_coefficients} 
\end{split}
\end{equation}
where the subscript ``$\mathrm{h}$'' identifies the holonomy corrected quantities \cite{Soares2023}. Let $D_{L}$, $D_{LS}$, and $D_{S}$ denote the observer--lens, lens--source, and observer--source angular diameter distances. For a signed image angle $\theta$ and an unlensed source angle $\beta$, the impact parameter is $b=D_{L}|\theta|$, and the thin lens equation reads \cite{VirbhadraEllis2000,SchneiderEhlersFalco1992}
\begin{equation} \beta=\theta-\frac{D_{LS}}{D_{S}}\operatorname{sgn}(\theta)\widehat{\alpha}\!\left(D_{L}|\theta|\right). \label{eq:holonomy_signed_lens_equation} \end{equation}
Introducing the angular scales
\begin{equation} \theta_{E}^{2}\equiv\frac{\mathcal{A}_{\mathrm{h}}D_{LS}}{D_{L}D_{S}}, \qquad \theta_{\mathrm{h}}^{3}\equiv\frac{\mathcal{B}_{\mathrm{h}}D_{LS}}{D_{L}^{2}D_{S}}, \label{eq:holonomy_lensing_scales} \end{equation}
transforms Eq.~\eqref{eq:holonomy_signed_lens_equation} into
\begin{equation} 
\beta = \theta-\frac{\theta_{E}^{2}}{\theta}-\operatorname{sgn}(\theta)\frac{\theta_{\mathrm{h}}^{3}}{\theta^{2}}+\mathcal{O}\!\left(\frac{M^{3}}{D_{L}^{3}|\theta|^{3}}\right). 
\label{eq:second_order_holonomy_lens_map} 
\end{equation}
At leading order, the positive-- and negative--parity images occupy the angular positions
\begin{equation} \theta_{\pm}^{(0)}=\frac{1}{2}\left(\beta\pm\sqrt{\beta^{2}+4\theta_{E}^{2}}\right). \label{eq:leading_holonomy_image_positions} \end{equation}
The next weak--field correction follows by perturbing the lens map about these roots,
\begin{equation} 
\theta_{\pm} = \theta_{\pm}^{(0)}+\operatorname{sgn}\!\left(\theta_{\pm}^{(0)}\right)\frac{\theta_{\mathrm{h}}^{3}}{\left[\theta_{\pm}^{(0)}\right]^{2}+\theta_{E}^{2}}+\mathcal{O}\!\left(\theta_{\mathrm{h}}^{6}\right), 
\label{eq:corrected_holonomy_images} 
\end{equation}
with $ b_{\pm}=D_{L}|\theta_{\pm}|$. The absolute magnification of each image is determined by the Jacobian of the angular map,
\begin{equation} 
\mu_{p} =  \left|\frac{\theta_{p}}{\beta}\frac{\mathrm{d}\theta_{p}}{\mathrm{d}\beta}\right|=\left|\frac{\theta_{p}}{\beta}\right|\left(1+\frac{\theta_{E}^{2}}{\theta_{p}^{2}}+\frac{2\theta_{\mathrm{h}}^{3}}{|\theta_{p}|^{3}}\right)^{-1}, \,\,\, p=\pm. 
\label{eq:holonomy_image_magnifications} 
\end{equation}
Both image positions and weights inherit the parameter $a$, even though the lens remains asymptotically flat and no conical contribution is present.

The relative phase between the two geometrical rays also contains a mass independent part. A reduced lensing potential reproducing Eq.~\eqref{eq:second_order_holonomy_lens_map} is
\begin{equation} 
\psi(\theta) = \theta_{E}^{2}\ln\!\left(\frac{|\theta|}{\theta_{\ast}}\right)-\frac{\theta_{\mathrm{h}}^{3}}{|\theta|}, \qquad \frac{\mathrm{d}\psi}{\mathrm{d}\theta}=\frac{\theta_{E}^{2}}{\theta}+\operatorname{sgn}(\theta)\frac{\theta_{\mathrm{h}}^{3}}{\theta^{2}}, 
\label{eq:holonomy_reduced_lensing_potential} 
\end{equation}
where the arbitrary scale $\theta_{\ast}$ cancels from every observable time difference. In units with $c=1$, the Fermat arrival time of image $p$ is
\begin{equation} 
T_{p}=\frac{D_{L}D_{S}}{D_{LS}}\left[\frac{1}{2}(\theta_{p}-\beta)^{2}-\psi(\theta_{p})\right].
\label{eq:holonomy_fermat_time}
\end{equation}
The geometrical optics propagation factor then takes the form \cite{SchneiderEhlersFalco1992,NakamuraDeguchi1999}
\begin{equation} 
G_{p}=\sqrt{\mu_{p}}\exp\!\left(i\,\Xi_{p}\right), \qquad \Xi_{p}\equiv E_{0}T_{p}-\frac{\pi n_{p}}{2}, 
\label{eq:holonomy_image_propagator} 
\end{equation}
where $n_{p}$ is the Morse index of the stationary path. The factor $\Xi_{p}$ is common to all mass eigenstates traveling along the same image and disappears from a single path flavor probability. It survives in the interference between different images.

At the leading lensing order, it is useful to define the dimensionless source position
$ u\equiv \beta/\theta_{E}$. The arrival time separation of the two images is then
\begin{equation} 
\begin{split}
& |\Delta T_{+-}|=\mathcal{A}_{\mathrm{h}}\left[\frac{|u|}{2}\sqrt{u^{2}+4}+\ln\!\left(\frac{\sqrt{u^{2}+4}+|u|}{\sqrt{u^{2}+4}-|u|}\right)\right], \label{eq:holonomy_two_image_delay} 
\end{split}
\end{equation}
where $\Delta T_{+-}\equiv T_{+}-T_{-}$. The same combination $4M+a$ that increases the Einstein ring also controls the dominant temporal separation of the paths. This link is important because the image delay determines whether the two neutrino packets can interfere at the detector.

Since the mass dependent phase on path $p$ was derived in Eq.~\eqref{eq:far_source_holonomy_phase}, for the nearly aligned geometry, $r_{D}\simeq D_{L}$, $r_{S}\simeq D_{LS}$, and $D_{S}\simeq r_{S}+r_{D}$, it is convenient to introduce the effective phase length
\begin{equation} 
\begin{split}
& \Lambda_{p}\equiv(r_{S}+r_{D})\left(1-\frac{b_{p}^{2}}{2r_{S}r_{D}}\right)+2M+\frac{a}{2}\ln\!\left(\frac{4r_{S}r_{D}}{b_{p}^{2}}\right), \\
&  \Phi_{i}^{(p)}=\frac{m_{i}^{2}}{2E_{0}}\Lambda_{p}. \label{eq:holonomy_effective_phase_length} 
\end{split}
\end{equation}
The phase governing interference between eigenstate $i$ on path $p$ and eigenstate $j$ on path $q$ is
\begin{equation} 
\Delta\Phi_{ij}^{pq}\equiv\Phi_{i}^{(p)}-\Phi_{j}^{(q)}. 
\label{eq:cross_path_mass_phase} 
\end{equation}
In order to highlight its physical content, we define
\begin{equation}
\begin{split}
& \Delta m_{ij}^{2}\equiv m_{i}^{2}-m_{j}^{2}, \qquad \Sigma m_{ij}^{2}\equiv m_{i}^{2}+m_{j}^{2}, \\
& \Delta b_{pq}^{2}\equiv b_{p}^{2}-b_{q}^{2}, \qquad \Sigma b_{pq}^{2}\equiv b_{p}^{2}+b_{q}^{2}. \label{eq:multipath_mass_impact_combinations} 
\end{split}
\end{equation}
The average phase length of the two paths is
\begin{equation} 
\begin{split}
\overline{\Lambda}_{pq} \equiv & \,\, \frac{\Lambda_{p} + \Lambda_{q}}{2} = r_{S} + r_{D} + 2M \\
& -\frac{(r_{S}+r_{D})\Sigma b_{pq}^{2}}{4r_{S}r_{D}}+\frac{a}{2}\ln\!\left(\frac{4r_{S}r_{D}}{b_{p}b_{q}}\right), \label{eq:average_holonomy_phase_length} 
\end{split} 
\end{equation}
while their difference is
\begin{equation} 
\Delta\Lambda_{pq}\equiv\Lambda_{p}-\Lambda_{q}=-\frac{(r_{S}+r_{D})\Delta b_{pq}^{2}}{2r_{S}r_{D}}+a\ln\!\left(\frac{b_{q}}{b_{p}}\right). 
\label{eq:difference_holonomy_phase_length} 
\end{equation}
The cross path phase separates into
\begin{equation} 
\Delta\Phi_{ij}^{pq}=\frac{\Delta m_{ij}^{2}}{2E_{0}}\overline{\Lambda}_{pq}+\frac{\Sigma m_{ij}^{2}}{4E_{0}}\Delta\Lambda_{pq}. 
\label{eq:decomposed_holonomy_cross_phase} 
\end{equation}
For $p=q$, we notice that $\Delta\Lambda_{pp}=0$, and Eq.~\eqref{eq:decomposed_holonomy_cross_phase} reduces to the usual dependence on $\Delta m_{ij}^{2}$. For $p\neq q$, the second term probes $\Sigma m_{ij}^{2}$ and the absolute neutrino mass scale \cite{CrockerGiuntiMortlock2004,AlexandreClough2018}. In the spacetime considered throughout the manuscript, this contribution contains not only the difference between $b_{p}^{2}$ and $b_{q}^{2}$ but also the genuinely holonomy induced term $a\ln(b_{q}/b_{p})$. The latter vanishes for identical paths and cannot be absorbed into a redefinition of the black hole mass.

Now, let us combine the lensed trajectories at the amplitude level, define the single image flavor amplitude
\begin{equation} 
\mathcal{A}_{\alpha\rightarrow\beta}^{(p)}\equiv\sum_{i=1}^{3}U_{\beta i}U_{\alpha i}^{*}\exp\!\left[-i\Phi_{i}^{(p)}\right]. 
\label{eq:single_image_flavor_amplitude} 
\end{equation}
If the two images remain fully coherent, their joint amplitude is
\begin{equation} \mathcal{A}_{\alpha\rightarrow\beta}^{\mathrm{lens}} = \frac{1}{\sqrt{\mathcal{Z}_{\alpha}}}\sum_{p=\pm}\sqrt{\mu_{p}}\exp\!\left(i\Xi_{p}\right)\mathcal{A}_{\alpha\rightarrow\beta}^{(p)}. \label{eq:coherent_lensed_flavor_amplitude} \end{equation}
A realistic description must also allow incomplete overlap of the image wave packets. We introduce $\Gamma_{pq}$, with $\Gamma_{pp} = 1$ and $\Gamma_{qp} = \Gamma_{pq}^{*}$, and write the reduced density operator in the mass basis as
\begin{equation} 
\begin{split}
\rho_{\alpha}(D) = & \frac{1}{\mathcal{Z}_{\alpha}}\sum_{p,q=\pm}\sum_{i,j=1}^{3}\sqrt{\mu_{p}\mu_{q}}\,\Gamma_{pq}\\
& \times \exp\!\left[i(\Xi_{p}-\Xi_{q})-i\Delta\Phi_{ij}^{pq}\right]U_{\alpha i}^{*}U_{\alpha j}|\nu_{i}\rangle\langle\nu_{j}|. \label{eq:lensed_neutrino_density_operator} 
\end{split}
\end{equation}
The normalization imposed by $\operatorname{Tr}\rho_{\alpha} = 1$ is
\begin{equation} 
\begin{split}
\mathcal{Z}_{\alpha} & = \sum_{p,q=\pm}\sqrt{\mu_{p}\mu_{q}}\,\,\Gamma_{pq}\exp\!\left[i(\Xi_{p}-\Xi_{q})\right] \\
& \times \sum_{i=1}^{3}|U_{\alpha i}|^{2}\exp\!\left[-i\left(\Phi_{i}^{(p)}-\Phi_{i}^{(q)}\right)\right]. \label{eq:lensed_state_normalization} 
\end{split}
\end{equation}
Hermiticity of $\Gamma_{pq}$ and the antisymmetry of the phase differences ensure that $\mathcal{Z}_{\alpha}$ is real. Projecting Eq.~\eqref{eq:lensed_neutrino_density_operator} onto the detected flavor gives the conditional transition probability
\begin{equation} 
\begin{split}
P_{\alpha\rightarrow\beta}^{\mathrm{lens}} = & \,\, \frac{1}{\mathcal{Z}_{\alpha}}\sum_{p,q=\pm}\sum_{i,j=1}^{3}\sqrt{\mu_{p}\mu_{q}}\,\,\Gamma_{pq}\\
& \times \exp\!\left[i(\Xi_{p}-\Xi_{q})-i\Delta\Phi_{ij}^{pq}\right]U_{\beta i}U_{\alpha i}^{*}U_{\beta j}^{*}U_{\alpha j}. \label{eq:holonomy_lensed_flavor_probability} 
\end{split}
\end{equation}
Unitarity of $U$ guarantees $\sum_{\beta}P_{\alpha\rightarrow\beta}^{\mathrm{lens}}=1$. The terms with $p=q$ contain the ordinary flavor interference associated with each image, whereas the terms with $p\neq q$ carry the joint geometrical, Morse, and mass dependent phases.

For Gaussian packets, the overlap of two image contributions may be represented by \cite{CrockerGiuntiMortlock2004}
\begin{equation} 
\Gamma_{pq}\simeq\exp\!\left[-\frac{(T_{p}-T_{q})^{2}}{8\sigma_{t}^{2}}-\frac{(\theta_{p}-\theta_{q})^{2}}{8\sigma_{\theta}^{2}}\right], 
\label{eq:gaussian_image_overlap} 
\end{equation}
where $\sigma_{t}$ is the temporal width of the packet and $\sigma_{\theta}$ characterizes the angular acceptance of the detector. Multipath interference requires
\begin{equation} 
|T_{p}-T_{q}|\lesssim\sigma_{t}, \qquad |\theta_{p}-\theta_{q}|\lesssim\sigma_{\theta}.
\label{eq:multipath_coherence_conditions} 
\end{equation}
Angularly unresolved images need not be coherent: a delay exceeding the packet duration removes the cross path terms even when the detector cannot separate the images on the sky.

Coherence among different mass eigenstates must survive as well. For a Gaussian packet with spatial width $\sigma_{x}$, the vacuum coherence length is approximately \cite{GiuntiKimLee1998}
\begin{equation} 
L_{\mathrm{coh},ij}\simeq\frac{4\sqrt{2}E_{0}^{2}\sigma_{x}}{|\Delta m_{ij}^{2}|}. 
\label{eq:neutrino_mass_coherence_length} 
\end{equation}
The proper length of a lensed path in the holonomy geometry is
\begin{equation}
\begin{split}
\ell_{p} = & \int_{r_{0,p}}^{r_{S}} \frac{\mathrm{d}r} {\sqrt{F(r)H(r)}  \sqrt{1-\dfrac{b_{p}^{2}F(r)}{r^{2}}}} \\
& + \int_{r_{0,p}}^{r_{D}} \frac{\mathrm{d}r} {\sqrt{F(r)H(r)}  \sqrt{1-\dfrac{b_{p}^{2}F(r)}{r^{2}}}},
\label{eq:proper_length_lensed_neutrino_path}
\end{split}
\end{equation}
and flavor interference remains appreciable when
$ \ell_{p}\lesssim L_{\mathrm{coh},ij}$. The loss of overlap suppresses the off diagonal components of the density operator but does not alter the phase accumulated by any surviving component. In the fully incoherent image limit, $\Gamma_{+-}=0$, Eq.~\eqref{eq:holonomy_lensed_flavor_probability} reduces to
\begin{equation} 
P_{\alpha\rightarrow\beta}^{\mathrm{inc}}=\frac{\mu_{+}P_{\alpha\rightarrow\beta}^{(+)}+\mu_{-}P_{\alpha\rightarrow\beta}^{(-)}}{\mu_{+}+\mu_{-}}. 
\label{eq:incoherent_image_probability} 
\end{equation}

At this stage, the holonomy parameter enters the observable pattern through four related channels: it changes the deflection coefficients and, as a natural consequence, the image positions, modifies the magnification weights, shifts the arrival time difference that controls path overlap, and adds a logarithmic contribution to the mass dependent cross path phase. It also enlarges the proper path length through $H(r)^{-1/2}$, which can move a given trajectory closer to the boundary of mass-eigenstate coherence. In the limit $a\rightarrow0$, the weak Schwarzschild lens map, the standard image delay, and the usual multipath neutrino probability are recovered. 

%%%%%%%%%%%%%%%%%%%%%%%%%%%%%%%%%%%%%%%%%%%%%%%%%%%%%%%%%%%%%%%%%%%%%%%%%%%%%%%%%%%%%%%%%%%%%%%%%%%%%%%%%%%%%%%%%%%%%%%%%%%%%%%%%%%%%%%%%%%%%%%%%%%%%%%%%%%%%%%%%%%%%%%%%%%%%%%%%%%%%%%%%%%%%%%%%%%%%%%%%%%%%%%%%%%%%%%%%%%%%%%%%%%%%%%%%%%%%%%%%%%%%%%%%%%%%%%%%%%%%%%%%%%%%%%%%%%%%%%%%%%%%%%%%%%%%%%%%%%%%%%%%%%%%%%%%%%%%%%%%%%%%%%%%%%%%%%%%%%%%%%%%%%%%%%%%%%%%%%%%%%%%%%%%%%%%%%%%%%%%%%%%%%%%%%%%%%%%%%%%%%%%%%%%%%%%%%%%%%%%%%%%%%%%%%%%%%%%%%%%%%%%%%%%%%%%%%%%%%

\section{Holonomy shaped coherence and flavor mode entanglement}
\label{sec:holonomy_quantum_correlations}

A single oscillating neutrino does not produce entanglement between different particles. Its quantum correlations arise instead from the coherent occupation of the electron, muon, and tau flavor modes \cite{BlasoneEtAl2008,BlasoneEtAl2009,BanerjeeEtAl2015}. Gravitational lensing enriches this structure because each flavor amplitude receives contributions from distinct spacetime trajectories. Once the unresolved path degree of freedom has been incorporated through the overlap factors $\Gamma_{pq}$ introduced in Eq.~\eqref{eq:gaussian_image_overlap}, the state arriving at the detector is described by the mass basis density operator $\rho_{\alpha}(D)$ of Eq.~\eqref{eq:lensed_neutrino_density_operator}.

The corresponding density operator in the flavor basis is
\begin{equation} 
\begin{split}
& \varrho_{\alpha}^{(f)}(D) =  \sum_{\beta,\gamma=e,\mu,\tau}\varrho_{\alpha;\beta\gamma}^{(f)}(D)|\nu_{\beta}\rangle\langle\nu_{\gamma}|, \\
& \varrho_{\alpha;\beta\gamma}^{(f)}(D)=\sum_{i,j=1}^{3}U_{\beta i}\rho_{\alpha;ij}(D)U_{\gamma j}^{*}. \label{eq:holonomy_flavor_density_operator} 
\end{split}
\end{equation}
Its diagonal entries are precisely the lensed flavor probabilities,
\begin{equation} 
\varrho_{\alpha;\beta\beta}^{(f)}(D)=P_{\alpha\rightarrow\beta}^{\mathrm{lens}}, \qquad \sum_{\beta=e,\mu,\tau}P_{\alpha\rightarrow\beta}^{\mathrm{lens}}=1, 
\label{eq:flavor_density_diagonal} 
\end{equation}
whereas the off diagonal elements retain the coherence between distinct flavor modes. To display the geometric content compactly, we define the path--mass kernel
\begin{equation} 
\mathcal{K}_{ij}^{pq}(a)\equiv\frac{\sqrt{\mu_{p}\mu_{q}}}{\mathcal{Z}_{\alpha}}\, \Gamma_{pq}\exp\!\left[i(\Xi_{p}-\Xi_{q})-i\Delta\Phi_{ij}^{pq}\right],
\label{eq:holonomy_path_mass_kernel} 
\end{equation}
in terms of which the flavor matrix elements become
\begin{equation} 
\varrho_{\alpha;\beta\gamma}^{(f)}(D) = \sum_{p,q=\pm}\sum_{i,j=1}^{3}\mathcal{K}_{ij}^{pq}(a)U_{\beta i}U_{\gamma j}^{*}U_{\alpha i}^{*}U_{\alpha j}. 
\label{eq:explicit_holonomy_flavor_matrix} 
\end{equation}
Eq.~\eqref{eq:explicit_holonomy_flavor_matrix} gathers every gravitational contribution into a single object: $a$ changes the image weights $\mu_{p}$, the Fermat phases $\Xi_{p}$, the overlap factors $\Gamma_{pq}$, and the cross path mass phases $\Delta\Phi_{ij}^{pq}$.

The three flavor modes may be represented as qubits restricted to the single excitation sector,
\begin{equation} 
|\nu_{e}\rangle\equiv|100\rangle, \qquad |\nu_{\mu}\rangle\equiv|010\rangle, \qquad |\nu_{\tau}\rangle\equiv|001\rangle. 
\label{eq:flavor_single_excitation_mapping} 
\end{equation}
Accordingly, the flavor density operator may be written as
\begin{equation} 
\varrho_{\alpha}^{(f)}(D)=\sum_{\beta,\gamma = e,\mu,\tau}\varrho_{\alpha;\beta\gamma}^{(f)}(D)|1_{\beta}\rangle\langle1_{\gamma}|, 
\label{eq:single_excitation_flavor_density} 
\end{equation}
where $|1_{\beta}\rangle$ denotes the configuration in which only mode $\beta$ is occupied. This mapping introduces no additional particle: it partitions the Hilbert space according to flavor occupation and thereby gives an operational meaning to flavor mode entanglement.

When the lensed packets remain fully coherent and no averaging over energy or source position is performed, $\varrho_{\alpha}^{(f)}$ is pure. The detected state can then be expressed as
\begin{equation} 
|\psi_{\alpha}^{\mathrm{lens}}\rangle=\sum_{\beta=e,\mu,\tau}a_{\alpha\beta}^{\mathrm{lens}}|1_{\beta}\rangle, \qquad \sum_{\beta=e,\mu,\tau}|a_{\alpha\beta}^{\mathrm{lens}}|^{2}=1, 
\label{eq:pure_lensed_flavor_state} 
\end{equation}
with amplitudes
\begin{equation} 
a_{\alpha\beta}^{\mathrm{lens}}=\frac{1}{\sqrt{\mathcal{Z}_{\alpha}}}\sum_{p=\pm}\sqrt{\mu_{p}}\exp(i\Xi_{p})\sum_{i=1}^{3}U_{\beta i}U_{\alpha i}^{*}\exp\!\left[-i\Phi_{i}^{(p)}\right]. 
\label{eq:pure_holonomy_flavor_amplitudes} 
\end{equation}
The measurable populations satisfy
$P_{\alpha\rightarrow\beta}^{\mathrm{lens}}=|a_{\alpha\beta}^{\mathrm{lens}}|^{2}$. Unlike a single path state, Eq.~\eqref{eq:pure_holonomy_flavor_amplitudes} combines the magnification, Morse phase, arrival time, and mass phase of both images before the flavor projection is performed.

For a pure state in the single mexcitation sector, an entropy based measure of tripartite flavor mode entanglement is obtained from the one mode reductions \cite{BlasoneEtAl2008,BanerjeeEtAl2015},
\begin{equation} 
\begin{split}
& \mathcal{E}_{3}^{(\alpha)}\equiv\frac{1}{2}\left[S(\varrho_{e}) +S(\varrho_{\mu})+S(\varrho_{\tau})\right], \\
& S(\varrho_{\beta})\equiv-\operatorname{Tr}\!\left(\varrho_{\beta}\log_{2}\varrho_{\beta}\right). \label{eq:tripartite_flavor_entropy} 
\end{split}
\end{equation}
Since the eigenvalues of $\varrho_{\beta}$ are $P_{\alpha\rightarrow\beta}^{\mathrm{lens}}$ and $1-P_{\alpha\rightarrow\beta}^{\mathrm{lens}}$, the measure reduces to
\begin{equation} 
\mathcal{E}_{3}^{(\alpha)}=\frac{1}{2}\sum_{\beta=e,\mu,\tau}h_{2}\!\left(P_{\alpha\rightarrow\beta}^{\mathrm{lens}}\right), 
\label{eq:tripartite_entanglement_probabilities}
\end{equation}
where $h_{2}(x)\equiv-x\log_{2}x-(1-x)\log_{2}(1-x)$. It vanishes when the neutrino occupies a definite flavor mode and reaches its upper value for an equally populated coherent state,
\begin{equation} 
0\leq\mathcal{E}_{3}^{(\alpha)}\leq\frac{3}{2}h_{2}\!\left(\frac{1}{3}\right)\simeq1.37744. 
\label{eq:tripartite_entanglement_bound}
\end{equation}

The probability representation in Eq.~\eqref{eq:tripartite_entanglement_probabilities} applies only to a pure flavor state. Temporal separation, finite angular resolution, mass wave packet separation, or averaging over an extended source generally produces a mixed state. Its degree of mixing is measured directly by the purity
\begin{equation} 
\mathcal{P}_{\alpha}\equiv\operatorname{Tr}\!\left[\left(\varrho_{\alpha}^{(f)}\right)^{2}\right]=\sum_{\beta=e,\mu,\tau}\left(P_{\alpha\rightarrow\beta}^{\mathrm{lens}}\right)^{2}+2\sum_{\beta<\gamma}\left|\varrho_{\alpha;\beta\gamma}^{(f)}\right|^{2},  
\label{eq:lensed_flavor_purity} 
\end{equation}
wuth $\frac{1}{3}\leq\mathcal{P}_{\alpha}\leq1$. For mixed states, the entropy average above must be extended through the convex roof,
\begin{equation}
\begin{split}
& \mathcal{E}_{3}^{\mathrm{cr}}\!\left(\varrho_{\alpha}^{(f)}\right)\equiv\inf_{\{w_{n},|\psi_{n}\rangle\}}\sum_{n}w_{n}\mathcal{E}_{3}(|\psi_{n}\rangle), \\
&\varrho_{\alpha}^{(f)}=\sum_{n}w_{n}|\psi_{n}\rangle\langle\psi_{n}|. \label{eq:convex_roof_flavor_entanglement} 
\end{split}
\end{equation}
This prescription prevents a classical statistical distribution over flavor outcomes from being mistaken for entanglement.

A quantity available without optimizing over pure state decompositions is the $l_{1}$ norm of coherence \cite{BaumgratzCramerPlenio2014},
\begin{equation} 
C_{l_{1}}\!\left(\varrho_{\alpha}^{(f)}\right)\equiv\sum_{\beta\neq\gamma}\left|\varrho_{\alpha;\beta\gamma}^{(f)}\right|=2\left(\left|\varrho_{\alpha;e\mu}^{(f)}\right|+\left|\varrho_{\alpha;e\tau}^{(f)}\right|+\left|\varrho_{\alpha;\mu\tau}^{(f)}\right|\right). 
\label{eq:l1_flavor_coherence} 
\end{equation}
For the pure state of Eq.~\eqref{eq:pure_lensed_flavor_state}, this expression becomes
\begin{equation} 
C_{l_{1}}^{(\alpha)}=2\left[\sqrt{P_{\alpha\rightarrow e}^{\mathrm{lens}}P_{\alpha\rightarrow\mu}^{\mathrm{lens}}}+\sqrt{P_{\alpha\rightarrow e}^{\mathrm{lens}}P_{\alpha\rightarrow\tau}^{\mathrm{lens}}}+\sqrt{P_{\alpha\rightarrow\mu}^{\mathrm{lens}}P_{\alpha\rightarrow\tau}^{\mathrm{lens}}}\right], 
\label{eq:pure_lensed_flavor_coherence} 
\end{equation}
with $0\leq C_{l_{1}}^{(\alpha)}\leq2$.  Complete flavor dephasing preserves the diagonal probabilities but drives $C_{l_{1}}$ to zero. The coherence measure separates a quantum superposition from an incoherent ensemble with identical flavor populations.

Pairwise flavor correlations are obtained by tracing over the unobserved mode. For two distinct flavors $\beta$ and $\gamma$, let $\delta$ denote the remaining flavor. In the ordered basis $\{|00\rangle,|01\rangle,|10\rangle,|11\rangle\}$, the reduced two mode state is
\begin{equation} 
\varrho_{\beta\gamma} = \begin{pmatrix}P_{\alpha\rightarrow\delta}^{\mathrm{lens}}&0&0&0\\0&P_{\alpha\rightarrow\gamma}^{\mathrm{lens}}&\varrho_{\alpha;\gamma\beta}^{(f)}&0\\0&\varrho_{\alpha;\beta\gamma}^{(f)}&P_{\alpha\rightarrow\beta}^{\mathrm{lens}}&0\\0&0&0&0\end{pmatrix}. 
\label{eq:reduced_two_flavor_mode_state} 
\end{equation}
The concurrence of this state is \cite{Wootters1998}
\begin{equation} 
\mathcal{C}_{\beta\gamma}=2\left|\varrho_{\alpha;\beta\gamma}^{(f)}\right|. 
\label{eq:pairwise_flavor_concurrence} 
\end{equation}
For a pure flavor state, it reduces to
\begin{equation} 
\mathcal{C}_{\beta\gamma}=2\sqrt{P_{\alpha\rightarrow\beta}^{\mathrm{lens}}P_{\alpha\rightarrow\gamma}^{\mathrm{lens}}}, 
\label{eq:pure_pairwise_flavor_concurrence} 
\end{equation}
and the associated entanglement of formation is
\begin{equation} 
\mathcal{E}_{\beta\gamma}=h_{2}\!\left(\frac{1+\sqrt{1-\mathcal{C}_{\beta\gamma}^{2}}}{2}\right). 
\label{eq:flavor_entanglement_of_formation} 
\end{equation}
An independent measure that remains applicable to the mixed two mode state is the negativity \cite{VidalWerner2002},
\begin{equation} 
\mathcal{N}_{\beta\gamma}=\frac{1}{2}\left[\sqrt{\left(P_{\alpha\rightarrow\delta}^{\mathrm{lens}}\right)^{2}+4\left|\varrho_{\alpha;\beta\gamma}^{(f)}\right|^{2}}-P_{\alpha\rightarrow\delta}^{\mathrm{lens}}\right]. 
\label{eq:pairwise_flavor_negativity} 
\end{equation}
The single excitation structure further implies the exact identity
\begin{equation} 
C_{l_{1}}\!\left(\varrho_{\alpha}^{(f)}\right)=\mathcal{C}_{e\mu}+\mathcal{C}_{e\tau}+\mathcal{C}_{\mu\tau}. 
\label{eq:coherence_concurrence_identity}
\end{equation}
This equality is special to the flavor mode partition of a single neutrino and does not hold for a generic three qubit density operator.

The Bell--CHSH inequality provides another test of pairwise nonclassicality \cite{ClauserEtAl1969}. Defining the correlation matrix by
\begin{equation} 
\left(T_{\beta\gamma}\right)_{mn}\equiv\operatorname{Tr}\!\left[\varrho_{\beta\gamma}(\sigma_{m}\otimes\sigma_{n})\right], \qquad m,n=1,2,3, 
\label{eq:flavor_chsh_correlation_matrix} 
\end{equation}
the Horodecki criterion gives the maximal Bell parameter \cite{HorodeckiEtAl1995},
\begin{equation} 
\mathcal{B}_{\beta\gamma}^{\max}=2\sqrt{4\left|\varrho_{\alpha;\beta\gamma}^{(f)}\right|^{2}+\max\!\left\{4\left|\varrho_{\alpha;\beta\gamma}^{(f)}\right|^{2},\left(2P_{\alpha\rightarrow\delta}^{\mathrm{lens}}-1\right)^{2}\right\}}. 
\label{eq:maximal_flavor_chsh_parameter} 
\end{equation}
Bell nonlocality occurs when $\mathcal{B}_{\beta\gamma}^{\max}>2$. A nonzero concurrence does not guarantee this violation, since entanglement and Bell nonlocality impose different conditions on mixed states.

The distribution of pairwise Bell correlations across the three flavor modes may be summarized by \cite{BanerjeeEtAl2015}
\begin{equation} 
\Sigma_{3}^{(\alpha)}\equiv\left(\mathcal{B}_{e\mu}^{\max}\right)^{2}+\left(\mathcal{B}_{e\tau}^{\max}\right)^{2}+\left(\mathcal{B}_{\mu\tau}^{\max}\right)^{2}, \qquad \Sigma_{3}^{(\alpha)}\leq12.
\label{eq:tripartite_bell_shareability} 
\end{equation}
A pronounced Bell violation in one flavor pair restricts the nonlocality that can be shared with the remaining mode. The quantity $\Sigma_{3}^{(\alpha)}$ tracks how lensing redistributes pairwise nonlocal correlations.

Notice that the holonomy dependence of these measures is inherited from the lensed density operator. At fixed impact parameter, the direct phase response follows from Eq.~\eqref{eq:holonomy_effective_phase_length},
\begin{equation} 
\left.\frac{\partial\Phi_{i}^{(p)}}{\partial a}\right|_{b_{p}}=\frac{m_{i}^{2}}{4E_{0}}\ln\!\left(\frac{4r_{S}r_{D}}{b_{p}^{2}}\right). 
\label{eq:direct_holonomy_phase_response} 
\end{equation}
The complete response also contains the $a$ dependence of $b_{p}$, $\mu_{p}$, $T_{p}$, and $\Gamma_{pq}$. For any correlation functional $\mathcal{Q}\in\{\mathcal{E}_{3}^{\mathrm{cr}},C_{l_{1}},\mathcal{C}_{\beta\gamma},\mathcal{E}_{\beta\gamma},\mathcal{N}_{\beta\gamma},\mathcal{B}_{\beta\gamma}^{\max},\Sigma_{3}^{(\alpha)}\}$, a useful feature is the holonomy residual
\begin{equation} 
\Delta_{a}\mathcal{Q}_{\alpha}\equiv\mathcal{Q}\!\left[\varrho_{\alpha}^{(f)}(a)\right]-\mathcal{Q}\!\left[\varrho_{\alpha}^{(f)}(0)\right]. 
\label{eq:holonomy_correlation_residual} 
\end{equation}
This subtraction isolates the deformation from the Schwarzschild background without confusing it with an overall normalization of the lensed flux.

The geometry does not create flavor mode entanglement when neutrino mixing is absent. It reshapes the amplitudes and phases through which the PMNS superposition is distributed among the three modes. If the holonomy induced displacement drives the coherent state toward a balanced flavor population, $\mathcal{E}_{3}^{(\alpha)}$ and $C_{l_{1}}^{(\alpha)}$ increase; if the modified image delay or proper path length destroys overlap, the off diagonal matrix elements, concurrence, negativity, and coherence decrease. Because these quantities depend nonlinearly on $\varrho_{\alpha}^{(f)}$, no universal multiplicative factor can convert their Schwarzschild values into the holonomy-corrected ones. One coment here is important for the sake of clarification: the current construction concerns correlations among flavor occupation modes after the unresolved trajectories have been incorporated. Entanglement between the path and flavor degrees of freedom requires retaining the path labels as an explicit quantum subsystem and constitutes a separate layer of the analysis.

%%%%%%%%%%%%%%%%%%%%%%%%%%%%%%%%%%%%%%%%%%%%%%%%%%%%%%%%%%%%%%%%%%%%%%%%%%%%%%%%%%%%%%%%%%%%%%%%%%%%%%%%%%%%%%%%%%%%%%%%%%%%%%%%%%%%%%%%%%%%%%%%%%%%%%%%%%%%%%%%%%%%%%%%%%%%%%%%%%%%%%%%%%%%%%%%%%%%%%%%%%%%%%%%%%%%%%%%%%%%%%%%%%%%%%%%%%%%%%%%%%%%%%%%%%%%%%%%%%%%%%%%%%%%%%%%%%%%%%%%%%%%%%%%%%%%%%%%%%%%%%%%%%%%%%%%%%%%%%%%%%%%%%%%%%%%%%%%%%%%%%%%%%%%%%%%%%%%%%%%%%%%%%%%%%%%%%%%%%%%%%%%%%%%%%%%%%%%%%%%%%%%%%%%%%%%%%%%%%%%%%%%%%%%%%%%%%%%%%%%%%%%%%%%%%%%%%%%%%%

\section{Trajectory flavor entanglement in the holonomy lens}
\label{sec:holonomy_trajectory_flavor_entanglement}

Weak gravitational lensing supplies the propagating neutrino with an external two level degree of freedom: the positive-- and negative--parity images define two distinguishable trajectories, while flavor remains an internal three level sector. If both images are coherently populated and the relative mass phases are not the same on the two rays, the complete state cannot be factorized into independent trajectory and flavor states. The resulting correlation is an intraparticle entanglement between commuting degrees of freedom of a single neutrino, analogous to path  internal state entanglement in matter wave interferometry; in other words, it does not describe nonlocality between different particles \cite{HasegawaEtAl2003,CrockerGiuntiMortlock2004,AlexandreClough2018}.

Before recombination at the detector, we represent the two weak--field images by the orthonormal path states $|+\rangle_{P}$ and $|-\rangle_{P}$, with $\langle p|q\rangle_{P}=\delta_{pq}$. Since the magnifications in Eq.~\eqref{eq:holonomy_image_magnifications} are absolute ones, their normalized weights are
\begin{equation} 
w_{p}\equiv\frac{\mu_{p}}{\mu_{+}+\mu_{-}}, \qquad w_{+}+w_{-}=1, \qquad p=\pm. 
\label{eq:holonomy_path_weights} 
\end{equation}
The flavor state transported along image $p$ may be written equivalently in the mass or flavor basis as
\begin{equation} 
\begin{split}
& |\chi_{\alpha}^{(p)}\rangle_{F} = \sum_{i=1}^{3}U_{\alpha i}^{*}\exp\!\left[-i\Phi_{i}^{(p)}\right]|\nu_{i}\rangle=\sum_{\beta=e,\mu,\tau}\mathcal{A}_{\alpha\rightarrow\beta}^{(p)}|\nu_{\beta}\rangle, \\
& \sum_{\beta=e,\mu,\tau}\left|\mathcal{A}_{\alpha\rightarrow\beta}^{(p)}\right|^{2}=1, \label{eq:path_conditioned_flavor_state} 
\end{split}
\end{equation}
where $\mathcal{A}_{\alpha\rightarrow\beta}^{(p)}$ is given by Eq.~\eqref{eq:single_image_flavor_amplitude}. In the ideal coherent limit, the joint trajectory--flavor state reaching the detector is
\begin{equation} 
\begin{split}
|\Psi_{\alpha}\rangle_{PF} = \sqrt{w_{+}}\exp(i\Xi_{+})|+\rangle_{P}\otimes|\chi_{\alpha}^{(+)}\rangle_{F}+\sqrt{w_{-}}\\
\times \exp(i\Xi_{-})|-\rangle_{P}\otimes|\chi_{\alpha}^{(-)}\rangle_{F}. \label{eq:pure_path_flavor_state} 
\end{split}
\end{equation}
The state in Eq.~\eqref{eq:pure_path_flavor_state} is normalized without an additional interference factor because the path labels are retained as orthogonal quantum modes. It is important to mention that the interference appears only when those modes are projected onto a common output channel.

The distinguishability of the internal states carried by the images is governed by their overlap
\begin{equation} 
\begin{split}
& \kappa_{\alpha}\equiv{}_{F}\langle\chi_{\alpha}^{(-)}|\chi_{\alpha}^{(+)}\rangle_{F}=\sum_{i=1}^{3}|U_{\alpha i}|^{2}\exp(-i\delta_{i}),  \label{eq:holonomy_flavor_path_overlap} 
\end{split}
\end{equation}
with $\delta_{i}\equiv\Phi_{i}^{(+)}-\Phi_{i}^{(-)}=\frac{m_{i}^{2}}{2E_{0}}\Delta\Lambda_{+-}$. Employing Eq.~\eqref{eq:difference_holonomy_phase_length}, the path induced phase of each mass eigenstate assumes the explicit form
\begin{equation} 
\delta_{i}=\frac{m_{i}^{2}}{2E_{0}}\left[-\frac{(r_{S}+r_{D})\Delta b_{+-}^{2}}{2r_{S}r_{D}}+a\ln\!\left(\frac{b_{-}}{b_{+}}\right)\right], 
\label{eq:holonomy_same_mass_path_phase} 
\end{equation}
with $\Delta b_{+-}^{2}\equiv b_{+}^{2}-b_{-}^{2}$.
Notice that the deformation leaves a direct logarithmic signature on the relative internal evolution, in addition to moving the image positions $b_{\pm}$. The modulus of the overlap is
\begin{equation} 
|\kappa_{\alpha}|^{2}=1-4\sum_{i<j}|U_{\alpha i}|^{2}|U_{\alpha j}|^{2}\sin^{2}\!\left(\frac{\delta_{i}-\delta_{j}}{2}\right),
\label{eq:holonomy_overlap_modulus} 
\end{equation}
where $\delta_{i}-\delta_{j}=\frac{\Delta m_{ij}^{2}}{2E_{0}}\Delta\Lambda_{+-}$. Only phase differences among mass components enter $|\kappa_{\alpha}|$. A common shift $m_{i}^{2}\mapsto m_{i}^{2}+m_{0}^{2}$ multiplies $\kappa_{\alpha}$ by an overall phase and leaves every pure-state trajectory--flavor entanglement measure unchanged. This is narrower than the absolute mass sensitivity of the multipath flavor probability in Eq.~\eqref{eq:decomposed_holonomy_cross_phase}, where different mass eigenstates may propagate along different images.

Tracing the pure state over flavor gives the reduced path operator
\begin{widetext}
\begin{equation} \rho_{P}^{(\alpha)}=\operatorname{Tr}_{F}\!\left(|\Psi_{\alpha}\rangle_{PF}{}_{PF}\langle\Psi_{\alpha}|\right)=\begin{pmatrix}w_{+}&\sqrt{w_{+}w_{-}}\exp[i(\Xi_{+}-\Xi_{-})]\kappa_{\alpha}\\\sqrt{w_{+}w_{-}}\exp[-i(\Xi_{+}-\Xi_{-})]\kappa_{\alpha}^{*}&w_{-}\end{pmatrix}, 
\label{eq:reduced_holonomy_path_state} 
\end{equation}
\end{widetext}
whereas tracing over the path sector yields
\begin{equation} 
\begin{split} \rho_{F}^{(\alpha)} & =  \operatorname{Tr}_{P}\!\left(|\Psi_{\alpha}\rangle_{PF}{}_{PF}\langle\Psi_{\alpha}|\right) \\
& = w_{+}|\chi_{\alpha}^{(+)}\rangle\langle\chi_{\alpha}^{(+)}|+w_{-}|\chi_{\alpha}^{(-)}\rangle\langle\chi_{\alpha}^{(-)}|. 
\label{eq:reduced_path_conditioned_flavor_state}
\end{split}
\end{equation}
Both reductions have the same nonzero eigenvalues,
\begin{equation} 
\lambda_{\pm}^{(\alpha)}=\frac{1}{2}\left[1\pm\sqrt{1-4w_{+}w_{-}\left(1-|\kappa_{\alpha}|^{2}\right)}\right]. 
\label{eq:path_flavor_schmidt_spectrum} 
\end{equation}
The entropy of either subsystem provides the entanglement entropy,
\begin{equation} 
\mathcal{E}_{PF}^{(\alpha)}=S\!\left(\rho_{P}^{(\alpha)}\right)=S\!\left(\rho_{F}^{(\alpha)}\right)=-\lambda_{+}^{(\alpha)}\log_{2}\lambda_{+}^{(\alpha)}-\lambda_{-}^{(\alpha)}\log_{2}\lambda_{-}^{(\alpha)}. 
\label{eq:holonomy_path_flavor_entropy} 
\end{equation}
For the pure $2\times3$ state, an algebraic measure is supplied by the I--concurrence \cite{RungtaEtAl2001,JakobBergou2007},
\begin{equation}
\mathcal{C}_{PF}^{(\alpha)}\equiv\sqrt{2\left\{1-\operatorname{Tr}\!\left[\left(\rho_{P}^{(\alpha)}\right)^{2}\right]\right\}}=2\sqrt{w_{+}w_{-}\left(1-|\kappa_{\alpha}|^{2}\right)}. 
\label{eq:holonomy_path_flavor_concurrence} 
\end{equation}
The negativity and logarithmic negativity carry the same Schmidt information in this ideal limit \cite{VidalWerner2002,Plenio2005},
\begin{equation} 
\begin{split}
& \mathcal{N}_{PF}^{(\alpha)}=\sqrt{w_{+}w_{-}\left(1-|\kappa_{\alpha}|^{2}\right)}=\frac{\mathcal{C}_{PF}^{(\alpha)}}{2}, \\
& \mathcal{E}_{\mathcal{N},PF}^{(\alpha)}=\log_{2}\!\left[1+2\mathcal{N}_{PF}^{(\alpha)}\right]. \label{eq:pure_path_flavor_negativities} 
\end{split}
\end{equation}

Combining Eqs.~\eqref{eq:holonomy_same_mass_path_phase}--\eqref{eq:holonomy_path_flavor_concurrence} makes the geometric dependence explicit:
\begin{widetext}
\begin{equation} \mathcal{C}_{PF}^{(\alpha)}=4\sqrt{w_{+}w_{-}\sum_{i<j}|U_{\alpha i}|^{2}|U_{\alpha j}|^{2}\sin^{2}\!\left\{\frac{\Delta m_{ij}^{2}}{4E_{0}}\left[-\frac{(r_{S}+r_{D})\Delta b_{+-}^{2}}{2r_{S}r_{D}}+a\ln\!\left(\frac{b_{-}}{b_{+}}\right)\right]\right\}}. 
\label{eq:explicit_holonomy_path_flavor_concurrence} 
\end{equation}
\end{widetext}
Eq.~\eqref{eq:explicit_holonomy_path_flavor_concurrence} separates the two ingredients required for entanglement. The prefactor $w_{+}w_{-}$ measures how evenly the lens populates the path modes, while the oscillatory kernel measures how far the transported flavor states depart from parallelism. Maximal entanglement, $\mathcal{C}_{PF}^{(\alpha)} = \mathcal{E}_{PF}^{(\alpha)}=1$, requires $w_{+}=w_{-}=1/2$ and $\kappa_{\alpha}=0$. Equal magnifications alone are insufficient: in the axial limit $b_{+}=b_{-}$, we get $\Delta\Lambda_{+-}=0$, $|\kappa_{\alpha}|=1$, and a separable state. Far from alignment the faint image becomes negligible, and the entanglement again tends to zero.

The Fermat phases $\Xi_{p}$ deserve a separate remark. Their difference governs the displacement of multipath fringes, but it does not occur in the Schmidt eigenvalues. Indeed, the transformation $|p\rangle_{P}\mapsto\exp(-i\Xi_{p})|p\rangle_{P}$ is local to the path subsystem and removes both phases from Eq.~\eqref{eq:pure_path_flavor_state}. The holonomy correction to the arrival time can, in this manner, change where an interference maximum is observed without changing the amount of pure trajectory--flavor entanglement. It affects the entanglement magnitude only indirectly, when the modified delay reduces the physical overlap of the packets.

This distinction is summarized by a complementarity identity. Defining the path predictability and flavor blind fringe visibility as \cite{Englert1996,JakobBergou2007}
\begin{equation}
\mathcal{P}_{P}^{(\alpha)}\equiv|w_{+}-w_{-}|, \qquad \mathcal{V}_{P}^{(\alpha)}\equiv2\sqrt{w_{+}w_{-}}|\kappa_{\alpha}|, 
\label{eq:holonomy_path_predictability_visibility} 
\end{equation}
we obtain
\begin{equation} 
\left(\mathcal{P}_{P}^{(\alpha)}\right)^{2} + \left(\mathcal{V}_{P}^{(\alpha)}\right)^{2}+\left(\mathcal{C}_{PF}^{(\alpha)}\right)^{2} = 1. 
\label{eq:holonomy_path_flavor_triality} 
\end{equation}
The reduction of fringe visibility has two coherent origins. An asymmetric lens reveals partial which path information through $\mathcal{P}_{P}^{(\alpha)}$, while path conditioned flavor evolution stores the remaining distinguishability as entanglement. In the same regime, the flavor purity obeys
\begin{equation} 
\operatorname{Tr}\!\left[\left(\rho_{F}^{(\alpha)}\right)^{2}\right] = 1 - 2w_{+}w_{-}\left(1-|\kappa_{\alpha}|^{2}\right)=1-\frac{1}{2}\left(\mathcal{C}_{PF}^{(\alpha)}\right)^{2}.
\label{eq:path_entanglement_flavor_purity} 
\end{equation}
Thereby, a mixed flavor state need not signal irreversible environmental noise. Even under unitary propagation, an observer who ignores the image label traces over a subsystem entangled with flavor. A coherent projection onto a superposition of $|+\rangle_{P}$ and $|-\rangle_{P}$ would instead select a conditional flavor state and could restore path sensitive oscillatory terms, which is the trajectory analogue of a quantum eraser arrangement \cite{HasegawaEtAl2003}.

Finite temporal overlap, angular resolution, and averaging over the source prevent the arriving state from remaining pure. Retaining the image labels while incorporating the overlap matrix $\Gamma_{pq}$ of Eq.~\eqref{eq:gaussian_image_overlap} gives
\begin{equation} 
\begin{split}
\rho_{PF}^{(\alpha)} = & \sum_{p,q=\pm}\sqrt{w_{p}w_{q}}\,\Gamma_{pq}\\
&\times \exp[i(\Xi_{p}-\Xi_{q})]|p\rangle_{P}{}_{P}\langle q|\otimes|\chi_{\alpha}^{(p)}\rangle_{F}{}_{F}\langle\chi_{\alpha}^{(q)}|,  \label{eq:mixed_holonomy_path_flavor_state} 
\end{split}
\end{equation}
with $\Gamma_{pp}=1$, and $\Gamma_{qp}=\Gamma_{pq}^{*}$. The entanglement of this mixed $2\times3$ state may be evaluated without choosing a pure state decomposition through the partial transpose measures
\begin{equation} 
\mathcal{N}_{PF}^{(\alpha)}=\frac{\left\|\left(\rho_{PF}^{(\alpha)}\right)^{T_{P}}\right\|_{1}-1}{2}, \qquad \mathcal{E}_{\mathcal{N},PF}^{(\alpha)}=\log_{2}\left\|\left(\rho_{PF}^{(\alpha)}\right)^{T_{P}}\right\|_{1}. 
\label{eq:mixed_path_flavor_negativity} 
\end{equation}
For $|\Gamma_{+-}|=1$, Eq.~\eqref{eq:mixed_holonomy_path_flavor_state} is locally equivalent to the pure state in Eq.~\eqref{eq:pure_path_flavor_state}. In the opposite limit, $\Gamma_{+-}=0$, it becomes
\begin{equation} 
\rho_{PF}^{(\alpha)}=w_{+}|+\rangle\langle+|\otimes|\chi_{\alpha}^{(+)}\rangle\langle\chi_{\alpha}^{(+)}|+w_{-}|-\rangle\langle-|\otimes|\chi_{\alpha}^{(-)}\rangle\langle\chi_{\alpha}^{(-)}|, 
\label{eq:incoherent_path_flavor_state} 
\end{equation}
which is separable. Different flavor probabilities in the two resolved or temporally separated images then encode classical trajectory--flavor correlations, not entanglement. Large flavor mode coherence, discussed in Sec.~\ref{sec:holonomy_quantum_correlations}, may survive in this limit and should not be used by itself as evidence for coherence between the images \cite{BittencourtEtAl2024}.

The parameter $a$ acts on the entanglement through three linked channels: it changes $w_{\pm}$ by shifting the weak--lensing map, modifies $\Delta\Lambda_{+-}$ through both $b_{\pm}(a)$ and the logarithmic term, and alters $\Gamma_{+-}$ through the arrival time separation. At fixed impact parameters, its direct phase response is
\begin{equation} 
\left.\frac{\partial(\delta_{i}-\delta_{j})}{\partial a}\right|_{b_{\pm}}=\frac{\Delta m_{ij}^{2}}{2E_{0}}\ln\!\left(\frac{b_{-}}{b_{+}}\right). 
\label{eq:direct_holonomy_path_entanglement_response} 
\end{equation}
These contributions may compete: a larger separation between the flavor states can raise the entanglement, whereas an increasingly unequal magnification pair or a loss of wave packet overlap suppresses it. No monotonic dependence on $a$ should be presumed in this way. In the Schwarzschild limit $a\rightarrow0$, the logarithmic phase disappears and the standard two image result is recovered. The geometry does not create the internal superposition by itself; instead, it regulates how neutrino mixing distributes quantum information between the trajectory and flavor sectors.

%%%%%%%%%%%%%%%%%%%%%%%%%%%%%%%%%%%%%%%%%%%%%%%%%%%%%%%%%%%%%%%%%%%%%%%%%%%%%%%%%%%%%%%%%%%%%%%%%%%%%%%%%%%%%%%%%%%%%%%%%%%%%%%%%%%%%%%%%%%%%%%%%%%%%%%%%%%%%%%%%%%%%%%%%%%%%%%%%%%%%%%%%%%%%%%%%%%%%%%%%%%%%%%%%%%%%%%%%%%%%%%%%%%%%%%%%%%%%%%%%%%%%%%%%%%%%%%%%%%%%%%%%%%%%%%%%%%%%%%%%%%%%%%%%%%%%%%%%%%%%%%%%%%%%%%%%%%%%%%%%%%%%%%%%%%%%%%%%%%%%%%%%%%%%%%%%%%%%%%%%%%%%%%%%%%%%%%%%%%%%%%%%%%%%%%%%%%%%%%%%%%%%%%%%%%%%%%%%%%%%%%%%%%%%%%%%%%%%%%%%%%%%%%%%%%%%%%%%%%

\section{Numerical analysis of holonomy on lensed neutrino observables}
\label{sec:holonomy_numerical_analysis}

We now examine the numerical content of the preceding results in a configuration designed to place the lensing geometry, flavor evolution, quantum correlations, and neutrino pair energy transfer on the same footing. Natural units, $\hbar = c = 1$, are adopted (as we did before). The black hole mass is fixed at $M=M_{\odot}$, the detector is located at $r_{D}=1\,\mathrm{AU}$, and the source is placed at $r_{S}=10^{5}\,\mathrm{AU}$. The energy measured with respect to the asymptotic Killing time is $E_{0}=10\,\mathrm{MeV}$. These distances keep both images in the weak--deflection regime while highlithing a sufficiently long phase baseline for the radial deformation of the geometry to become visible \cite{CrockerGiuntiMortlock2004,AlexandreClough2018}.

Writing the four mixing parameters as $\boldsymbol{\vartheta}=(\theta_{12},\theta_{13},\theta_{23},\delta_{\mathrm{CP}})$, the normal ordering (NO) inputs are
\begin{equation}
\begin{split}
& \boldsymbol{\vartheta}_{\mathrm{NO}} =(33.44^{\circ},8.57^{\circ},49.2^{\circ},197^{\circ}), \\
& (\Delta m_{21}^{2},\Delta m_{31}^{2})_{\mathrm{NO}}=(7.42\times10^{-5},\,2.517\times10^{-3})\,\mathrm{eV}^{2}, \label{eq:holonomy_numerical_NO_inputs} 
\end{split}
\end{equation}
whereas inverted ordering (IO) is represented by
\begin{equation} 
\begin{split}
& \boldsymbol{\vartheta}_{\mathrm{IO}}=(33.44^{\circ},8.60^{\circ},\,49.5^{\circ},282^{\circ}), \\
&(\Delta m_{21}^{2},|\Delta m_{31}^{2}|)_{\mathrm{IO}}=(7.42\times10^{-5},\,2.498\times10^{-3})\,\mathrm{eV}^{2}. \label{eq:holonomy_numerical_IO_inputs} 
\end{split}
\end{equation}
The lightest mass is set to zero in both orderings. Notice that this choice fixes the absolute mass component of the general cross path phase but does not influence the pure trajectory--flavor entanglement, which depends only on mass squared differences through Eq.~\eqref{eq:explicit_holonomy_path_flavor_concurrence}.

The source angle is sampled over $10^{-7}\leq|\beta|\leq3\times10^{-3}$. For each value of $|\beta|$, the two signed image positions are found from the weak lens equation including the term of order $M^{2}/b^{2}$, with the deflection coefficients $\mathcal{A}_{\mathrm{h}}=4M+a$ and $\mathcal{B}_{\mathrm{h}} = 3\pi(5+2\delta + \delta^{2})M^{2}/4$, where $\delta=a/(2M) = \lambda^{2}/(1+\lambda^{2})$ \cite{Soares2023}. The resulting impact parameters determine the magnifications, Fermat times, and neutrino phases without imposing the leading Einstein image approximation after the initial root estimate. The probability and entanglement curves use $600$ equal subintervals; the coherence and deposition calculations use $240$ and $180$ subintervals, respectively.

All dimensional quantities are converted to $\mathrm{eV}^{-1}$ before numerical evaluation. The computation is performed with $80$ digit working precision, accuracy and precision goals equal to $45$, and an enlarged internal precision reserve. Oscillatory phases are reduced modulo $2\pi$ before exponentiation, while every density matrix is Hermitized and normalized by its trace. The weak--field phase of Eq.~\eqref{eq:holonomy_effective_phase_length} is used in the plotted curves, and the exact radial integral containing $H(r)^{-1/2}$ is retained. At $|\beta| = 10^{-3}$ and $\lambda = 0.3$, raising the working precision from $40$ to $100$ digits leaves the displayed results unchanged:
\begin{equation}
\begin{split}
& P_{\nu_{e}\rightarrow\nu_{\mu}}^{\mathrm{lens}} = 0.2903625103986768, \\
& C_{PF}^{(e)} = 0.0494820769785711, \\
& \mathcal{P}_{P}^{2}+\mathcal{V}_{P}^{2}+\left(C_{PF}^{(e)}\right)^{2} - 1 = 0. \label{eq:holonomy_numerical_convergence} 
\end{split}
\end{equation}
The last equality also checks the normalization of the two path weights and the flavor overlap independently of the transition probability.

For the oscillation and trajectory--flavor panels, we compare (for $\lambda\in\{0,0.1,0.2,0.3\}$)
\begin{equation} 
\frac{a}{2M}=\frac{\lambda^{2}}{1+\lambda^{2}}\in\{0,0.00990099,0.0384615,0.0825688\}. 
\label{eq:holonomy_numerical_lambda_values} 
\end{equation}
The dependence on $\lambda$ is even, and the nonnegative part is therefore sufficient. The value $\lambda=0$ supplies the Schwarzschild reference. For $\lambda\neq0$, the numerical response combines the displacement of the image positions, the change in their magnifications, the modification of the arrival time difference, and the explicit logarithmic contribution $a\ln(b_{q}/b_{p})$ to the phase length. As we argued in the last sections, these effects cannot be represented by an overall rescaling of a Schwarzschild oscillation curve.

The probability plots are evaluated with $\Gamma_{+-}=0$. This path decohered prescription suppresses interference between the two images but preserves the coherent superposition of mass eigenstates along each image. Accordingly, the result is the magnification weighted sum in Eq.~\eqref{eq:incoherent_image_probability}. Such a regime is appropriate when the images remain angularly unresolved but their temporal separation exceeds the duration of the incident wave packet.

\begin{figure}[!htbp]
\centering
\makebox[\columnwidth][l]{
    \hspace*{-1.2cm}
    \includegraphics[width=1.15\columnwidth]{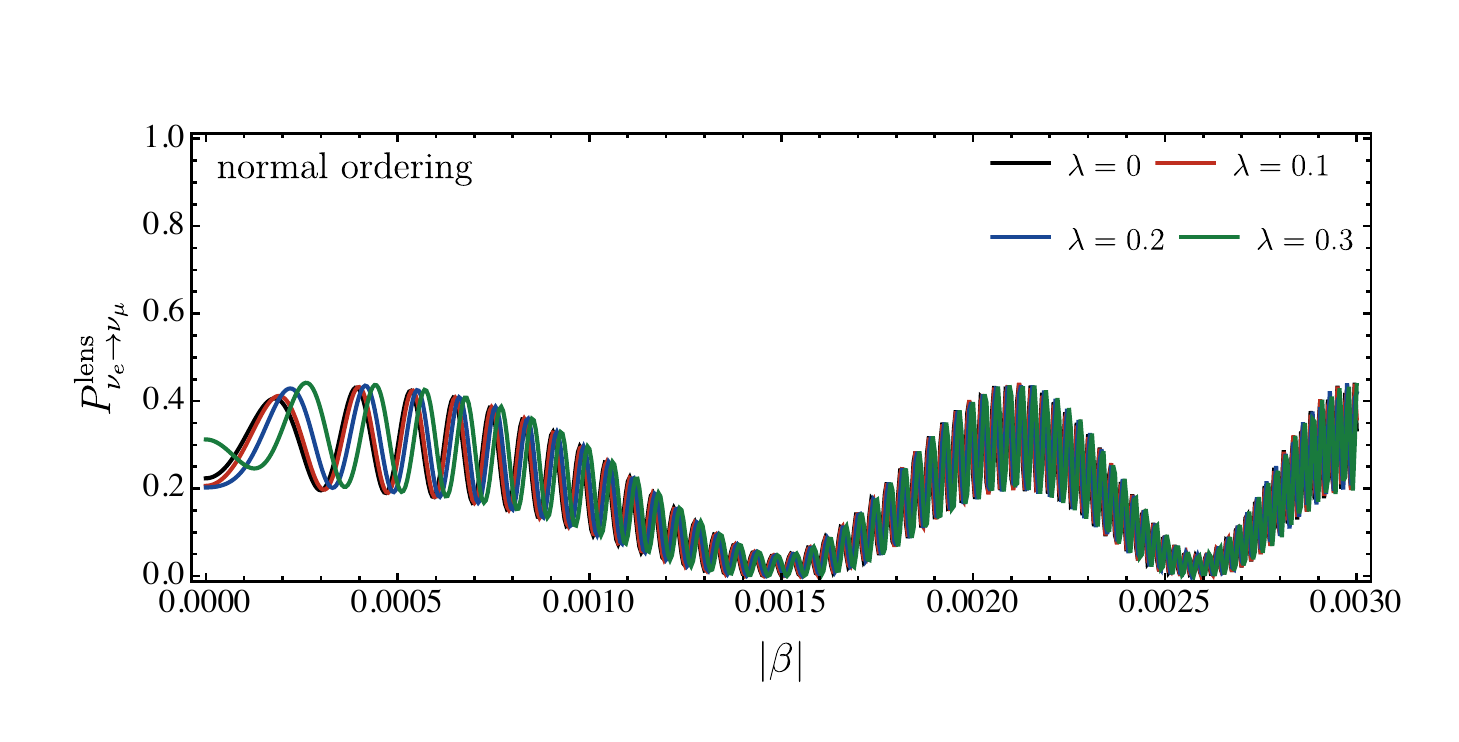}}
\caption{Weakly lensed transition probability $P_{\nu_{e}\rightarrow\nu_{\mu}}^{\mathrm{lens}}$ for normal ordering, $\Gamma_{+-}=0$, and the four holonomy values in Eq.~\eqref{eq:holonomy_numerical_lambda_values}.}
\label{fig:holonomy_probability_NO}
\end{figure}

Fig.~\ref{fig:holonomy_probability_NO} displays the electron to muon channel for normal ordering. The probability develops a sequence of increasingly rapid oscillations superposed on a slowly varying envelope. Its first group of maxima reaches approximately $0.43$, the signal is strongly suppressed around $|\beta|\simeq1.5\times10^{-3}$, and a broad oscillatory band reappears near $|\beta|\simeq2.1\times10^{-3}$. A second near zero envelope occurs around $2.55\times10^{-3}$. This modulation is generated by the nonlinear variation of the two impact parameters with the source angle. The holonomy parameter primarily changes the registration of the fringes: increasing $\lambda$ moves the extrema while leaving the broad envelope almost intact. The effect is particularly transparent in the low $|\beta|$ region, where the first maximum is displaced without being uniformly amplified.

\begin{figure}[!htbp]
\centering  \makebox[\columnwidth][l]{
    \hspace*{-1.5cm}
    \includegraphics[width=1.15\columnwidth]{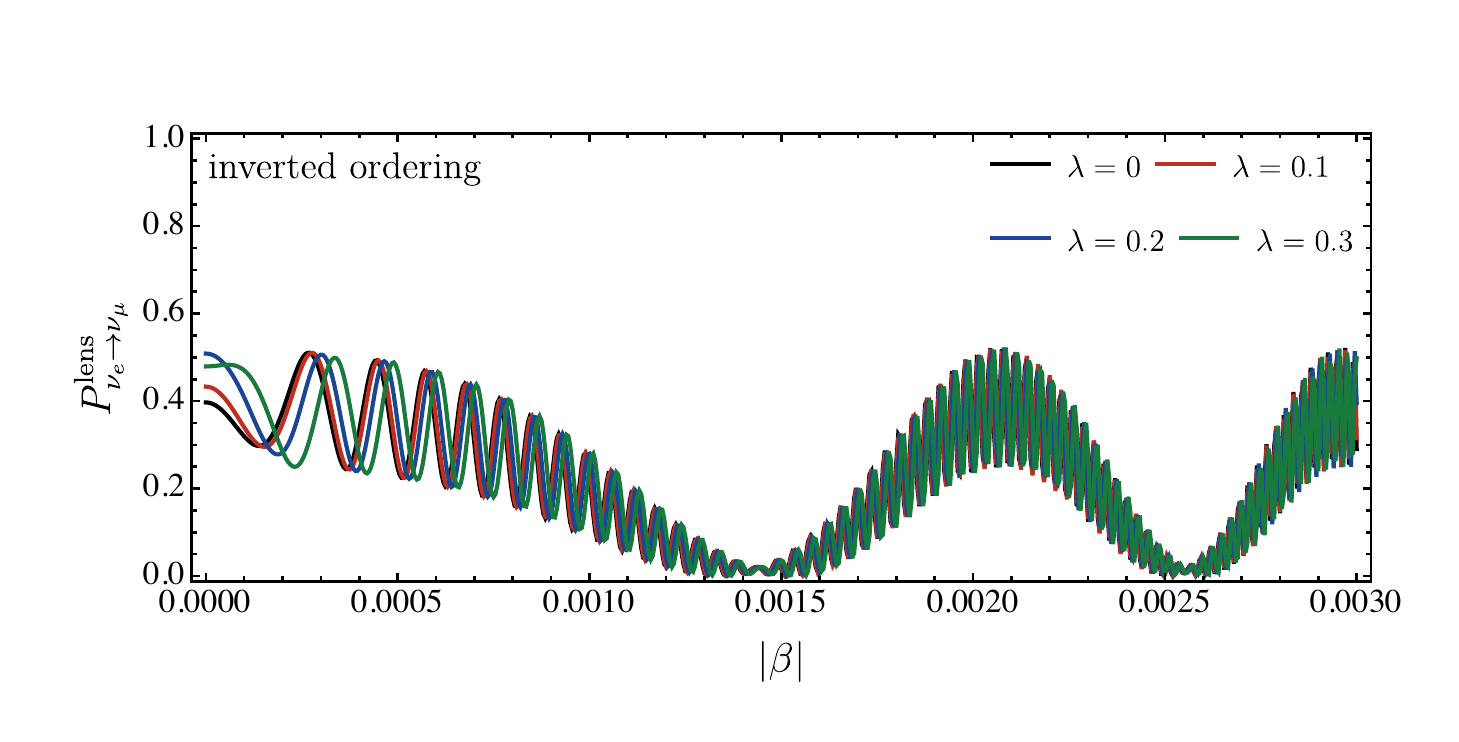}} \captionof{figure}{Electron--to--muon lensing probability for inverted ordering. The geometric configuration, $\lambda$ values, and path overlap prescription coincide with those of Fig.~\ref{fig:holonomy_probability_NO}.}     \label{fig:holonomy_probability_IO} 
\end{figure}

The inverted ordering pattern in Fig.~\ref{fig:holonomy_probability_IO} retains the same envelope nodes, since they are controlled mainly by the lens geometry, its low angle conversion begins at a larger value and the first oscillatory band reaches approximately $0.5$. The four $\lambda$ curves are again separated mainly by phase. A comparison with Fig.~\ref{fig:holonomy_probability_NO} shows that the mass ordering and the spacetime deformation leave distinguishable numerical fingerprints: changing the ordering redistributes the flavor weights within the envelope, whereas changing $\lambda$ moves the fine structure through the altered path lengths. Nevertheless, an uncertainty in $|\beta|$ can imitate part of the holonomy induced phase shift. A meaningful constraint on $\lambda$ would require either several resolved fringes or an independent reconstruction of the image geometry.

\begin{figure}[!htbp]
\centering
\makebox[\columnwidth][l]{
    \hspace*{-1.1cm}
    \includegraphics[width=1.15\columnwidth]{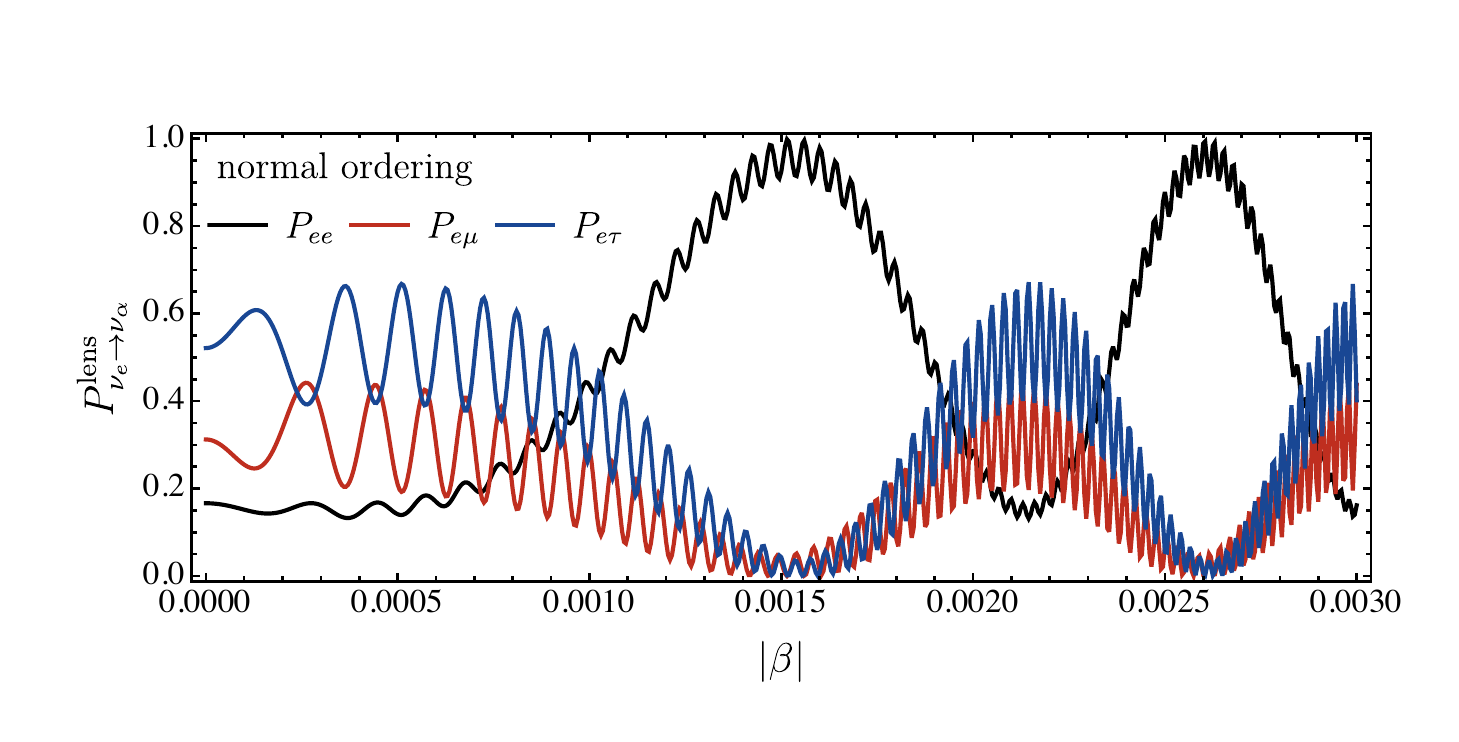}}
\caption{Complete electron--neutrino flavor budget for normal ordering at $\lambda=0.3$ and $\Gamma_{+-}=0$. The black, red, and blue curves represent $P_{ee}$, $P_{e\mu}$, and $P_{e\tau}$, respectively.}
\label{fig:holonomy_channels_NO}
\end{figure}

The three active channels at $\lambda=0.3$ are shown in Fig.~\ref{fig:holonomy_channels_NO}. For normal ordering, the small angle region is dominated by conversion into $\nu_{\tau}$, while the survival probability remains close to $0.15$ and $P_{e\mu}$ oscillates around an intermediate contribution. The character of the signal reverses as the source moves away from the optical axis: $P_{ee}$ approaches unity near $|\beta|\simeq1.55\times10^{-3}$, decreases to a broad minimum around $2.1\times10^{-3}$, and returns to a second survival maximum near $2.6\times10^{-3}$. The appearance channels jointly compensate each survival minimum, including the high frequency ripples. At every source position,
\begin{equation}
P_{ee}^{\mathrm{lens}}+P_{e\mu}^{\mathrm{lens}}+P_{e\tau}^{\mathrm{lens}}=1, 
\label{eq:holonomy_flavor_probability_closure} 
\end{equation}
so the geometry reorganizes the flavor content without attenuating the total neutrino probability.

\begin{figure}[!htbp]
\centering
\makebox[\columnwidth][l]{
    \hspace*{-1.2cm}
    \includegraphics[width=1.15\columnwidth]{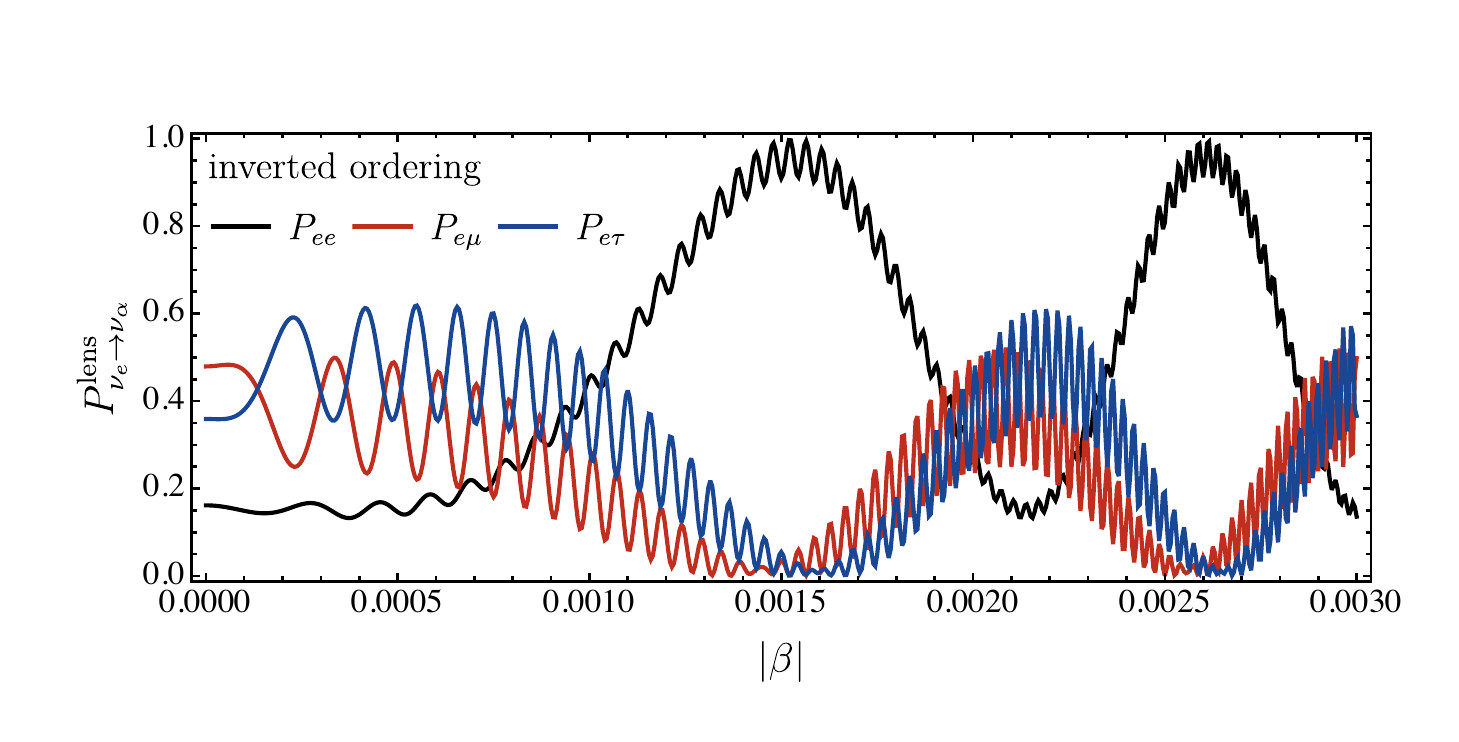}}
\caption{Three channel decomposition for inverted ordering at $\lambda=0.3$ and vanishing cross image overlap.}
\label{fig:holonomy_channels_IO}
\end{figure}

Fig.~\ref{fig:holonomy_channels_IO} gives the corresponding IO decomposition. The locations of the broad survival maxima remain close to their NO counterparts, while the division of the appearance signal changes markedly at small $|\beta|$: the muon channel is increased and the tau channel is reduced. In the central conversion band, both appearance probabilities exhibit comparable rapid modulations, although their local maxima need not coincide. Notice that this contrast illustrates why a three channel measurement is more informative than a single transition probability. A geometric phase shift moves the oscillatory structure of every channel while preserving Eq.~\eqref{eq:holonomy_flavor_probability_closure}; the ordering also changes how the converted component is shared between $\nu_{\mu}$ and $\nu_{\tau}$.

The next four figures concern a different operational limit. The two image labels are retained as a coherent quantum subsystem, so the state is the pure trajectory--flavor state of Eq.~\eqref{eq:pure_path_flavor_state}. This corresponds to $|\Gamma_{+-}|=1$ before any uncontrolled path information is discarded. It is not the same regime used in Figs.~\ref{fig:holonomy_probability_NO}--\ref{fig:holonomy_channels_IO}: the two sets deliberately separate the robust flavor signal of incoherent images from the more demanding quantum correlations of an overlapping image pair.

\begin{figure}[!htbp]
\centering
\makebox[\columnwidth][l]{
    \hspace*{-0.85cm}
    \includegraphics[width=1.2\columnwidth]{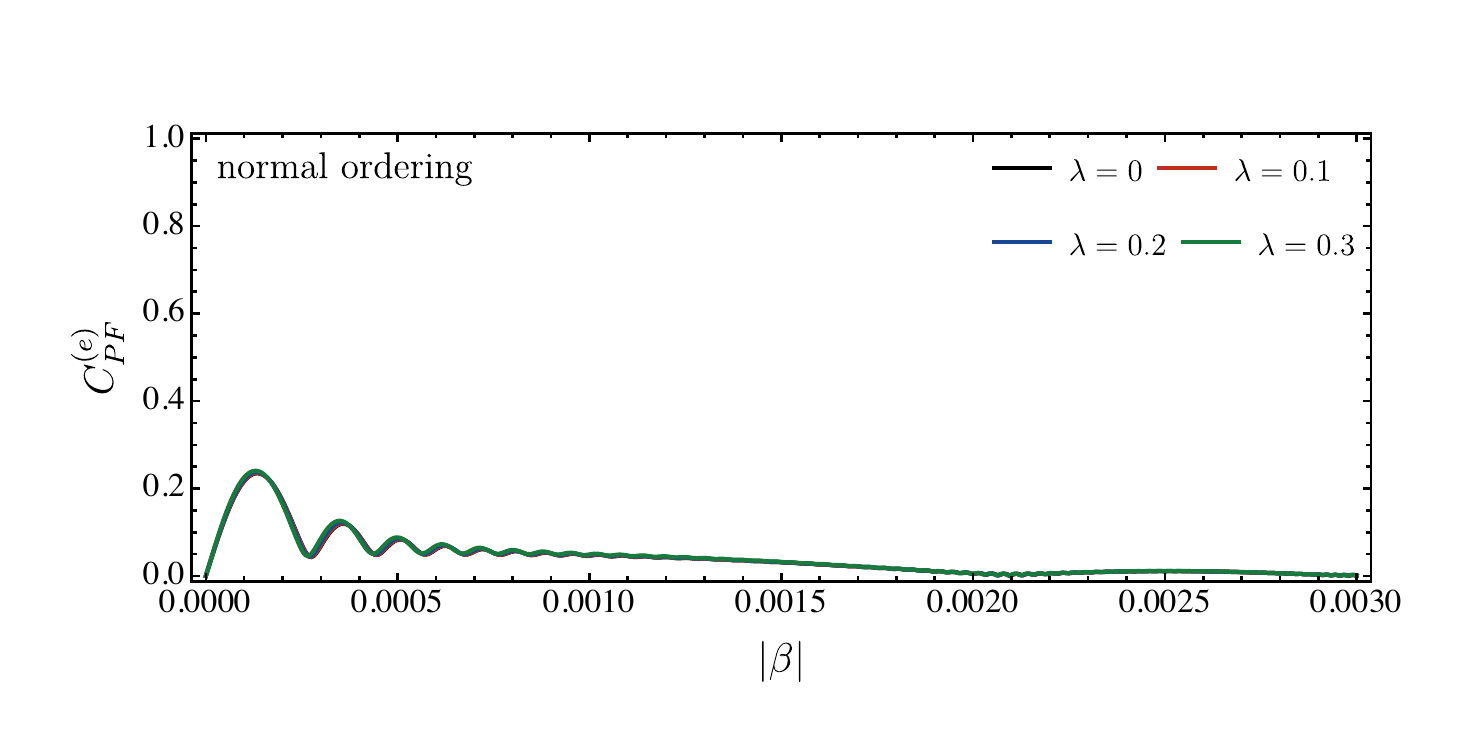}}
\caption{Trajectory--flavor I--concurrence $C_{PF}^{(e)}$ for an initially electronic neutrino with normal mass ordering. Both lensed trajectories are retained coherently.}
\label{fig:holonomy_concurrence_NO}
\end{figure}

The normal ordering concurrence in Fig.~\ref{fig:holonomy_concurrence_NO} vanishes in the axial limit and rises to a principal maximum of approximately $0.23$ at $|\beta|\simeq1.3\times10^{-4}$. The subsequent peaks decay as the negative parity image loses weight, although a low amplitude tail persists until the first broad zero near $|\beta|\simeq2.0\times10^{-3}$. This profile follows from the competition displayed in Eq.~\eqref{eq:explicit_holonomy_path_flavor_concurrence}. Close to alignment, $b_{+}=b_{-}$ and the two rays transport the same flavor state, so equal image weights do not produce entanglement. A small displacement separates the path dependent mass phases while both magnifications remain appreciable. At larger $|\beta|$, the factor $w_{+}w_{-}$ decreases and eventually dominates over the continuing phase oscillations.

The four holonomy curves in Fig.~\ref{fig:holonomy_concurrence_NO} almost coincide. However, this does not mean that the phases are insensitive to $\lambda$, as Figs.~\ref{fig:holonomy_probability_NO} and \ref{fig:holonomy_probability_IO} demonstrate. The concurrence depends on the modulus of the flavor overlap and on the product of the image weights. A holonomy induced phase displacement may be partly compensated by the simultaneous movement and reweighting of the images.

\begin{figure}[!htbp]
\centering
\makebox[\columnwidth][l]{
    \hspace*{-1.4cm}
    \includegraphics[width=1.15\columnwidth]{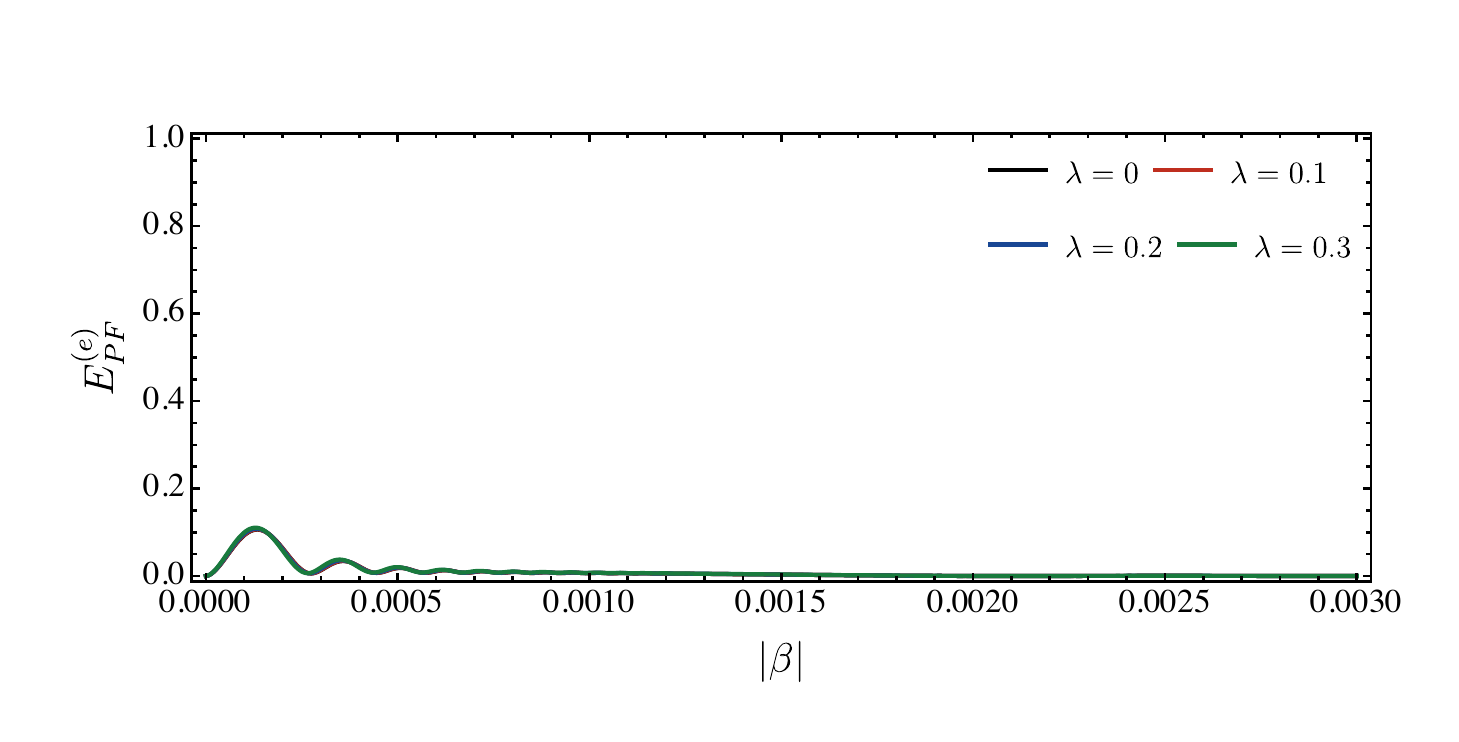}}
\caption{Entropy of trajectory--flavor entanglement for normal ordering and the same coherent image pairs used in Fig.~\ref{fig:holonomy_concurrence_NO}.}
\label{fig:holonomy_entropy_NO}
\end{figure}

The entropy in Fig.~\ref{fig:holonomy_entropy_NO} reproduces the concurrence extrema through the Schmidt spectrum in Eq.~\eqref{eq:path_flavor_schmidt_spectrum}. Its first maximum is close to $0.10$ bit, followed by a rapidly damped sequence of smaller peaks. The result remains well below the one bit value available to a maximally entangled $2\times3$ pure state. The lens does not realize the two necessary conditions simultaneously: the image weights are most balanced near alignment, whereas orthogonality of the transported flavor states requires a finite difference between the path phases.

\begin{figure}[!htbp]
\centering
\makebox[\columnwidth][l]{
    \hspace*{-1.35cm}
    \includegraphics[width=1.15\columnwidth]{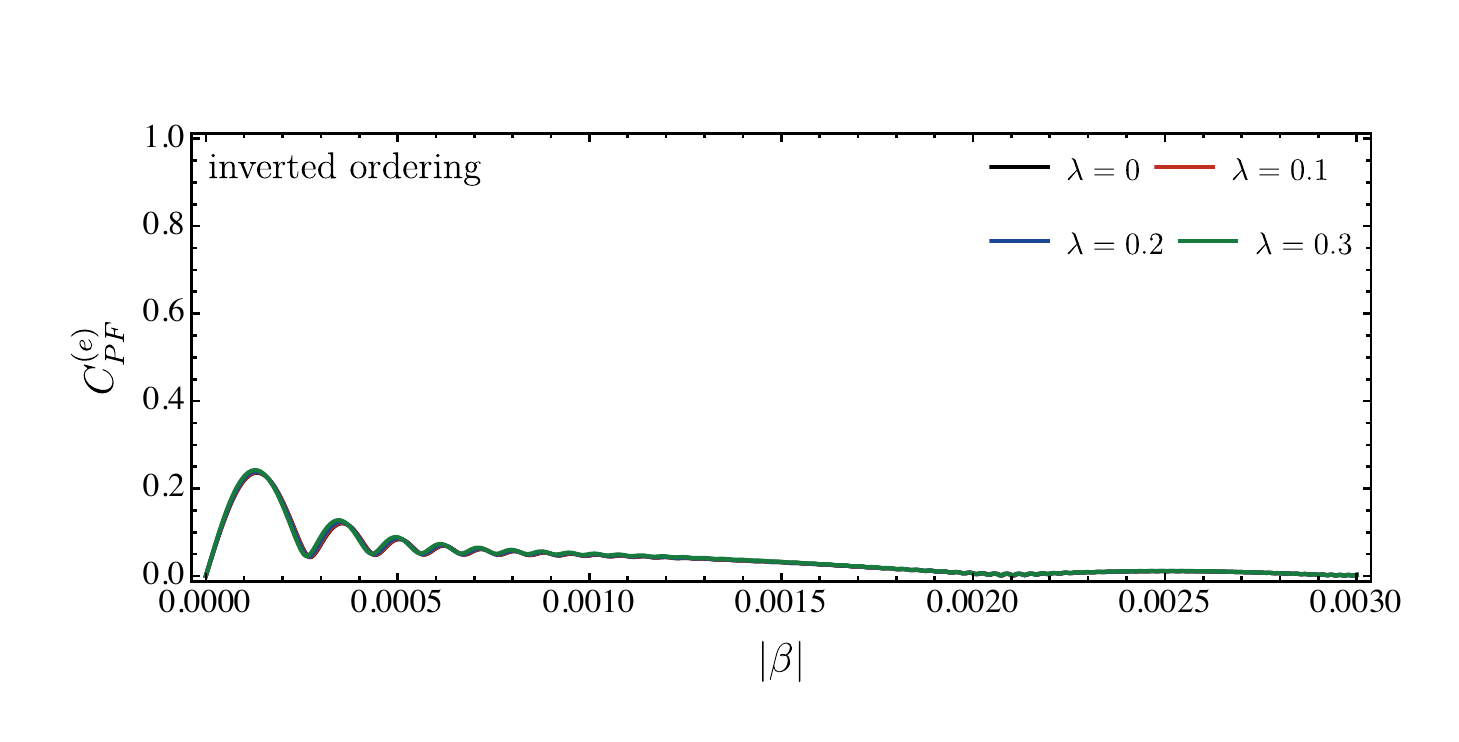}}
\caption{Electron--neutrino trajectory--flavor concurrence for inverted ordering and the holonomy grid of Eq.~\eqref{eq:holonomy_numerical_lambda_values}.}
\label{fig:holonomy_concurrence_IO}
\end{figure}

The IO concurrence in Fig.~\ref{fig:holonomy_concurrence_IO} is visually very close to the NO result. This near degeneracy has a direct origin. For an initially electronic neutrino, $\kappa_{e}$ is controlled by the weights $|U_{ei}|^{2}$; these depend on $\theta_{12}$ and $\theta_{13}$ but not on $\theta_{23}$ or $\delta_{\mathrm{CP}}$. The two orderings in Eqs.~\eqref{eq:holonomy_numerical_NO_inputs} and \eqref{eq:holonomy_numerical_IO_inputs} use the same $\theta_{12}$, nearly equal $\theta_{13}$, and closely spaced atmospheric mass splittings. In this manner, their largest differences in $\theta_{23}$ and $\delta_{\mathrm{CP}}$ affect the channel probabilities more strongly than $C_{PF}^{(e)}$.

\begin{figure}[!htbp]
\centering
\makebox[\columnwidth][l]{
    \hspace*{-1.2cm}
    \includegraphics[width=1.15\columnwidth]{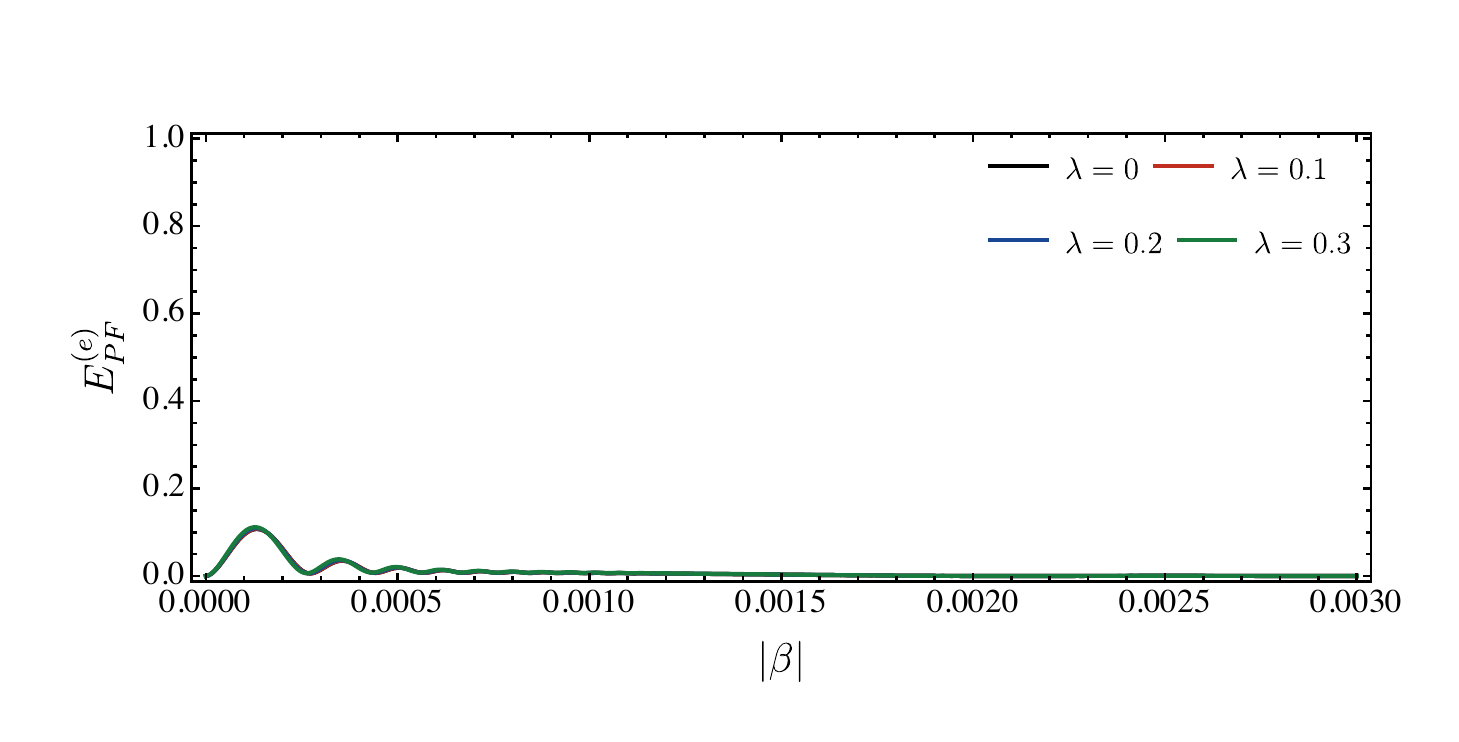}}
\caption{Trajectory--flavor entanglement entropy for inverted ordering.}
\label{fig:holonomy_entropy_IO}
\end{figure}

Fig.~\ref{fig:holonomy_entropy_IO} carries this degeneracy into the entropy. Its leading peak again approaches $0.10$ bit, and neither the ordering change nor the interval $0\leq\lambda\leq0.3$ produces a visible separation among most curves. The entropy and concurrence are nevertheless useful as coherence features: a nonzero value proves that the path labels and the internal flavor state cannot be described independently. An important remark is in order: their weak response to $\lambda$ in the present setup should not be confused with an absence of holonomy in the joint state.

\begin{figure}[!htbp]
\centering
\makebox[\columnwidth][l]{
    \hspace*{-1.2cm}
    \includegraphics[width=1.15\columnwidth]{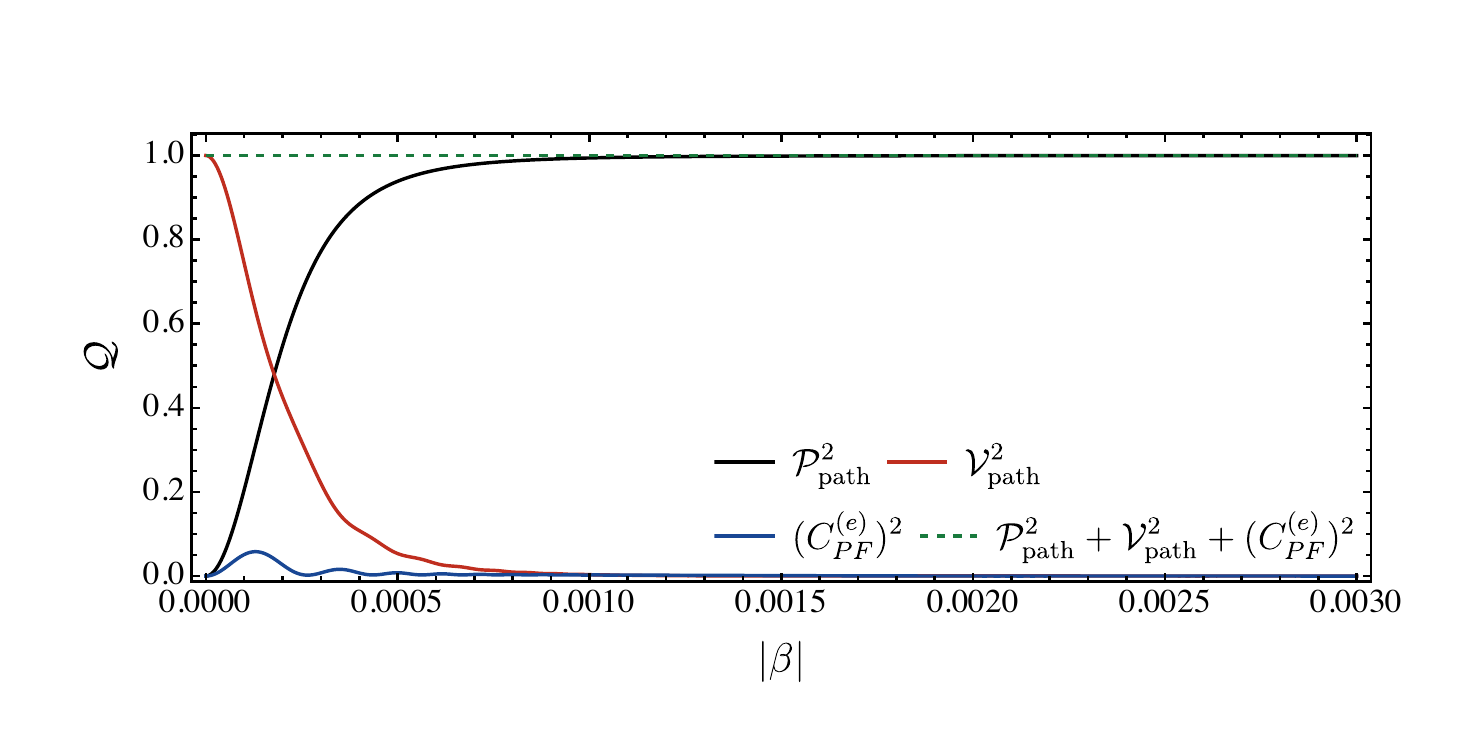}}
\caption{Squared path predictability, squared visibility, squared trajectory--flavor concurrence, and their sum for normal ordering at $\lambda=0.3$.}
\label{fig:holonomy_complementarity}
\end{figure}

The complementarity balance is resolved in Fig.~\ref{fig:holonomy_complementarity}. In the alignment limit, the image weights are equal and the transported flavor states coincide; then, $\mathcal{P}_{P}^{2}\rightarrow0$, $\mathcal{V}_{P}^{2}\rightarrow1$, and $(C_{PF}^{(e)})^{2}\rightarrow0$. As $|\beta|$ grows, the lens increasingly selects one path: predictability approaches unity and visibility falls toward zero. The squared concurrence forms a small intermediate peak, reaching about $0.05$ where neither the path imbalance nor the internal state overlap is dominant. The dashed curve remains fixed at unity across the entire interval, numerically saturating Eq.~\eqref{eq:holonomy_path_flavor_triality} \cite{Englert1996,JakobBergou2007}. In this case, reduced visibility in a coherent experiment cannot be assigned to decoherence alone; part of it may have been converted into which path information or stored as trajectory--flavor entanglement.

\begin{figure}[!htbp]
\centering
\makebox[\columnwidth][l]{
    \hspace*{-1.35cm}
    \includegraphics[width=1.15\columnwidth]{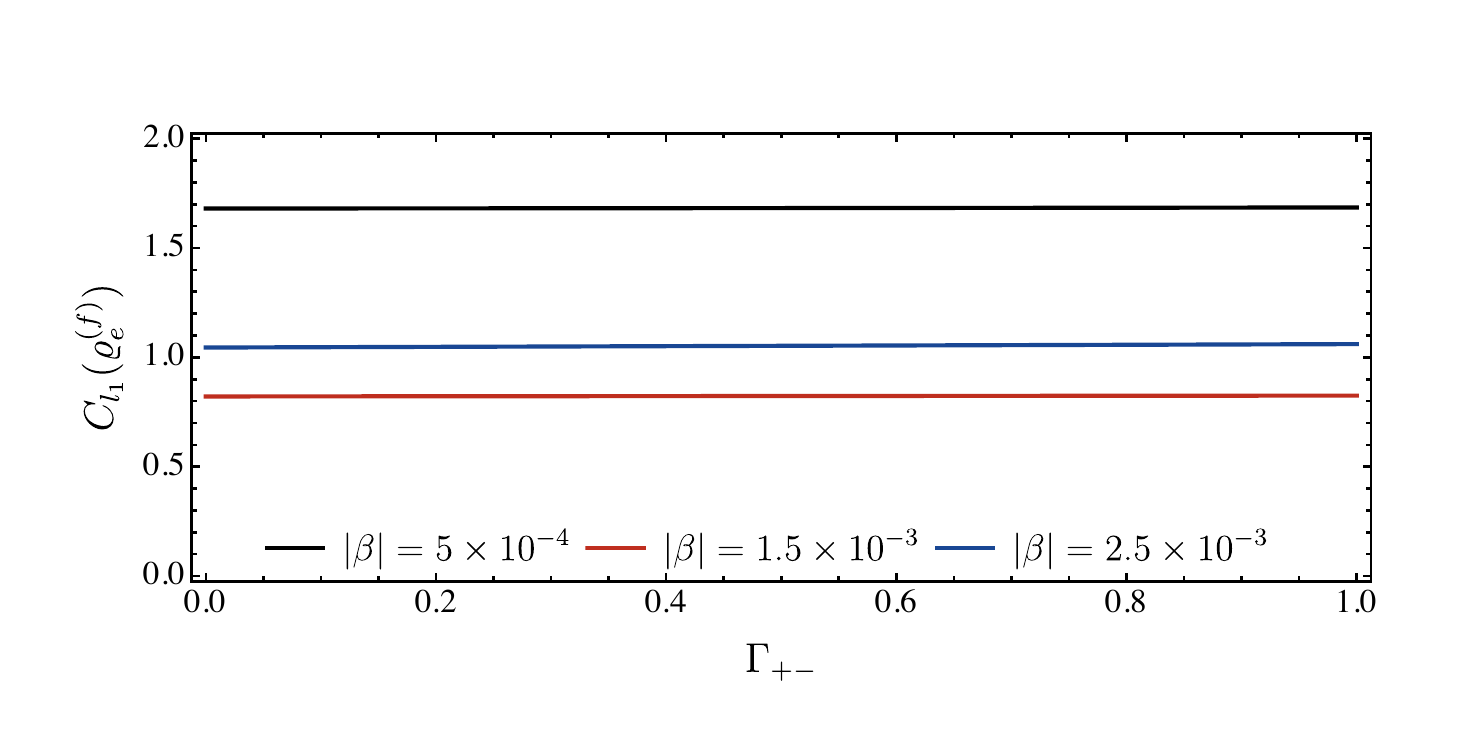}}
\caption{$l_{1}$ coherence of the reduced flavor state as a function of $\Gamma_{+-}$ for $|\beta|=5\times10^{-4}$, $1.5\times10^{-3}$, and $2.5\times10^{-3}$, with normal ordering and $\lambda=0.3$.}
\label{fig:holonomy_l1_coherence}
\end{figure}

Fig.~\ref{fig:holonomy_l1_coherence} interpolates between distinguishable and fully overlapping images. For the three selected angles, $C_{l_{1}}(\varrho_{e}^{(f)})$ remains near $1.67$, $0.80$, and $1.05$, respectively, and changes only slightly as $\Gamma_{+-}$ varies from zero to unity. The dominant difference is produced by the source position. The diagonal path contributions already carry coherent mass superpositions, while the rapidly phased off diagonal terms modify the density matrix and its normalization together. This numerical behavior is specific to the chosen lens and energy. It also reinforces a conceptual distinction: flavor mode coherence can remain large when the joint path--flavor state is separable, and it cannot by itself certify trajectory--flavor entanglement \cite{BlasoneEtAl2008,BittencourtEtAl2024}.

The annihilation calculation explores a broader holonomy interval,
\begin{equation} 
\begin{split}
& \lambda\in\{0,0.5,1,2\}, \qquad \frac{a}{2M}\in\{0,0.2,0.5,0.8\}, \\
&0.01\leq\frac{M}{R_{\nu}}\leq\frac{1}{3}. \label{eq:holonomy_deposition_numerical_grid} 
\end{split}
\end{equation}
Here $R_{\nu}$ denotes the radius of the effective neutrinosphere. The lapse $F(r)$, the Tolman temperature factor, and the angular aperture of the emitting surface retain their Schwarzschild forms. Holonomy enters the integrated rate through the radial proper volume factor $H(r)^{-1/2}=(1-a/r)^{-1/2}$ in Eq.~\eqref{eq:holonomy_deposition_integral}. Unlike a constant radial rescaling, this contribution varies over the integration domain and becomes stronger as the emitting surface approaches the compact object \cite{SalmonsonWilson1999,LambiaseMastrototaro2020}.

\begin{figure}[!htbp]
\centering
\makebox[\columnwidth][l]{
    \hspace*{-1.2cm}
    \includegraphics[width=1.15\columnwidth]{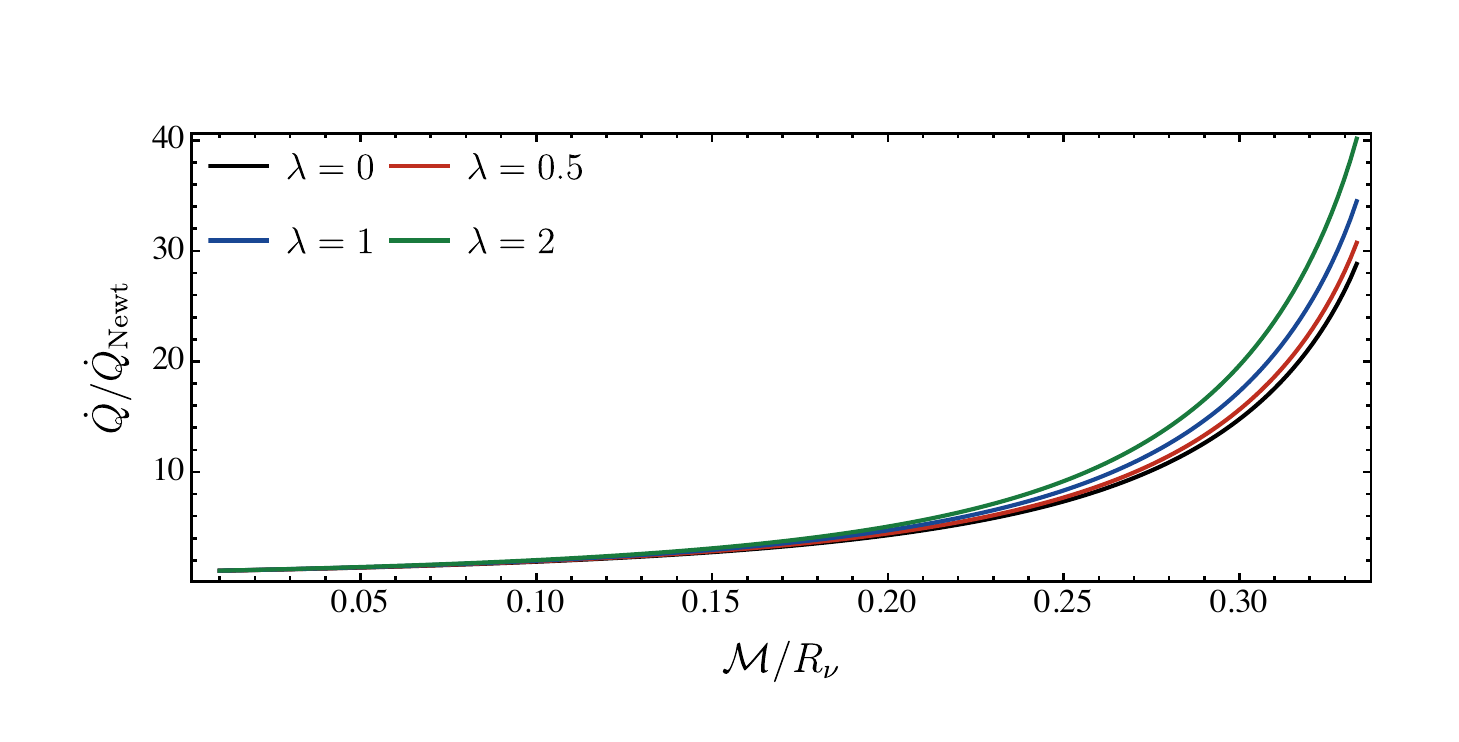}}
\caption{Neutrino--antineutrino annihilation power relative to the Newtonian rate as a function of the compactness $M/R_{\nu}$ for four values of $\lambda$.}
\label{fig:holonomy_deposition_newtonian}
\end{figure}

Fig.~\ref{fig:holonomy_deposition_newtonian} shows that curvature increases the deposited power even in the Schwarzschild limit and that the radial holonomy correction reinforces this enhancement. At $M/R_{\nu}=0.20$, the four rates $\dot{Q}/\dot{Q}_{\mathrm{Newt}}$ for $\lambda=0,0.5,1,2$ are approximately $4.27$, $4.43$, $4.71$, and $5.06$. As the compactness approaches $1/3$, they rise to approximately $28.82$, $30.73$, $34.49$, and $40.15$. The separation is not uniform: it is modest for a dilute source and widens rapidly in the compact regime because $a/(yR_{\nu})$ grows throughout the region that contributes most strongly to the integral.

\begin{figure}[!htbp]
\centering
\makebox[\columnwidth][l]{
    \hspace*{-1.2cm}
    \includegraphics[width=1.15\columnwidth]{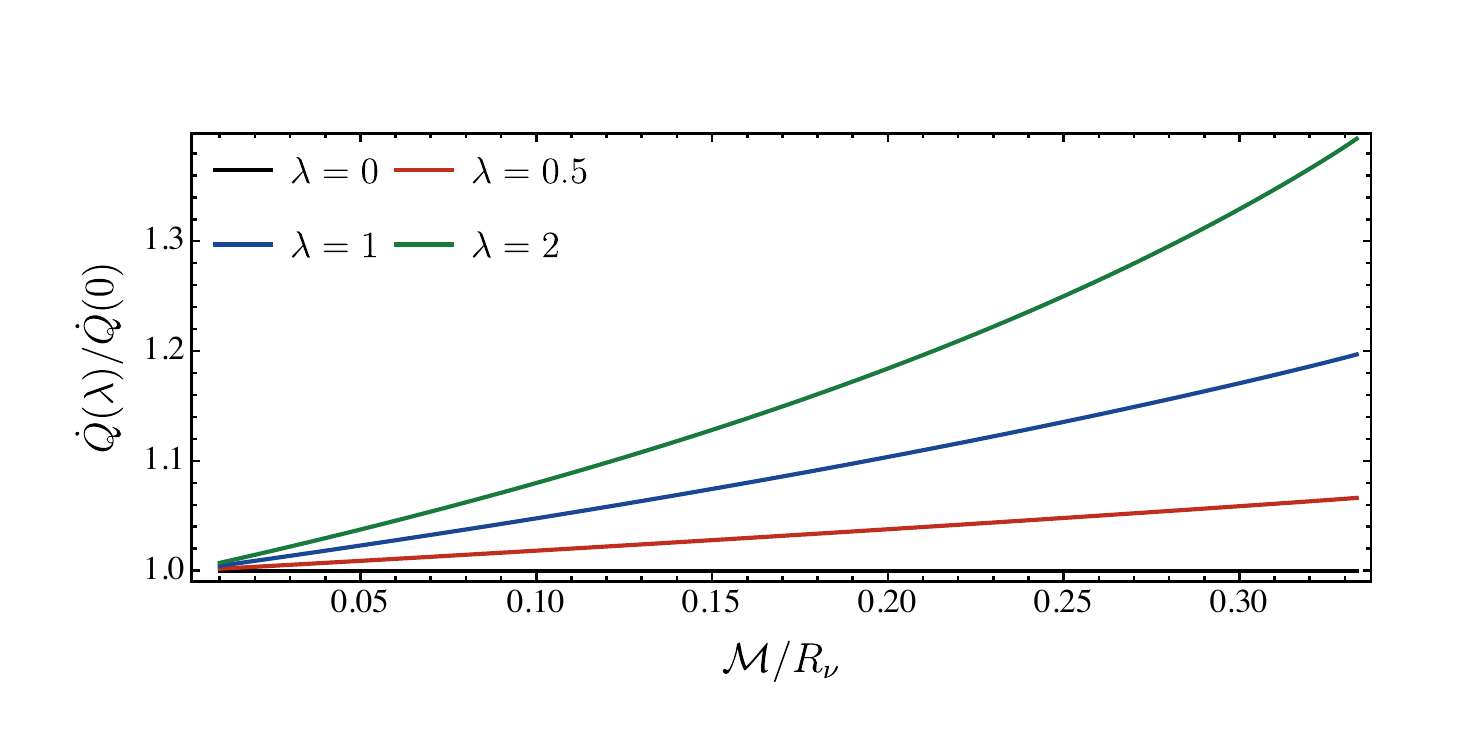}}
\caption{Holonomy enhancement of the annihilation power relative to the Schwarzschild value, $\dot{Q}(\lambda)/\dot{Q}(0)$, for the same compactness and parameter ranges as in Fig.~\ref{fig:holonomy_deposition_newtonian}.}
\label{fig:holonomy_deposition_schwarzschild}
\end{figure}

The purely holonomy driven part is isolated in Fig.~\ref{fig:holonomy_deposition_schwarzschild}. At $M/R_{\nu}=0.20$, the corrections for $\lambda=0.5,1,2$ are approximately $3.78\%$, $10.36\%$, and $18.41\%$. At $M/R_{\nu}=1/3$, the corresponding enhancements become $6.63\%$, $19.69\%$, and $39.33\%$. These values confirm the strict inequality in Eq.~\eqref{eq:holonomy_schwarzschild_ratio}: for $0<a<2M$, the positive weight of every radial shell is multiplied by a factor larger than unity. They also show why the annihilation observable contains information absent from a constant normalization. Its response depends jointly on $\lambda$ and compactness, giving the family of curves a changing separation.

\begin{figure}[!htbp]
\centering
\makebox[\columnwidth][l]{
    \hspace*{-0.85cm}
    \includegraphics[width=1.2\columnwidth]{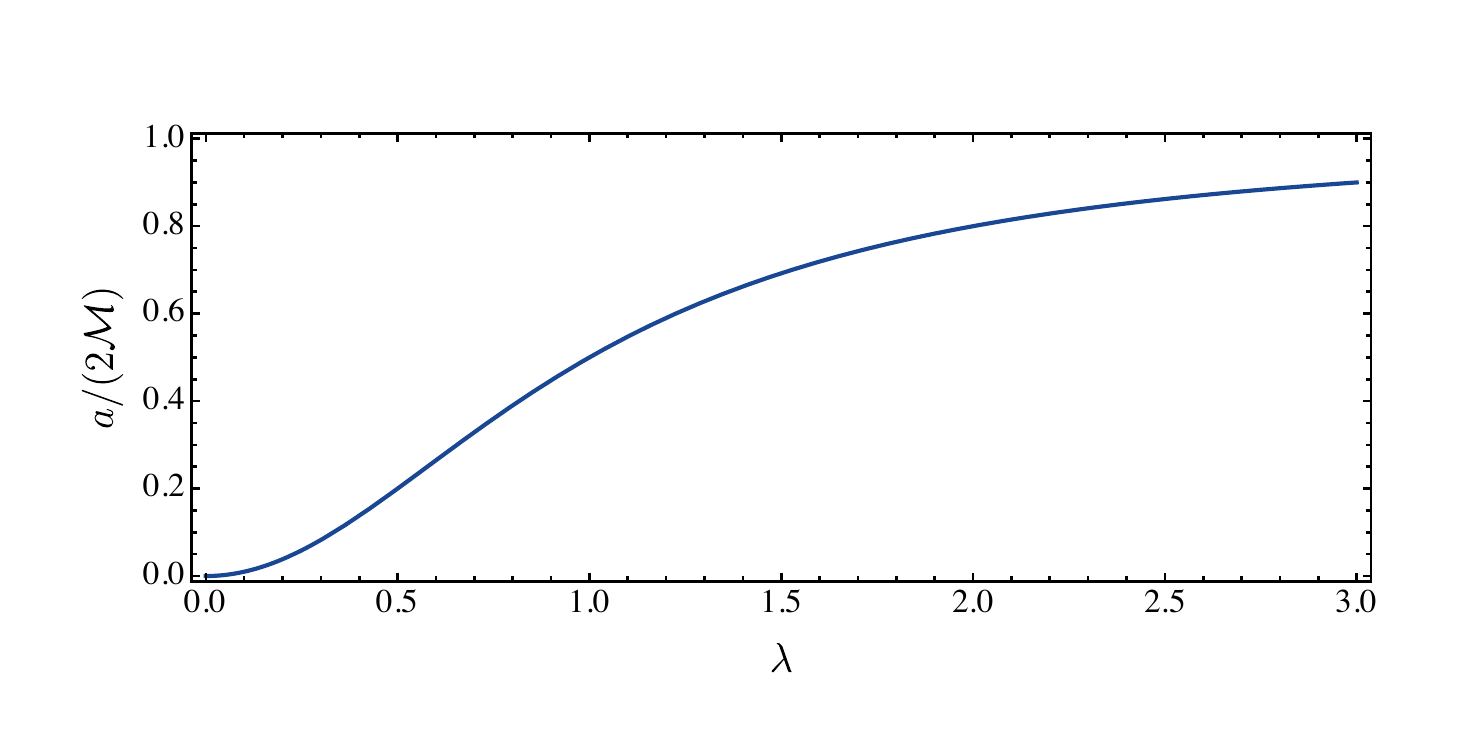}}
\caption{Bounded holonomy ratio $a/(2M)=\lambda^{2}/(1+\lambda^{2})$ as a function of $\lambda$.}
\label{fig:holonomy_parameter_map}
\end{figure}

Fig.~\ref{fig:holonomy_parameter_map} makes the nonlinear parameter map explicit. For $\lambda\ll1$, we have $a/(2M)\simeq\lambda^{2}$, whereas $a/(2M)$ approaches unity as $\lambda$ becomes large. The representative values $\lambda=0.5$, $1$, $2$, and $3$ correspond to $a/(2M)=0.2$, $0.5$, $0.8$, and $0.9$. Therefore, equal increments in $\lambda$ do not represent equal displacements of the geometry. The saturation also guarantees $a<2M$ for every finite $\lambda$, placing the zero of $H(r)$ at or inside the Schwarzschild horizon while leaving the exterior region regular in the parameter range considered \cite{AlonsoBardaji2022PLB,AlonsoBardaji2022PRD}.

Finally, the numerical results expose three complementary responses to the same deformation. Flavor probabilities are phase sensitive: the logarithmic term in the phase length and the displaced image geometry translate the fringes without producing a universal change in their envelope. Trajectory--flavor entanglement is more selective, because it retains only the internal state distinguishability weighted by the balance of the two images; in the present electron--neutrino case, this suppresses most of the visible $\lambda$ dependence and nearly removes the ordering dependence. The annihilation power is nonoscillatory and grows monotonically with $a$, acquiring its strongest relative enhancement near the largest compactness.

%%%%%%%%%%%%%%%%%%%%%%%%%%%%%%%%%%%%%%%%%%%%%%%%%%%%%%%%%%%%%%%%%%%%%%%%%%%%%%%%%%%%%%%%%%%%%%%%%%%%%%%%%%%%%%%%%%%%%%%%%%%%%%%%%%%%%%%%%%%%%%%%%%%%%%%%%%%%%%%%%%%%%%%%%%%%%%%%%%%%%%%%%%%%%%%%%%%%%%%%%%%%%%%%%%%%%%%%%%%%%%%%%%%%%%%%%%%%%%%%%%%%%%%%%%%%%%%%%%%%%%%%%%%%%%%%%%%%%%%%%%%%%%%%%%%%%%%%%%%%%%%%%%%%%%%%%%%%%%%%%%%%%%%%%%%%%%%%%%%%%%%%%%%%%%%%%%%%%%%%%%%%%%%%%%%%%%%%%%%%%%%%%%%%%%%%%%%%%%%%%%%%%%%%%%%%%%%%%%%%%%%%%%%%%%%%%%%%%%%%%%%%%%%%%%%%%%%%%%%%%%%%%%%%%%

\section{Neutrino bounds on the holonomy parameter}
\label{hb:section}

The neutrino observables derived above provide two different routes to the holonomy parameter: the displacement of an oscillation phase and the increase of the volume integrated annihilation power. Their conversion into a bound requires a specified uncertainty on the measured quantity and independent control of the source geometry. Since the numerical configurations considered here are not fitted to an observed neutrino source, we derive conditional upper limits and identify the assumptions under which they could become observational constraints.

We retain the parametrization 
\begin{equation} 
\delta=\frac{a}{2M} = \frac{\lambda^2}{1+\lambda^2}, \qquad 0\leq\delta<1. 
\label{hb:parameters} 
\end{equation} 
Here, $M$ is the parameter appearing in $F(r)=1-2M/r$. For this metric, the ADM mass is $M_{\mathrm{ADM}}=M+a/2$, whereas $M$ is the asymptotic Komar mass~\cite{Moreira2023}; an external mass measurement must therefore be matched to the appropriate definition. All comparisons below hold $M$ fixed. A limit $\delta\leq\delta_{\max}<1$ implies \begin{equation} 
a\leq 2M\delta_{\max}, \qquad |\lambda|\leq\sqrt{\frac{\delta_{\max}}{1-\delta_{\max}}}. 
\label{hb:conversion} 
\end{equation} 
The restriction $a<2M$ specifies the model domain; it is not an empirical neutrino bound. We use $G=\hbar=c=1$ except in numerical conversions.

The annihilation channel admits a direct inversion because its holonomy response is monotonic. Let $\mathcal C=M/R_\nu$ and define an average with the positive Schwarzschild weight obtained above, 
\begin{align} 
\langle f\rangle_{\mathcal C} &=\frac{\int_1^\infty \mathcal W_{\mathcal C}(y)f(y)\,dy} {\int_1^\infty \mathcal W_{\mathcal C}(y)\,dy}, 
\label{hb:average}\\ 
\mathcal W_{\mathcal C}(y) &=\frac{y^2(1-x)^4(x^2+4x+5)}{(1-2\mathcal C/y)^5}, \nonumber\\ 
x^2&=1-\frac{1-2\mathcal C/y}{y^2(1-2\mathcal C)}.
\nonumber 
\end{align} 
We take the nonnegative root for $x$ and $0<\mathcal C\leq1/3$, as in the spherical emission model used in the numerical analysis. At fixed $M$, $R_\nu$, and $L_\infty$, the exact enhancement is \begin{equation} \mathcal R_Q(\delta;\mathcal C) \equiv\frac{\dot Q(a)}{\dot Q(0)} =\left\langle \left(1-\frac{2\mathcal C\delta}{y}\right)^{-1/2} \right\rangle_{\mathcal C}. \label{hb:ratio} \end{equation} Its derivative satisfies 
\begin{equation} 
\frac{\partial\mathcal R_Q}{\partial\delta} =\mathcal C\left\langle \frac{1}{y}\left(1-\frac{2\mathcal C\delta}{y}\right)^{-3/2} \right\rangle_{\mathcal C}>0.
\label{hb:monotonic} 
\end{equation} 
In this manner, an allowed fractional excess $\mathcal R_Q-1\leq\epsilon_Q$ gives a unique upper endpoint through \begin{equation} \mathcal R_Q(\delta_{\max};\mathcal C)=1+\epsilon_Q, \label{hb:inversion} \end{equation} provided $1+\epsilon_Q<\mathcal R_Q(1^-;\mathcal C)$. Otherwise, the measurement does not restrict the physical interval $0\leq\delta<1$.

An analytic estimate follows by introducing $\kappa_n(\mathcal C)=\langle y^{-n}\rangle_{\mathcal C}$: \begin{equation} \mathcal R_Q =1+\mathcal C\kappa_1\delta +\frac{3}{2}\mathcal C^2\kappa_2\delta^2 +\mathcal O(\delta^3). \label{hb:expansion} \end{equation} For a small allowed excess, $\delta_{\max}\simeq\epsilon_Q/(\mathcal C\kappa_1)$ and $a_{\max}\simeq2R_\nu\epsilon_Q/\kappa_1$. Convexity also gives a conservative upper bound without truncating the holonomy factor: \begin{align} \mathcal R_Q&\geq(1-2\mathcal C\delta\kappa_1)^{-1/2}, \nonumber\\ \delta&\leq \frac{1-(1+\epsilon_Q)^{-2}}{2\mathcal C\kappa_1}, \label{hb:jensen} \end{align} with the model restriction $\delta<1$ imposed separately.

The numerical bounds follow by inverting the exact integral, without using the small-$\delta$ expansion. At $\mathcal C=1/3$, we find $\kappa_1=0.902641$ and $\mathcal R_Q(0.8;1/3)=1.393274$, reproducing the enhancement at $\lambda=2$. Allowing a $10\%$ excess then yields \begin{equation} \begin{aligned} \delta&\leq0.2879,\qquad |\lambda|\leq0.6358,\\ a&\leq0.8502\,\frac{M}{M_\odot}\,\mathrm{km}. \end{aligned} \label{hb:benchmark} \end{equation} At the same compactness, allowances of $1\%$, $5\%$, and $20\%$ give, respectively, $\delta_{\max}=0.03274$, $0.15436$, and $0.50595$, with $|\lambda|_{\max}=0.18397$, $0.42725$, and $1.01198$. For the metric parameter $M=M_\odot$, these correspond to $a_{\max}=0.09668$, $0.45587$, and $1.49421\,\mathrm{km}$; the length limits scale linearly with $M$. Reducing the compactness to $\mathcal C=1/5$ weakens the $5\%$ and $10\%$ limits to $\delta_{\max}=0.26019$ and $0.48513$, or $|\lambda|_{\max}=0.59303$ and $0.97070$, respectively. Their corresponding lengths are $0.76839(M/M_\odot)\,\mathrm{km}$ and $1.43272(M/M_\odot)\,\mathrm{km}$. These bounds are conditional on fixed source parameters and the stated fractional allowances; no observational error or confidence level is assigned to those allowances.

The decrease in sensitivity at lower compactness eventually prevents a useful restriction of the holonomy interval. For $\mathcal C=0.1$, even the limit $\delta\to1$ increases the power by only $10.31\%$; an allowance of $20\%$ gives no additional bound. The enhancement approaches $59.80\%$ at $\mathcal C=1/3$ and $\delta\to1$. Thereby, the $39.3\%$ increase displayed at $\lambda=2$ is not the maximum permitted by the model.

The comparison with an astrophysical energy budget requires the same definition of power on both sides. The quantity $\dot Q$ above integrates the local deposition rate over proper volume, following the annihilation calculation~\cite{SalmonsonWilson1999,LambiaseMastrototaro2020}. For deposition at rest in the static frame, the corresponding rate of Killing energy is instead 
\begin{equation} 
\dot Q_{\mathrm{dep},\infty} = \int F(r)\dot q(r)\,\mathrm{d}V_{\mathrm{prop}}. 
\label{hb:powerinfinity} 
\end{equation} 
The two factors of $\sqrt{F}$ account for energy redshift and clock rate. For this quantity, Eqs.~\eqref{hb:ratio}--\eqref{hb:jensen} apply with $\mathcal W_{\mathcal C}\to(1-2\mathcal C/y)\mathcal W_{\mathcal C}$, and the numerical bounds change accordingly. For example, at $\mathcal C = 1/3$ the same $10\%$ allowance on the ratio of Killing-energy deposition rates gives $\delta\leq0.2929$, or $a\leq0.8650(M/M_\odot)\,\mathrm{km}$. An escaping jet luminosity additionally requires the escape fraction, outflow dynamics, and radiative efficiency. Calculations of neutrino powered outflows show why these source properties cannot be eliminated by identifying the deposited power with the observed gamma ray luminosity~\cite{LengGiannios2014,JustEtAl2016}.

The phase channel probes the same radial deformation through the accumulated mass dependent phase. For an outward radial trajectory with independently specified areal endpoints $r_D>r_S\gg a$, the leading holonomy displacement is 
\begin{equation} 
\begin{aligned} 
\Delta\Phi_{ij}^{\mathrm{rad}}(a) -\Delta\Phi_{ij}^{\mathrm{rad}}(0) &=\frac{\Delta m_{ij}^2a}{4E_0}\ln\!\left(\frac{r_D}{r_S}\right)\\ &\quad+\mathcal O\!\left(\frac{|\Delta m_{ij}^2|a^2}{E_0r_S}\right). 
\end{aligned} 
\label{hb:radialshift} 
\end{equation} 
If a resolved spectrum limits this displacement to $\epsilon_\Phi$ radians on a phase, which is continuously connected to Schwarzschild, namely, 
\begin{equation} 
a_{\max}^{\mathrm{phase}} 
\simeq\frac{4E_0\epsilon_\Phi} {|\Delta m_{ij}^2|\ln(r_D/r_S)}. 
\label{hb:phasebound} 
\end{equation} 
For illustration, $E_0=10\,\mathrm{MeV}$, $|\Delta m_{31}^2|=2.517\times10^{-3}\,\mathrm{eV}^2$, $r_D/r_S=10^5$, and an assumed $\epsilon_\Phi=0.1$ give $a_{\max}^{\mathrm{phase}}\simeq27.2\,\mathrm m$. This is a radial sensitivity part, distinct from the two image configuration of the numerical plots. A probability measured at one energy does not provide this bound because of oscillation periodicity and its degeneracy with the baseline and mass splitting. Moreover, the cancellation of $H(r)$ in the local proper oscillation length excludes a direct identification of laboratory oscillation precision with a limit on $a$.

For the lensed signal, the impact parameters must also vary with $a$. Differentiating the weak--field phase length at fixed $M$, $r_S$, $r_D$, and source angle gives 
\begin{equation} 
\left. \frac{\mathrm{d}\Lambda_p}{\mathrm{d}a}\right|_0 = \frac12 \ln\!\left(\frac{4r_Sr_D}{b_{p0}^2}\right) - \frac{(r_S + r_D)  b_{p0}}{r_Sr_D} \left. \frac{\mathrm{d}b_p}{\mathrm{d}a} \right|_0, 
\label{hb:lensresponse} 
\end{equation} 
where $b_{p0} = b_p(a=0)$ is obtained from the lens equation. The same fit must include changes in magnification and image overlap. At leading deflection order, lensing measures $4M+a$, which makes an independent determination of $M$ necessary~\cite{Soares2023}.

Energy averaging supplies a stronger restriction for the long baseline used in the plots. In a narrow energy bin, a Gaussian distribution of true energies with width $\sigma_E$ suppresses an oscillatory factor by \begin{equation} 
\begin{aligned} 
\left\langle e^{i\Delta\Phi_{ij}}\right\rangle_E &\simeq e^{i\Delta\Phi_{ij}(E_0)}\\ &\quad\times\exp\!\left[-\frac12 \left(\frac{\sigma_E}{E_0}\Delta\Phi_{ij}(E_0)\right)^2\right]. 
\end{aligned} 
\label{hb:averaging} 
\end{equation} 
The condition for retaining contrast is $(\sigma_E/E_0)|\Delta\Phi_{ij}|\lesssim1$. With $r_S=10^5\,\mathrm{AU}$, $r_D=1\,\mathrm{AU}$, and $E_0=10\,\mathrm{MeV}$, the leading atmospheric phase is approximately $9.54\times10^{12}$, requiring $\sigma_E/E_0\lesssim1.05\times10^{-13}$. Wave packet separation imposes an independent condition. The coherence length quoted above requires an effective spatial width $\sigma_x\gtrsim6.66\,\mathrm{cm}$ for this atmospheric mode and baseline~\cite{GiuntiKimLee1998}. For comparison, microscopic production estimates for supernova neutrinos give widths of order $10^{-11}\,\mathrm{cm}$~\cite{KerstenSmirnov2016}. These conditions show why the monochromatic curves alone cannot support an observational phase bound.

When the images are mutually incoherent and the mass oscillations along each image are fully averaged, the normalized flavor probability becomes \begin{equation} 
\overline P_{\alpha\to\beta}^{\mathrm{inc}}  = \sum_i|U_{\alpha i}|^2|U_{\beta i}|^2, 
\label{hb:averagedprobability} 
\end{equation} independent of $a$ for the same emitted flavor state on both paths. Lensing may still change the total flux. Coherent crosspath interference requires a separate assessment of packet overlap and the energy variation of the full Fermat and mass phase~\cite{CrockerGiuntiMortlock2004}. The trajectory--flavor concurrence and entropy do not furnish additional empirical limits without measurements retaining the relevant path coherence; their weak variation in the configuration studied here further limits their parameter sensitivity.

The mixing parameters entering a future source fit can be constrained with reactor data. JUNO finds $\Delta m_{21}^2 =(7.50\pm0.12) \times 10^{-5}\,\mathrm{eV}^2$ and $\sin^2\theta_{12} = 0.3092\pm0.0087$~\cite{JUNO2026FirstOscillation}, while the final gadolinium capture sample of Daya Bay gives $\Delta m_{32}^2 =(2.466\pm0.060)\times10^{-3}\,\mathrm{eV}^2$ for normal ordering~\cite{DayaBay2023FinalOscillation}. These results supply external oscillation information. In particular, $\Delta m_{32}^2$ must be converted to the $\Delta m_{31}^2$ convention before it is used with the numerical inputs of this work.

The deposition bounds also require an uncertainty on the Schwarzschild prediction. At fixed $M$ and $R_\nu$, its luminosity dependence gives \begin{equation} \frac{\Delta\dot Q(0)}{\dot Q(0)} \simeq\frac94\frac{\Delta L_\infty}{L_\infty}. \label{hb:luminosityuncertainty} \end{equation} A $10\%$ luminosity uncertainty  produces a roughly $22.5\%$ normalization uncertainty already at first order. A $10\%$ allowance on the holonomy enhancement requires substantially better source calibration, including $R_\nu$, compactness, spectral shape, and the conversion from deposition to the measured energy channel. The $1\%$ allowance considered above is consequently a precision target, not an established capability of neutrino source modeling.

An observational analysis should convolve the flavor probabilities and absolute flux with the source distribution and detector response, and profile over these nuisance parameters. Writing them collectively as $\eta$, an upper endpoint would be obtained from \begin{equation} 
\Delta\chi^2(\delta) = \min_\eta\chi^2(\delta,\eta) -\min_{0\leq\delta'<1,\eta}\chi^2(\delta',\eta),
\label{hb:profile} 
\end{equation} 
with coverage calibrated for the boundary $\delta=0$ and any disconnected oscillation solutions. No confidence level can be assigned to the illustrative allowances without that likelihood. Within the present emission model, the monotonic deposition integral supplies the direct route to an upper bound; the phase channel becomes constraining only when a measurable interference pattern survives averaging and its geometric degeneracies are controlled.

%%%%%%%%%%%%%%%%%%%%%%%%%%%%%%%%%%%%%%%%%%%%%%%%%%%%%%%%%%%%%%%%%%%%%%%%%%%%%%%%%%%%%%%%%%%%%%%%%%%%%%%%%%%%%%%%%%%%%%%%%%%%%%%%%%%%%%%%%%%%%%%%%%%%%%%%%%%%%%%%%%%%%%%%%%%%%%%%%%%%%%%%%%%%%%%%%%%%%%%%%%%%%%%%%%%%%%%%%%%%%%%%%%%%%%%%%%%%%%%%%%%%%%%%%%%%%%%%%%%%%%%%%%%%%%%%%%%%%%%%%%%%%%%%%%%%%%%%%%%%%%%%%%%%%%%%%%%%%%%%%%%%%%%%%%%%%%%%%%%%%%%%%%%%%%%%%%%%%%%%%%%%%%%%%%%%%%%%%%%%%%%%%%%%%%%%%%%%%%%%%%%%%%%%%%%%%%%%%%%%%%%%%%%%%%%%%%%%%%%%%%%%%%%%%%%%%%%%%%%%%%%%%%%%%%

\section{Conclusion}

We investigated three flavor neutrino dynamics in an effective holonomy corrected Schwarzschild geometry, where the radial deformation modified propagation despite the unchanged lapse function. The weak--deflection expansion through second post--Minkowskian order showed stronger focusing for a positive holonomy scale and determined the corresponding changes in the image geometry. The exact radial phase and the leading nonradial expression revealed logarithmic contributions with explicit dependence on the propagation endpoints and impact parameters. These contributions altered the accumulated phase between fixed areal radii, while the local oscillation length retained its standard form in terms of proper distance and locally measured energy.

The two image formulation incorporated magnifications, Fermat phases, and wave packet overlap into a normalized flavor probability. Coherent cross path terms contained information about the absolute neutrino mass scale, whereas the incoherent image limit retained only the dependence on squared mass differences. For a solar mass lens and an energy of $10\,\mathrm{MeV}$, the numerical probabilities in the latter regime exhibited displaced oscillation extrema and a redistribution among the three active flavors. Over $0\leq\lambda\leq0.3$, the holonomy correction primarily shifted the fine oscillatory structure and left the broad envelope nearly intact. The mass ordering affected the division of the appearance signal between the muon and tau channels, while the total flavor probability remained unity.

The quantum correlation analysis distinguished flavor mode coherence from entanglement between trajectory and flavor. In the coherent image limit, the electron-neutrino $I$-concurrence reached approximately $0.23$, and the corresponding entropy approached $0.10$ bit. Both quantities varied weakly with the holonomy parameter and changed little between the two mass orderings in the selected configuration. Their principal maxima occurred at intermediate source offsets, where both images retained appreciable weights and transported different flavor states. The saturated predictability visibility concurrence relation accounted for the reduction of interference visibility through path imbalance and trajectory--flavor entanglement. Flavor mode coherence survived the loss of cross image overlap, even after the joint trajectory--flavor state became separable.

Neutrino--antineutrino annihilation exhibited a monotonic increase in the integrated deposition power. At fixed mass, neutrinosphere radius, and asymptotic luminosity, the thermal redshift and angular emission aperture preserved their Schwarzschild forms. The holonomy correction entered exclusively through the radial proper-volume factor $H(r)^{-1/2}$, which increased the contribution of each emitting shell without altering its local deposition density. For $\lambda=2$, the enhancement relative to Schwarzschild approached $39.3\%$ as the compactness approached $M/R_\nu=1/3$. Finally, at fixed source parameters and $M/R_\nu=1/3$, an assumed maximum excess of $10\%$ in the annihilation power yielded the conditional bounds $a/(2M)\leq0.2879$ and $|\lambda|\leq0.6358$.

%%%%%%%%%%%%%%%%%%%%%%%%%%%%%%%%%%%%%%%%%%%%%%%%%%%%%%%%%%%%%%%%%%%%%%%%%%%%%%%%%%%%%%%%%%%%%%%%%%%%%%%%%%%%%%%%%%%%%%%%%%%%%%%%%%%%%%%%%%%%%%%%%%%%%%%%%%%%%%%%%%%%%%%%%%%%%%%%%%%%%%%%%%%%%%%%%%%%%%%%%%%%%%%%%%%%%%%%%%%%%%%%%%%%%%%%%%%%%%%%%%%%%%%%%%%%%%%%%%%%%%%%%%%%%%%%%%%%%%%%%%%%%%%%%%%%%%%%%%%%%%%%%%%%%%%%%%%%%%%%%%%%%%%%%%%%%%%%%%%%%%%%%%%%%%%%%%%%%%%%%%%%%%%%%%%%%%%%%%%%%%%%%%%%%%%%%%%%%%%%%%%%%%%%%%%%%%%%%%%%%%%%%%%%%%%%%%%%%%%%%%%%%%%%%%%%%%%%%%%%%%%%%%%%%%%%%%%%%%%%%%%%%%%%%%%%%%%%%%%%%%%%%%

\section*{Acknowledgments}
\hspace{0.5cm} A. A. Araújo Filho is supported by Conselho Nacional de Desenvolvimento Cient\'{\i}fico e Tecnol\'{o}gico (CNPq) -- [150223/2025-0].

%%%%%%%%%%%%%%%%%%%%%%%%%%%%%%%%%%%%%%%%%%%%%%%%%%%%%%%%%%%%%%%%%%%%%%%%%%%%%%%%%%%%%%%%%%
\section*{Data Availability Statement}

Data associated with this study consist of the analytical expressions and numerical figures presented in the manuscript. No additional dataset is required to reproduce the analytical results. The code used to generate the numerical results and figures is available from the corresponding author upon reasonable request.

\bibliographystyle{apsrev4-2}
\bibliography{main}

\end{document}